\documentclass[%
superscriptaddress,
amsmath,amssymb,
 aps,prd,
 twocolumn,
 tightenlines,
 nofootinbib
]{revtex4-1}

\usepackage{graphicx,url,amssymb,amsmath,longtable,rotating,color,units,wasysym,subfigure,epsfig,hyperref,epstopdf}
\usepackage[dvipsnames]{xcolor}
\usepackage{enumitem}

\usepackage{graphicx}
\usepackage{dcolumn}
\usepackage{bm}
\usepackage{url,hyperref}
\usepackage{array,multirow}
\usepackage{color} 
\usepackage{aas_macros} 
\usepackage{ifthen}
\usepackage{etoolbox}

\usepackage{siunitx} 
\definecolor{darkgreen}{rgb}{0.0, 0.5, 0.0}

\newtoggle{fullauthorlist}
\toggletrue{fullauthorlist}

\newtoggle{endauthorlist}
\toggletrue{endauthorlist}

\usepackage{xr}
\begin{document}

\newcommand{\params}{\mathbf{\Theta}}


\def\UpperLimitAlphaZero{{\color{black}\ensuremath{5.8\times 10^{-9}}}}
\def\UpperLimitAlphaTwoThirds{{\color{black}\ensuremath{3.4\times 10^{-9}}}}
\def\UpperLimitAlphaThree{\ensuremath{{\color{black}3.9\times 10^{-10}}}}
\def\UpperLimitAlphaMarg{\ensuremath{{\color{black}6.6\times 10^{-9}}}}


\def\UpperLimitAlphaZeroUniform{{\color{black}\ensuremath{1.7\times 10^{-8}}}}
\def\UpperLimitAlphaTwoThirdsUniform{{\color{black}\ensuremath{1.2\times 10^{-8}}}}
\def\UpperLimitAlphaThreeUniform{\ensuremath{{\color{black}1.3\times 10^{-9}}}}
\def\UpperLimitAlphaMargUniform{\ensuremath{{\color{black}2.7\times 10^{-8}}}}


\def\UpperLimitAlphaZeroOTwoLog{\ensuremath{3.5\times 10^{-8}}}
\def\UpperLimitAlphaTwoThirdsOTwoLog{\ensuremath{3.0\times 10^{-8}}}
\def\UpperLimitAlphaThreeOTwoLog{\ensuremath{5.1\times 10^{-9}}}
\def\UpperLimitAlphaMargOTwoLog{\ensuremath{3.4\times 10^{-8}}}


\def\UpperLimitAlphaZeroOTwoUniform{\ensuremath{6.0\times 10^{-8}}}
\def\UpperLimitAlphaTwoThirdsOTwoUniform{\ensuremath{4.8\times 10^{-8}}}
\def\UpperLimitAlphaThreeOTwoUniform{\ensuremath{7.9\times 10^{-9}}}
\def\UpperLimitAlphaMargOTwoUniform{\ensuremath{1.1\times 10^{-7}}}

\def\ImprovementAlphaZero{{\color{black}6.0}}
\def\ImprovementAlphaTwoThirds{{\color{black}8.8}}
\def\ImprovementAlphaThree{{\color{black}13.1}}
\def\ImprovementAlphaMarg{{\color{black} 5.1}}

\def\ImprovementAlphaZeroUniform{{\color{black} 3.6}}
\def\ImprovementAlphaTwoThirdsUniform{{\color{black} 4.0}}
\def\ImprovementAlphaThreeUniform{{\color{black} 5.9}}
\def\ImprovementAlphaMargUniform{{\color{black} 4.1}}

\def\TLimit{{\color{black} \ensuremath{6.4 \times 10^{-9}}}}
\def\SLimit{{\color{black} \ensuremath{2.1 \times 10^{-8}}}}
\def\VLimit{{\color{black} \ensuremath{7.9 \times 10^{-9}}}}

\def\ImprovementTensor{\ensuremath{5.0}}
\def\ImprovementScalar{\ensuremath{2.9}}
\def\ImprovementVector{\ensuremath{3.7}}

\def\NonGRToGRBF{{\ensuremath{\color{black} -0.2 }}}
\def\SignalToNoiseNonGRLike{{\ensuremath{\color{black} -0.4 }}}
\def\SignalToNoiseTensorLike{{\ensuremath{\color{black} -0.3}}}


\def\PtEstErrBarAlphaZeroHL{{\color{black}\ensuremath{-2.1 \pm 8.2}}}
\def\PtEstErrBarAlphaZeroHV{{\color{black}\ensuremath{229 \pm 98}}}
\def\PtEstErrBarAlphaZeroLV{{\color{black}\ensuremath{-134 \pm 63}}}
\def\PtEstErrBarAlphaZero{{\color{black}\ensuremath{1.1 \pm 7.5}}}

\def\PtEstErrBarAlphaTwoThirdsHL{{\color{black}\ensuremath{-3.4 \pm 6.1}}}
\def\PtEstErrBarAlphaTwoThirdsHV{{\color{black}\ensuremath{145 \pm 60 }}}
\def\PtEstErrBarAlphaTwoThirdsLV{{\color{black}\ensuremath{-82 \pm 40}}}
\def\PtEstErrBarAlphaTwoThirds{{\color{black}\ensuremath{-0.2 \pm 5.6}}}

\def\PtEstErrBarAlphaThreeHL{{\color{black}\ensuremath{-1.3 \pm 0.9}}}
\def\PtEstErrBarAlphaThreeHV{{\color{black}\ensuremath{9.1 \pm 4.1}}}
\def\PtEstErrBarAlphaThreeLV{{\color{black}\ensuremath{-4.9 \pm 3.1}}}
\def\PtEstErrBarAlphaThree{{\color{black} \ensuremath{-0.6 \pm 0.8}}}

\def\DeltaFNinetyNineAlphaZeroHL{{\color{black} 76.1}}
\def\DeltaFNinetyNineAlphaTwoThirdsHL{{\color{black} 90.2}}
\def\DeltaFNinetyNineAlphaThreeHL{{\color{black} 282.8}}

\def\DeltaFNinetyNineAlphaZeroHV{{\color{black} 97.7}}
\def\DeltaFNinetyNineAlphaTwoThirdsHV{{\color{black} 117.8}}
\def\DeltaFNinetyNineAlphaThreeHV{{\color{black} 375.8}}

\def\DeltaFNinetyNineAlphaZeroLV{{\color{black} 88.0}}
\def\DeltaFNinetyNineAlphaTwoThirdsLV{{\color{black} 107.3}}
\def\DeltaFNinetyNineAlphaThreeLV{{\color{black} 388.0}}

\def\DeltaFNinetyNineAlphaZero{{\color{black} 76.6}}
\def\DeltaFNinetyNineAlphaTwoThirds{{\color{black} 90.6}}
\def\DeltaFNinetyNineAlphaThree{{\color{black} 291.6}}

\def\chiSquaredHL{{\color{black} 0.99}}
\def\chiSquaredHV{{\color{black} 0.97}}
\def\chiSquaredLV{{\color{black} 0.96}}
\def\chiSquaredAll{{\color{black} 0.98}}

\def\percentNotchedAllHL{{\color{black} 13.3\%}}
\def\percentNotchedMidFHL{{\color{black} 3.2\%}}
\def\percentNotchedAllHV{{\color{black} 21.5\%}}
\def\percentNotchedMidFHV{{\color{black} 9.3\%}}
\def\percentNotchedAllLV{{\color{black} 18.9\%}}
\def\percentNotchedMidFLV{{\color{black} 5.9\%}}

\def\percentGatedH{{\color{black} 0.4\%}} 
\def\percentGatedL{{\color{black} 1\%}} 

\def\TimeHL{{\color{black} 205.4 days}}
\def\TimeHV{{\color{black} 187.5 days}}
\def\TimeLV{{\color{black} 195.4 days}}

\def\percentLivetimeLostHL{{\color{black} 17.9\%}}
\def\percentLivetimeLostHV{{\color{black} 22.1\%}}
\def\percentLivetimeLostLV{{\color{black} 21.9\%}}

\def\CalHOOne{4.8\%}
\def\CalLOOne{5.4\%}
\def\CalHOTwo{2.6\%}
\def\CalLOTwo{3.9\%}
\def\CalH{{\color{black} 7.0\%}}
\def\CalL{{\color{black} 6.4\%}}
\def\CalV{{\color{black}5\%}}

\def\stdAlpha{{\color{black} 3.5}} 

\def\kappaH{\ensuremath{\color{black}  3.0 \times 10^{-25}\ {\rm pT^{-1}} }}
\def\kappaL{\ensuremath{\color{black}   9.0 \times 10^{-24}\ {\rm pT^{-1}} }}
\def\kappaV{\ensuremath{\color{black}   1.3 \times 10^{-24}\ {\rm pT^{-1}} }}
\def\betaH{\ensuremath{\color{black}   -2.2 }}
\def\betaL{\ensuremath{\color{black}    -4.6 }}
\def\betaV{\ensuremath{\color{black}   -1.6 }}

\def\SchumannVsNoiseLogBayesFactor{{\color{black}\ensuremath{-0.03}}}
\def\SchumannGWVsNoiseLogBayesFactor{{\color{black}\ensuremath{-0.3}}}
\def\SchumannGWVsNoiseHLLogBayesFactor{{\color{black} \ensuremath{-0.3}}}

\def\NumberOThreeADetections{{\color{black}} 39}
\def\NumberOfTotalBBHs{{\color{black}}44}

\title{Upper Limits on the Isotropic Gravitational-Wave Background from Advanced LIGO's and Advanced Virgo's Third Observing Run}
\date{\today}

\iftoggle{endauthorlist}{
  %
  %
  \let\mymaketitle\maketitle
  \let\myauthor\author
  \let\myaffiliation\affiliation
  \author{The LIGO Scientific Collaboration}
  \author{The Virgo Collaboration}
  \author{The KAGRA Collaboration}
  \email{Full author list given at the end of the article.}
\noaffiliation
}{
  %
  %
  \iftoggle{fullauthorlist}{
  }{
    \author{The LIGO Scientific Collaboration}
    \affiliation{LSC}
    \author{The Virgo Collaboration}
    \affiliation{Virgo}

  }
}

\begin{abstract}
We report results of a search for an isotropic gravitational-wave background (GWB) using data from Advanced LIGO's and Advanced Virgo's third observing run (O3) combined with upper limits from the earlier O1 and O2 runs. 
Unlike in previous observing runs in the advanced detector era, we include Virgo in the search for the GWB. 
The results of the search are consistent with uncorrelated noise, and therefore we place upper limits on the strength of the GWB. 
We find that the dimensionless energy density $\Omega_{\rm GW}\leq \UpperLimitAlphaZero$ at the $95\%$ credible level for a flat (frequency-independent) GWB, using a prior which is uniform in the log of the strength of the GWB, with 99\% of the sensitivity coming from the band 20-\DeltaFNinetyNineAlphaZero~Hz;  $\Omega_{\rm GW}(f)\leq \UpperLimitAlphaTwoThirds$ at 25 Hz for a power-law GWB with a spectral index of 2/3 (consistent with expectations for compact binary coalescences), in the band 20-\DeltaFNinetyNineAlphaTwoThirds~Hz; and $\Omega_{\rm GW}(f)\leq \UpperLimitAlphaThree$ at 25~Hz for a spectral index of 3, in the band 20-\DeltaFNinetyNineAlphaThree~Hz. 
These upper limits improve over our previous results by a factor of \ImprovementAlphaZero{} for a flat GWB, \ImprovementAlphaTwoThirds{} for a spectral index of 2/3, and \ImprovementAlphaThree{} for a spectral index of 3.  
We also search for a GWB arising from scalar and vector modes, which are predicted by alternative theories of gravity; we do not find evidence of these, and place upper limits on the strength of GWBs with these polarizations.
We demonstrate that there is no evidence of correlated noise of magnetic origin by performing a Bayesian analysis that allows for the presence of both a GWB and an effective magnetic background arising from geophysical Schumann resonances.
We compare our upper limits to a fiducial model for the GWB from the merger of compact binaries, updating the model to use the most recent data-driven population inference from the systems detected during O3a. 
Finally, we combine our results with observations of individual mergers and show that, at design sensitivity, this joint approach may yield stronger constraints on the merger rate of binary black holes at $z\gtrsim2$ than can be achieved with individually resolved mergers alone.
\end{abstract}

\maketitle

\section{Introduction}
\label{sec:Introduction}

The gravitational-wave background (hereafter referred to as the GWB or the background) is a superposition of gravitational-wave (GW) sources that is best characterized statistically~\cite{Cornish:2015pda}. There are many possible astrophysical and cosmological contributions to the background, including
distant compact binary coalescences (CBCs) that cannot be resolved individually 
\citep{PhysRevD.84.084004, 
2011ApJ...739...86Z, 
PhysRevD.84.124037, 
PhysRevD.85.104024, 
2013MNRAS.431..882Z}, 
core collapse supernovae \citep{ Buonanno:2004tp, 
2004MNRAS.351.1237H,
Sandick:2006sm,
2009MNRAS.398..293M, 
2010MNRAS.409L.132Z}, 
rotating neutron stars \citep{Ferrari:1998jf,
Regimbau:2001kx,
2011MNRAS.410.2123H,  
2011ApJ...729...59Z,  
2011MNRAS.411.2549M, 
Rosado:2012bk, 
WuEA_2013,
PhysRevD.87.063004}, 
stellar core collapses
\cite{2015PhRvD..92f3005C,
2017PhRvD..95f3015C},
cosmic strings \citep{1976JPhA....9.1387K,
2002PhLB..536..185S,
2005PhRvD..71f3510D,
2007PhRvL..98k1101S,
O1_cosmic_strings}, 
primordial black holes \cite{SasakiEA_2016,
MandicEA_2016, 
Wang:2016ana},
superradiance of axion clouds around black holes \cite{Brito:2017wnc, 
Brito:2017zvb, 
Fan:2017cfw,
Tsukada:2018mbp}, 
phase transitions in the early universe \cite{Lopez:2013mqa,
Dev:2016feu,
Marzola:2017jzl,
vonHarling:2019gme},
and GWs produced during inflation \citep{1979JETPL..30..682S,
1997PhRvD..55..435T, 
1994PhRvD..50.1157B}
or in a preheating phase at the end of inflation~\cite{2006JCAP...04..010E,
2007PhRvL..99v1301E}.
While some sources of the GWB, such as slow roll inflation, have a fundamentally stochastic character, others like the background from CBCs are a superposition of deterministic sources. 

The LIGO Scientific Collaboration and Virgo Collaboration have previously placed upper limits on isotropic \cite{stoch_O2} and anisotropic \cite{stoch_dir_O2} GWBs using data from the first two observing runs, in the frequency range 20-1726 Hz.
The searches were performed by calculating the cross correlation between pairs of detectors.
An extension of this method has been applied to
searching for a background of non-tensor modes~\cite{TestingGR_stoch,stoch_nongr_O1,stoch_O2};
see \cite{RomanoCornish,Christensen_2018} for recent reviews. 
Cross-correlation methods have also been applied to publicly released LIGO data \cite{Abbott:2019ebz} by other groups, who have obtained similar upper limits \cite{Renzini:2018vkx,Renzini:2018nee,Renzini:2019vmt}. A new method that does not rely on the cross-correlation technique and targets the background from CBCs was proposed in~\cite{tbs_methods}.

In this work we apply the cross-correlation based method used in previous analyses to Advanced LIGO's \cite{aLIGO_2015} and Advanced Virgo's \cite{aVirgo_2015} first three observing runs (O1, O2, and O3).  
We do not find evidence for the GWB, and therefore place an upper limit on the strength. 
Unlike in previous observing runs, in this work we present the headline results using a log uniform prior~\cite{JeffreysPrior}.
We find two advantages to using a log uniform prior.
First, a log uniform prior gives equal weight to different orders of magnitude of the strength of the GWBs, which is appropriate given our current state of knowledge.
Second, a log uniform prior is agnostic as to which power we raise the strain data.
It is not clear whether one should put a uniform prior on the strain amplitude, or the strength of the GWB, which scales like the square of the strain. 
On the other hand, the log uniform prior does not depend on the exponent of the strain data. 
For completeness, we also present results with a uniform prior on the strength of the GWB in Section~\ref{sec:Results}.
Results with any other prior can be obtained by reweighing the posterior samples available at \cite{stoch_o3_data_release}.

There are several new features in our analysis of the O3 data. 
First, we incorporate Virgo, by cross correlating the three independent baselines in the LIGO-Virgo network and combining them in an optimal way \cite{Allen_Romano_1999}.  
Second, in order to handle a large rate of loud glitches in O3, we analyze data where these artifacts have been removed via gating~\cite{gatingDocument,y_stochasticGatingDocument}. 
Third, we perform a careful analysis of correlated magnetic noise that could impact the search. In addition to constructing a correlated magnetic noise budget, as in past runs, we use a Bayesian statistical framework developed in \cite{Meyers:2020qrb} to constrain the presence of magnetic noise. 

Perhaps the most interesting source of an astrophysical GWB, given the current network sensitivity, is the GWB from CBCs. 
Previous studies have shown that this GWB may be detectable with Advanced LIGO and Advanced Virgo running at design sensitivity \cite{gw150914_stoch,gw170817_stoch}, and the ability to detect such a background has been confirmed with mock data challenges~\cite{Regimbau:2012ir,Regimbau:2014uia,Meacher:2015iua}. 
Therefore  in this work we carefully consider the implications of our results for the CBC population.
We estimate the GWB using the most up-to-date information from  observations during O3 \cite{O3aCatalog,gw_190412,gw_190425,gw_190521,
gw_190521_implications,gw_190814} and compare with the sensitivity of the current and future detector networks. We show that an upgrade of the current Advanced LIGO facilities, known as A+ \cite{aplus_design}, could dig into a substantial part of the expected parameter space for the GWB at its target sensitivity.
Furthermore, we apply the methods of~\cite{Callister:2020arv} to constrain the merger rate as a function of redshift for binary black holes (BBHs) by combining the GWB upper limits with information about individually resolvable events. 
We find that the cross-correlation analysis can provide complementary information at large redshifts, compared to the population analysis using individually detectable events alone \cite{O3aRatesAndPop}. 
We make the results of our cross correlation analysis available~\cite{stoch_o3_data_release}, enabling further detailed studies of the GWB from CBCs and other models.

The rest of this work is organized as follows. In Section~\ref{sec:Methods}, we review the method of the cross-correlation search. We discuss the data quality procedures and studies we performed in Section~\ref{sec:DataQuality}. We present the main results of the search in Section~\ref{sec:Results}: we derive upper limits on the GWB in Section~\ref{sec:UpperLimits}, put constraints on the presence of scalar- and vector-polarized backgrounds in Section~\ref{sec:Non-GR}, and in Section~\ref{sec:JointGWMag} we extend these results by simultaneously fitting for an astrophysical GWB and an effective GWB arising from magnetic correlations of terrestrial origin. We compare our upper limits with a fiducial model for the GWB from CBCs in Section~\ref{sec:CBCFiducial}, and derive constraints on the BBH merger rate using the upper limits on the GWB and observations of individual CBCs in Section~\ref{sec:CBCMergerRate}. We conclude in Section~\ref{sec:Conclusions}.

\section{Methods}
\label{sec:Methods}

A GWB that is Gaussian, isotropic, unpolarized, and stationary is fully characterized by a spectral energy density. It is standard to express the spectrum in terms of the dimensionless quantity $\Omega_{\rm GW}(f)  $, which is the GW energy density ${\rm d} \rho_{\rm GW}$ contained in the frequency interval $f$ to $f+{\rm d}f$, multiplied by the GW frequency and divided by ${\rm d}f$ times the critical energy density $\rho_c$ needed to have a flat Universe
\begin{equation}
\Omega_{\rm GW}(f) = \frac{f}{\rho_c} \frac{{\rm d} \rho_{\rm GW}}{{\rm d} f},
\end{equation}
where $\rho_c=3 H_0^2 c^2/(8\pi G)$, $c$ is the speed of light, and $G$ is Newton's constant. For consistency with other GW measurements (for example those of \cite{O3aCatalog}), we take the Hubble constant from Planck 2015 observations to be $H_0=67.9\ {\rm km\ s^{-1}\ Mpc^{-1}}$ \cite{Planck_2015}.

\subsection{Cross correlation spectra}
Let us label the GW detectors in the LIGO-Hanford, LIGO-Livingston, and Virgo (HLV) network by the index $I=\{H,L,V\}$. We denote the time-series output of the detectors by $s_I(t)$, and the Fourier transform by $\tilde{s}_I(f)$. Following \cite{Allen_Romano_1999,RomanoCornish}, we define the cross-correlation statistic for the baseline $IJ$ as
\begin{equation}
        \label{eq:methods:bin_by_bin_estimator}
\hat{C}^{IJ}(f) =\frac{2}{T}  \frac{{\rm Re}[\tilde{s}_I^\star(f) \tilde{s}_J(f)]}{\gamma_{IJ}(f) S_0(f)},
\end{equation}
where $\gamma_{IJ}(f)$ is the normalized overlap reduction function \cite{christensen92,Allen_Romano_1999,Mingarelli2019} for the baseline $IJ$, the function $S_0(f)$ is given by
$S_0(f) = (3 H_0^2)/(10\pi^2f^3)$, and $T$ is the observation time. In practice, because the noise is non-stationary, we break the data into segments, and then take $T$ to be the segment duration. We then average together segments using inverse noise weighting \cite{Allen_Romano_1999}. If the noise were stationary, this average would reproduce Eq.~\ref{eq:methods:bin_by_bin_estimator}.
This estimator is normalized so that $\langle \hat{C}^{IJ}(f) \rangle = \Omega_{\rm GW}(f)$ in the absence of correlated noise. In the small signal-to-noise ratio limit, the variance can be estimated as
\begin{equation}
\label{eq:def_sigma}
\sigma_{IJ}^2(f) \approx \frac{1}{2T \Delta f}  \frac{P_I(f) P_J(f)}{\gamma_{IJ}^2(f) S_0^2(f)},
\end{equation}
where $\Delta f$ is the frequency resolution, and $P_I(f)$ is the one-sided power spectral density in detector $I$. Note that $T\Delta f$ need not equal one if several frequency bins are coarse grained around the central frequency $f$ to produce the estimator in Eq.~\ref{eq:methods:bin_by_bin_estimator}.

While we have expressed the cross-correlation estimator in terms of the GW strain channel, in fact this analysis can be applied to \emph{any} pair of instruments. Following \cite{Meyers:2020qrb}, in Sections~\ref{sec:MagneticBudget} and~\ref{sec:JointGWMag} we will also employ these techniques to cross correlate magnetometer channels to search for correlated magnetic noise.

\subsection{Optimal filtering}
Strictly speaking, the optimal estimator for a given signal includes both auto-correlation and cross-correlation terms \cite{RomanoCornish}. We only use the cross correlation, and not auto-correlation, in the search because the noise power spectral density is not known precisely enough to be subtracted accurately, and therefore in practice the cross correlation is nearly optimal. With this caveat, we can construct an optimal estimator to search for a GWB of any spectral shape by combining the cross-correlation spectra from different frequency bins with appropriate weights
\begin{eqnarray}
\label{eq:optimal-filter}
\hat C^{IJ} &=& \frac{\sum_k w(f_k) \hat C^{IJ}(f_k)\sigma^{-2}_{IJ}(f_k)}{\sum_k w(f_k)^{2}\sigma^{-2}_{IJ}(f_k)}, \nonumber \\
\sigma_{IJ}^{-2} &=& \sum_k w(f_k)^{2} \sigma^{-2}_{IJ}(f_k),
\end{eqnarray}
where $f_k$ are a discrete set of frequencies, and the optimal weights for spectral shape $\Omega_{\rm GW}(f)$ are given by
\begin{equation}
w(f) = \frac{\Omega_{\rm GW}(f)}{\Omega_{\rm GW}(f_{\rm ref})}.
\label{eq:optimal-w}
\end{equation}
Here, $f_{\rm ref}$ is a fixed reference frequency. For ease of comparison with previous observing runs, we choose the reference frequency to be $f_{\rm ref}=25\ {\rm Hz}$. This is approximately the start of the most sensitive frequency band for the isotropic search as described in \cite{stoch_O2}. 
This analysis is very flexible and can be applied to a GWB of any spectral shape. We will report results for a power law GWB of the form
\begin{equation}
\Omega_{\rm GW}(f) = \Omega_{\rm ref} \left(\frac{f}{f_{\rm ref}}\right)^\alpha.
\end{equation}
Our final estimator combines information from all baselines optimally using the sum
\begin{equation}
\hat{C} = \frac{\sum_{IJ} \hat{C}^{IJ} \sigma_{IJ}^{-2}}{\sum_{IJ} \sigma_{IJ}^{-2}},\ \ \sigma^{-2} = \sum_{IJ} \sigma_{IJ}^{-2},
\end{equation}
where $\sum_{IJ}$ is a shorthand notation meaning a sum over all \emph{independent} baselines $IJ$.
We can also include cross correlation results from previous observing runs in a natural way by including them in this sum as separate baselines. More concretely, we combine HL-O1, HL-O2, HL-O3, HV-O3 and LV-O3.

\subsection{Parameter estimation}
\label{sec:Methods-PE}

In order to estimate parameters of a specific model of the GWB, we combine the spectra from each baseline $IJ$ to form the likelihood \cite{StochPE}
\begin{eqnarray}
\label{eq:likelihood}
p(\hat{C}^{IJ}_k | \params) &\propto&\exp\left[  -\frac{1}{2} \sum_{IJ} \sum_k \left(\frac{\hat{C}^{IJ}_k - \Omega_{\rm M}(f_k|\params)}{\sigma^2_{IJ}(f_k)}\right)^2  \right], \nonumber \\
\end{eqnarray}
where $\hat{C}^{IJ}_k \equiv \hat{C}^{IJ}(f_k)$, and
where we assume that the $\hat{C}^{IJ}_k$ are Gaussian-distributed in the absence of a signal. 
The term $\Omega_{\rm M}(f|\params)$ describes the model for the GWB, characterized by the set of parameters $\params$.  This hybrid frequentist-Bayesian approach has been shown to be equivalent to a fully Bayesian analysis in \cite{Matas:2020roi}.

Equation~\eqref{eq:likelihood} assumes that cross-correlation spectra measured between different baselines are uncorrelated.
This is not strictly true, as different baselines share detectors in common.
Correlations between baselines, however, enter at $\mathcal{O}(\Omega^2)$ and so can be neglected in the small-signal limit~\cite{Allen_Romano_1999}.

In this work we shall consider several different models:
\begin{itemize}
\item \textsc{Noise} (N): $\Omega_{\rm N}(f)=0$. We implicitly include uncorrelated Gaussian noise as part of every model that follows.
\item \textsc{Power Law} (PL): $\Omega_{\rm PL}(f)=\Omega_{\rm ref}\left(\frac{f}{f_{\rm ref}}\right)^\alpha$. 
The parameters $\mathbf{\Theta}_{\rm PL}$ are the amplitude $\Omega_{\rm ref}$ and spectral index $\alpha$. 
We will consider cases in which $\alpha$ is allowed to vary as well as those in which it is fixed.
\item \textsc{Scalar-Vector-Tensor Power Law} (SVT-PL): This model contains tensor polarizations, as allowed in general relativity (GR), and vector and scalar polarizations, which are forbidden in GR but generically appear in alternative theories of gravity.
We define ${\rm p}$ to be an index referring to polarization, ${\rm p}=\{{\rm T,V,S}\}$, where ${\rm T}$, ${\rm V}$, and ${\rm S}$ refer to tensor, vector, and scalar polarized GWs, respectively.
We assume the GWB for each polarization can be described by a power law, which may be different for each polarization. 
Thus there are six parameters $\mathbf{\Theta}_{\rm SVT-PL}$, given by the amplitudes $\Omega_{\rm ref}^{\rm (p)}$ and spectral indices $\alpha_{\rm p}$ for each polarization.
The model is given by the sum $\Omega_{\rm SVT-PL}(f)=\sum_{\rm p} \beta_{IJ}^{\rm (p)}(f) \Omega_{\rm ref}^{\rm (p)}\left(\frac{f}{f_{\rm ref}}\right)^{\alpha_{\rm p}}$, where $\beta_{IJ}^{\rm (p)}(f)=\gamma^{\rm (p)}_{IJ}(f)/\gamma_{IJ}(f)$ is the ratio of the overlap reduction function for polarization ${\rm p}$ and baseline $IJ$ to the standard (tensor) overlap reduction function for that baseline~\cite{TestingGR_stoch}. 
\item \textsc{Magnetic} (MAG): $\Omega_{\rm MAG}(f)$ describes correlations between two detectors induced by large-scale coherent magnetic fields, which can appear as an \emph{effective} background. We model this effective background in terms of magnetometer correlations and a transfer function between the local magnetic field and the strain channel of the detectors. The free parameters $\mathbf{\Theta}_{\rm MAG}$ describe the coupling function, as described in Section~\ref{sec:JointGWMag}.
\item \textsc{CBC}: $\Omega_{\rm CBC}(f)$ is determined by an underlying parametrized model for the mass distribution of compact binaries and their merger rate $R(z)$ as a function of redshift. The parameters of this model are discussed in Section~\ref{sec:CBCImplications}.
\end{itemize}
We will also consider combinations of these models, for example $\Omega_{\rm M}(f) = \Omega_{\rm PL}(f) + \Omega_{\rm MAG}(f)$. Given the likelihood, we form a posterior using Bayes theorem, $p(\params|C^{IJ}_k)\propto p(C^{IJ}_k|\params)p(\params)$, where $p(\params)$ is the prior distribution on the parameters $\params$. We will consider different prior choices for each model we consider below.

Finally, it is often of interest to combine upper limits on the amplitude of the GWB with other observations, in order to obtain the best possible constraints on a given model. For example, such a strategy can be used to combine measurements across a range of frequency bands as in \cite{Lasky:2015lej}. Denoting data from the other observations as $\hat{D}$, we can consider a factorized likelihood
\begin{equation}
\label{eq:factorized-likelihood}
p(\hat{C}^{IJ}(f_k),\hat{D}|\params) = p(\hat{C}^{IJ}(f_k)|\params)\, p(\hat{D} |\params).
\end{equation}
In Section~\ref{sec:CBCMergerRate}, we will apply this method to combine the upper limits on the GWB with observations of individual BBH events from \cite{O3aCatalog}, similar to that performed in \cite{Callister:2020arv}.

\section{Data Quality}
\label{sec:DataQuality}

\subsection{Data}

We analyze strain data taken during O3 by the LIGO-Hanford, LIGO-Livingston, and Virgo detectors. 
The O3 run is divided into two sets. 
The first, O3a, began April 1, 2019, 15:00 UTC, and continued until October 1, 2019 15:00 UTC, while O3b ran from November 1 2019, 15:00 UTC, to March 27, 2020 17:00 UTC. 
The HL baseline had \TimeHL{} of coincident livetime, HV \TimeHV{}, and LV \TimeLV{}, before applying any data quality vetoes.

We look for correlated magnetic noise using magnetometers located at the sites. 
Each LIGO detector has installed two low-noise LEMI-120 magnetometers \cite{lemi}. 
The Virgo detector has two low-noise MFS-06 magnetometers by Metronix~\cite{metronix}. 
In order to allow a comparison between the magnetic and GW searches, we apply the same data processing to the magnetometer channels that we do to the strain channels, except where otherwise stated.

The data are first downsampled. 
For the GW data, we decimate the data from the original sampling rate 16384 Hz to 4096 Hz. 
The maximum frequency that we analyze is 1726 Hz, which is sufficiently below the Nyquist frequency to avoid aliasing effects. 
Since we only analyze magnetic data up to 100 Hz, we downsample the LEMI magnetometers from 16384 Hz to 512 Hz, and the Metronix magnetometers from 2000 Hz to 512 Hz. 
Then the data are high-pass filtered using a 16th-order Butterworth filter with a knee frequency of 11 Hz, which is constructed using second-order sections. 
We divide the original data stream into time segments of duration 192 s which are Hann-windowed and overlapped by 50\%, then compute a discrete Fourier transform on each of these segments, and coarse-grain the resulting spectrum to a frequency resolution of 1/32 Hz. 
We perform the cross-correlation search with a publicly available implementation~\cite{x_stochastic_public} of the algorithm described in Section~\ref{sec:Methods} using \textsc{Matlab} \cite{MATLAB:2020}.  

As an end-to-end test of the entire system, we added stochastic signals in the Hanford and Livingston detectors by actuating the test masses, following the procedures described in \cite{Biwer:2016oyg}. We injected the same realization of the stochastic background with a flat power law index and strength of $4.3\times 10^{-5}$ in two 15-minute segments of data. We found for the first injection $\Omega_{\rm GW}=(4.6 \pm 0.4)\times 10^{-5}$ and the second $\Omega_{\rm GW}=(4.7\pm 0.4)\times 10^{-5}$; both recoveries are statistically consistent with the injected signal.

\subsection{Time and frequency domain cuts}

For each baseline, we require that both detectors in the baseline are in observing mode, and that there are no critical issues with the detector hardware, as defined by category 1 vetoes described in \cite{O1_CBC_DQ}. 
As in previous runs, we apply a non-stationarity cut by removing times where the square root of the variance in Eq~\ref{eq:optimal-filter} is found to vary by more than 20\% between segments. 
We take the union of the cuts for $\alpha=\{-5,0,3,5\}$; each power law is sensitive to a different frequency band. 
While we use $\alpha=2/3$ for the search, we do not include it in the cut since it does not provide significantly new information for the non-stationarity cut, because the frequency range is very similar to the one probed by $\alpha=0$.
We remove Hanford data from April 1--April 16 2019 due to non-stationarity arising from calibration lines at 35.9 and 36.7 Hz. 
These lines were moved below 20 Hz on April 16, 2019. 

In principle, the CBC signals known to be present in the data contribute to the integrated cross-correlation. A simple estimate by using the median values for the masses and redshifts in O3a from \cite{O3aCatalog}, the livetime for O3a, and the inspiral approximation, in Eq. 16 of \cite{Meacher:2015iua}, yields $\Omega_{\rm GW}(25\ {\rm Hz})\lesssim {\color{black} 10^{-10}}$, which is well below the O3 sensitivity. 
Therefore we do not remove the observed CBCs from the data. 

After applying the category 1 vetoes and non-stationarity cut, we found that \percentLivetimeLostHL{} of available livetime was lost in the HL baseline, \percentLivetimeLostHV{} in the HV baseline, and \percentLivetimeLostLV{} in the LV  baseline.

We do not analyze frequency bins where there is evidence of coherence between instruments that is determined to be caused by the instruments themselves. The coherence between two channels,
\begin{equation}
\Gamma_{IJ}(f) = \frac{\langle | \tilde{s}_I^\star (f) \tilde{s}_J (f) |^2 \rangle }{\langle |\tilde{s}_I(f)|^2\rangle \langle |\tilde{s}_J(f)|^2 \rangle },
\end{equation}
is a useful measure to determine when correlations in a given frequency bin exceed what is expected from uncorrelated data. In the above expression, the angle brackets $\langle \rangle$ refer to an average over analysis segments.
The coherence between the strain and auxiliary channels at a given site can also be used to identify an instrumental source of contamination \cite{CWStochDetchar}. We removed \percentNotchedAllHL{} of the frequency band in the HL baseline, \percentNotchedAllHV{} of the frequency band in the HV baseline, and \percentNotchedAllLV{} of the frequency band in the LV baseline. However, we only removed \percentNotchedMidFHL{} from HL, \percentNotchedMidFHV{} from HV, and \percentNotchedMidFLV{} from LV below 300 Hz, where the search is most sensitive. 
In O3, we found many 1 Hz harmonics which were coherent between Hanford and Virgo. 
We also observed a large coherent line in the HL baseline at 33.2 Hz, which was likely due to the beating of two different calibration lines at Hanford and Livingston, and therefore did not appear in linear coherences between the strain and auxiliary channels.
Generally speaking, line mitigation efforts were particularly effective at the LIGO-Livingston detector, and the HL and LV baselines had many fewer coherent lines.
The full list of frequencies removed from the analysis is available online~\cite{stoch_o3_data_release}.

\subsection{Gating}

\begin{figure*}
\includegraphics[width=\textwidth]{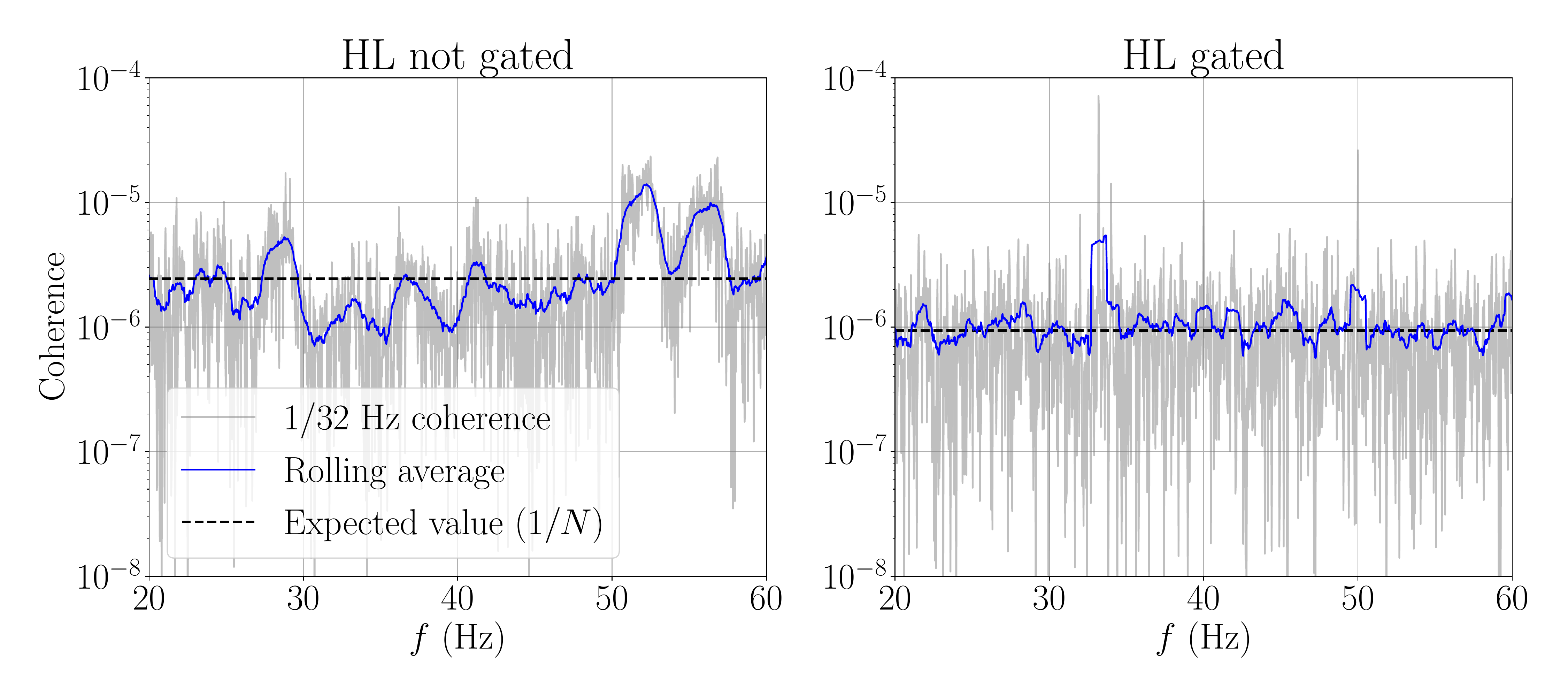}
\caption{Coherence spectra for the HL baseline without (left panel) and with (right panel) gating applied. In both cases, we have applied the non-stationarity cut described in the main text. We show the coherence spectrum at 1/32 Hz (the same frequency resolution used for the isotropic search) as a gray line, a rolling average with a 1 Hz resolution as a blue line, and the average value expected for uncorrelated, Gaussian noise which is given by 1 divided by the number of averages used to make the coherence spectrum as a black dashed line. The gated coherence is consistent with the expectation from Gaussian noise, while the spectrum without gating is not. Additionally, we see that more segments are used for averages in the gated spectrum after applying the non-stationarity cut described in the text. This is because without gating, many segments are removed due to a single glitch biasing the power spectrum estimate of the segment. Note that in this figure, we have not applied any frequency notching; the large line visible at 33.2 Hz is due to a beat note in the calibration lines at both H1 and L1 arising from non-linear couplings.}
\label{fig:coherences}
\end{figure*}

In O3, we found a much higher rate of loud glitches compared to previous observing runs. 
A naive application of the standard non-stationarity cut used in previous searches led to losing $> 50\%$ of the data when running with 192-s data segments. 
In order to reduce the amount of data lost to the non-stationarity cut, and thus improve the sensitivity of the search, we pre-conditioned the data by applying a gating procedure. 
This procedure involves first identifying data from the Hanford and Livingston baselines that contain a glitch, and then zeroing out these data. 
We defined segments containing a glitch when the root-mean-squre (RMS) value of the whitened strain channel in the 25-50 Hz band or 70-110 Hz band exceeded a threshold value. 
We then removed the glitches from the time series by multlipying the data in these segments by an inverse Tukey window.
We found that a total of \percentGatedH{}  of Hanford data was gated in the data that we analyzed, and \percentGatedL{} of Livingston data for each baseline. 
We refer the interested reader to \cite{gatingDocument} for further details of the procedure, including the whitened channels and precise thresholds used. 
This was not necessary for Virgo data due to the lower rate of large glitches. 
The impact of gating can be clearly seen on the coherence spectra, as we show in Figure~\ref{fig:coherences}. 
Compared to the non-gated data, many more segments are analyzed after applying non-stationarity cuts, and the spectrum is much closer to what is expected from uncorrelated Gaussian noise. 
It was discovered that from April 20-25 a 1/120-Hz comb was visible in the Livingston data around large calibration lines. 
The comb was caused by an inadvertently running diagnostic camera clicking at regular two minute intervals. 
To be cautious, we removed this period of time from the analysis. 
We have verified with a mock data challenge that applying this gating procedure to simulated data did not affect our ability to recover a GWB. 
This check is described further in~\cite{y_stochasticGatingDocument}.

\subsection{Correlated magnetic noise budget}
\label{sec:MagneticBudget}

In order to be able to claim detection of a GWB, one must understand and control environmental sources of correlated noise. Some magnetic fields are expected to be correlated between sites and are monitored with sensitive magnetometers placed away from the buildings. For example, Schumann resonances are electromagnetic modes of the Earth-ionosphere resonant cavity \cite{Schumann_theory}. They are coherent on a global scale \cite{Schumann_4}, so if they couple to the interferometer and produce noise in the GW channel, they will cause correlations between the outputs of detectors on different continents \cite{Schumann_1, Schumann_2} . If these effects are large enough, they can be a source of confusion noise for cross-correlation searches. In this section we show that there is no evidence for correlated magnetic noise in the O3 GW strain data. 

As in past runs \cite{stoch_O1,stoch_O2}, following \cite{Schumann_1, Schumann_2} we create a budget for the magnetic correlations
\begin{equation}
        \hat{C}_{\textrm{mag}, IJ}(f) = \frac{2}{T} \frac{|T_I(f)||T_J(f)| \textrm{Re}[\tilde m_I^\star(f)\tilde m_J(f)]}{\gamma_{IJ}(f) S_0(f)},
        \label{eq:magnetic:model}
\end{equation}
where $\tilde{m}_I(f)$ are Fourier transforms of the magnetometer channels.  The coupling functions $T_{I}(f)$ are estimated by injecting an oscillating magnetic field of a known frequency and amplitude at different locations near each detector, and measuring the resulting output in the GW strain channel. 
Weekly injections were performed to study the time-dependence of the magnetic coupling~\cite{alog:WeeklyMagneticInjections}. 

Potential differences in the strength of the magnetic field at the magnetometers located around the detector versus the strength of the field at the ``true'' coupling location mean that these measurements are only rough estimates, and are susceptible to large uncertainties. 
This uncertainty is estimated by comparing injections at different locations at each site; to account for this, we include a factor of two uncertainty in the coupling function of each detector~\cite{z_PEMCouplingUncertainty}.

Another possible source of error in the coupling function measurement is that the low-noise magnetometers are located outside, far from the local magnetic noise associated with the buildings, but the weekly injections described above are performed inside. One may worry that ferromagnetic material in the buildings can amplify the outside-to-inside magnetic coupling. However, additional measurements at Handford suggest that the coupling function from outside to inside the building is less than one.
Injections were performed around the corner station using seven frequencies ranging from 11 to 444 Hz, and the magnetic field was measured inside and outside the building at the same distance from the injection coil. A power-law fit to the ratio of the magnetic field measured inside to the field measured outside as a function of frequency indicates that the magnetic coupling is suppressed by up to a factor of 2 in the frequency range 10-100 Hz, however with variation depending on the orientation of the field. To be conservative, we assume the inside-to-outside magnetic coupling is equal to one.

\begin{figure}
\includegraphics[width=0.49\textwidth]{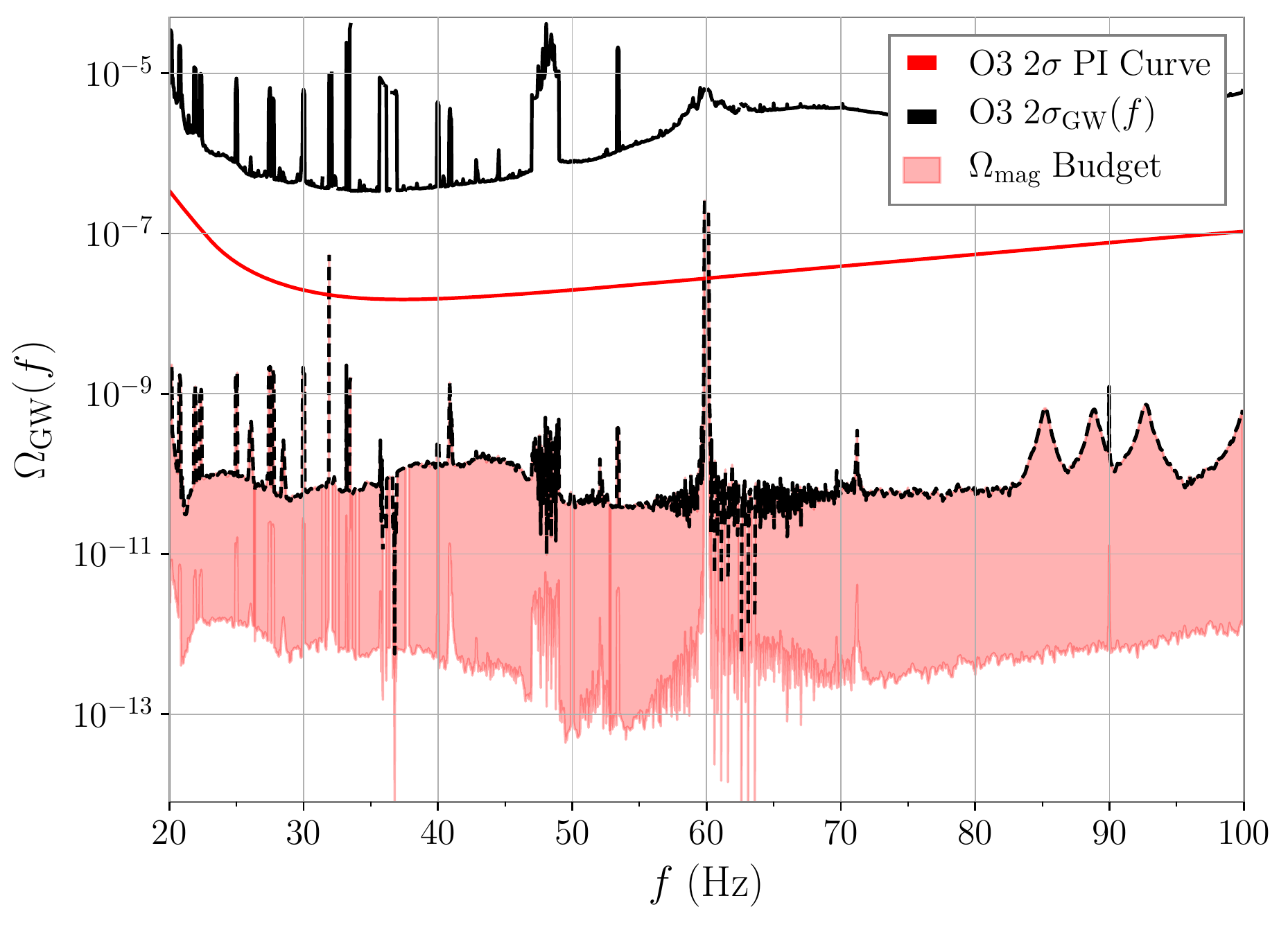}
\caption{Correlated magnetic noise budget, as described in the main text. The band shows the expected range of magnetic contamination, using weekly measurements of the coupling function at each site, and accounting for the uncertainty in the coupling function measurements. We display the O3 sensitivity in two ways. First, we plot the power-law integrated (PI) curve as a red line. This indicates the sensitivity of the search to power-law backgrounds and includes the effect of integrating over frequencies, and should be compared with the overall trend of the red band. Second, we plot the square root of the variance, $\sigma(f)$, as a black line, which gives the sensitivity of the search to narrowband features. This can be compared with narrow features in the upper range of the noise budget, which we show as a black dashed line. We conclude that the O3 sensitivity  is well above the level of correlated magnetic noise estimated in O3.}
\label{fig:magnetic:budget_plot}
\end{figure}

To construct the budget, we first compute a linear interpolation for the coupling function as a function of frequency measured at each detector in each week. 
For weeks where a coupling function was not measured, we use the coupling function that was nearest in time. 
For each baseline, and each week, we then multiply the coupling functions for each detector by the magnetic cross-correlation spectrum for that baseline, to form a budget.
We use the pair of directions that gives the largest coherence.
Studies based on shorter stretches of data indicate that the coherence of the magnitude of the magnetic field can be up to a factor of two larger than the coherence of the worst-case components; therefore to be conservative we multiply the coherence in each detector baseline by a factor of two.
We combine the budgets across baselines by using the error bars from the GW channels as weights to account for the relative sensitivity of each baseline, $\hat{C}_{\rm mag}=\sum_{IJ} w_{IJ}(f) \hat{C}_{{\rm mag},IJ}(f)$, where $w_{IJ}(f)=(\sigma_{IJ}(f)/\sigma(f))^{-2}$.
We show an estimate of the correlated magnetic noise compared to the O3 sensitivity curve in Figure~\ref{fig:magnetic:budget_plot}, combining all three baselines. 
The red band shows the range of budgets we obtain accounting for the combined weekly magnetic coupling function measurements, as well as the overall factor of two uncertainty in each detector's coupling function described above.
The overall trend of the red band should be compared with the O3 power-law integrated (PI) curve \cite{locus}, which shows the sensitivity of our search to power law backgrounds, accounting for integration over frequency.
The black dotted line shows the upper range of the budget.
Narrowband features should be compared with $\sigma(f)$, shown as a black solid line, which shows the sensitivity to a GWB in every frequency bin.
The measurements at Hanford were sampled at a fine frequency resolution due to the use of broadband injections with a large coil \cite{q_alog:LargeCoil}. This allowed us to see fine-grained features in the coupling function, such as the broad resonances visible between 80 Hz and 100 Hz in Figure~\ref{fig:magnetic:budget_plot}.
While the exact origin of these resonances is presently unknown, they are correlated with excess motion of test masses in the power recycling cavity~\cite{alog:MagneticCouplingResonances}.
The final budget indicates that the non-observation of correlated magnetic noise is expected given the coupling function measurements.

\section{Results}
\label{sec:Results}

\subsection{Upper Limits on the GWB}
\label{sec:UpperLimits}
In Table~\ref{tab:results} we report the point estimate and 1-$\sigma$ error bar from O3 obtained from each baseline independently, as well as combining all three baselines together with the HL baseline results available from O1 and O2, using an optimal filter for three different power law models
\begin{itemize}
\item $\alpha=0$ approximately characterizes cosmic string \citep{1976JPhA....9.1387K,2002PhLB..536..185S,2005PhRvD..71f3510D,2007PhRvL..98k1101S,O1_cosmic_strings},  and slow-roll inflation GWBs \cite{1979JETPL..30..682S,1997PhRvD..55..435T,1994PhRvD..50.1157B} in the LIGO-Virgo frequency band. 
\item $\alpha=2/3$ describes the CBC GWB when contributions from the inspiral dominate the GWB, which is a very good approximation in the LIGO-Virgo frequency band \cite{Regimbau_2011}. However, this approximation may not be valid for mergers of binaries arising from Population III stars \cite{Perigois:2020ymr}, or from heavy BBH mergers with masses above the pair-instability mass gap \cite{Ezquiaga:2020tns}.
\item $\alpha=3$ is a fiducial choice used in past searches which approximately describes some astrophysical sources such as supernovae \cite{2009MNRAS.398..293M}-\cite{PhysRevD.73.104024}, and corresponds to a GWB that is flat in the strain power, $S_h(f) \propto f^{-3} \Omega_{\rm GW}(f)$ \cite{Allen_Romano_1999}. 
\end{itemize}
While we use the entire band 20-1726 Hz to compute the point estimate and error bar, we also show  $f^{IJ}_{99\%}$, which is the upper frequency of the band starting at 20 Hz that contains 99\% of the sensitivity in baseline $IJ$.

\begin{figure}
\includegraphics[width=0.49\textwidth]{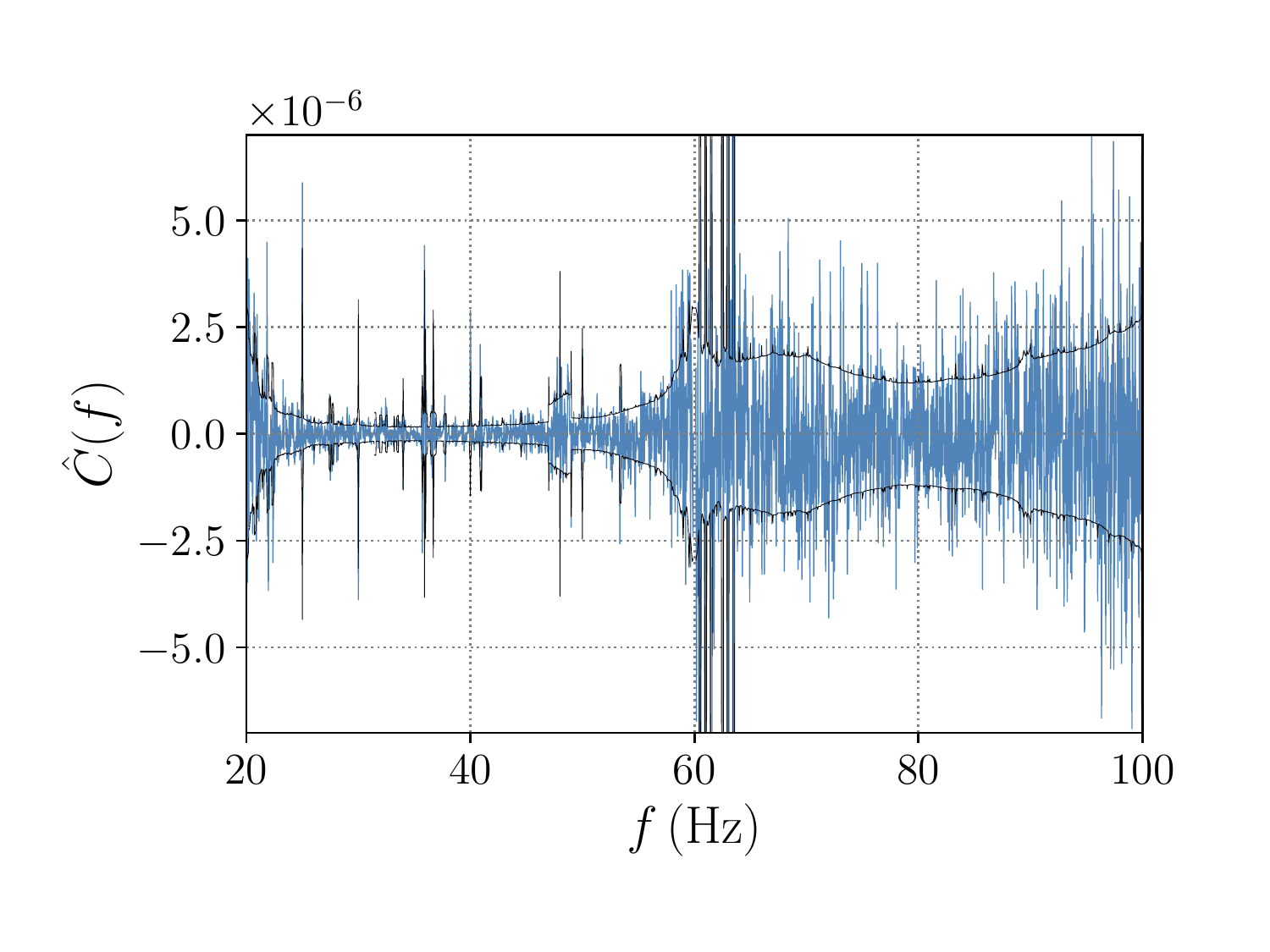}
\caption{Cross-correlation spectra combining data from all three baselines in O3, as well as the HL baseline in O1 and O2. As described in the main text, the spectrum is consistent with expectations from uncorrelated, Gaussian noise. }
\label{fig:Ysig}
\end{figure}

The HL baseline contributes most to the sensitivity. 
The contributions from the baselines that include Virgo are relatively more important at higher frequencies and especially relevant to searches for larger power laws. 
 We note that the point estimates for HV and LV are approximately $2\sigma$ away from zero, however we do not interpret this as evidence of a signal given that the point estimate of the much more sensitive HL baseline is consistent with zero to within $1\sigma$.
The combined spectrum is shown in Figure~\ref{fig:Ysig}. 
From this figure, one can see that the point estimate fluctuates roughly symmetrically around zero, consistent with expectations from Gaussian noise. 
Additionally, by comparing with Figure 1 of \cite{stoch_O2}, it is clear that the addition of Virgo data compensates for a zero in the HL overlap reduction function at around 64 Hz. 
After having applied the data quality cuts described in Section~\ref{sec:DataQuality}, data are consistent with uncorrelated, Gaussian noise. 
The spectra have a $\chi^2$-per-degree-of-freedom value of \chiSquaredAll{}. 

Since we do not find evidence of a GWB, we place upper limits on the PL model, combining the O3 spectra with the results from previous runs. 
We report upper limits using both a prior that is uniform in the log of the strength of the GWB, and a prior that is uniform in the strength.
We choose to report the upper limit obtained with the log uniform prior as our headline result, because a log uniform prior is a more natural choice for a scaling parameter, and also is more sensitive to small signals. 
However, since upper limits computed with a uniform prior are more conservative, we present results for the uniform prior as well.
For both cases, we choose the upper bound of the prior to be large enough that there is no posterior support at the upper end of the prior range.
For the log uniform prior, the upper limit depends mildly on the lower bound of the prior range, which cannot be taken to be zero.
Following \cite{stoch_O2}, we choose the lower bound to be $\Omega_{\rm ref}\geq 10^{-13}$.
This choice enables a direct comparison with previous upper limits, and is the same order of magnitude as the expected reach of next-generation ground-based detectors \cite{Regimbau:2016ike, Sachdev:2020bkk, Martinovic:2020hru}.

For the spectral index, we compute upper limits by fixing $\alpha$ to the three values $\{0,2/3,3\}$ discussed earlier, as well as allowing $\alpha$ to vary. For the latter case, we assume a Gaussian prior on $\alpha$ with zero mean and standard deviation \stdAlpha. 
This prior on $\alpha$ is very similar to the triangular prior on $\alpha$ we used in the O2 analysis \cite{stoch_O2}, however it does not vanish for large values of $|\alpha|$. Therefore in principle, this prior allows us to probe extreme power laws if the data support them.
We have checked that the Gaussian prior gives posterior distributions that are nearly identical to those produced using the triangular prior. 

\begin{figure}
\includegraphics[width=0.49\textwidth]{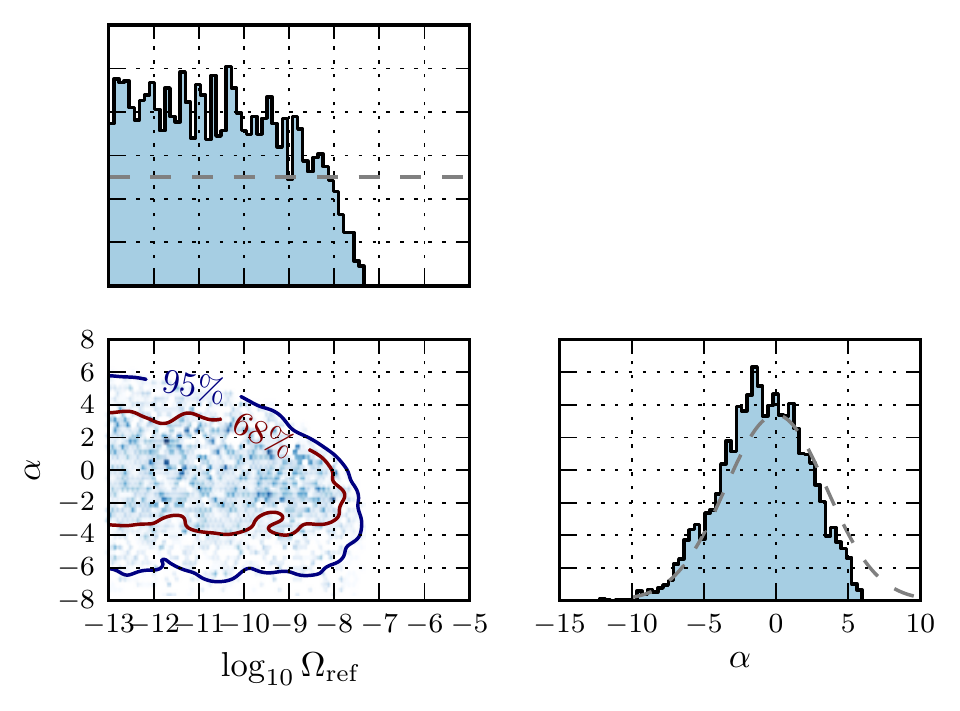}
\caption{Posteriors for the strength $\Omega_{\rm ref}$ and spectral index $\alpha$ for the power law model described in the main text, using a prior uniform in the log of $\Omega_{\rm ref}$. The top and right panels show marginalized posteriors for $\Omega_{\rm ref}$ and $\alpha$, while the center plot shows the 2D posterior density. The dashed, gray lines indicate the prior distributions.}
\label{fig:Omega_alpha_posterior}
\end{figure}

We marginalize over calibration uncertainty following the methods in \cite{StochCalUncertainty}. 
We use an amplitude calibration uncertainty of \CalH{} for Hanford, \CalL{} for Livingston, and \CalV{} for Virgo \cite{Sun:2020wke}; this is a conservative choice describing the worst case over the entire run. 
We use the same amplitude calibration uncertainty factors for O1 and O2 as in previous analysis \cite{stoch_O2}.
In all cases, phase uncertainty is negligible. 
The results are given in Table~\ref{tab:ULs}. 
We also show the posterior in the $\Omega_{\rm ref}$-$\alpha$ plane in Figure~\ref{fig:Omega_alpha_posterior}. 

At the 95\% credible level, using a log-uniform (uniform) prior, we find that $\Omega_{\rm GW}(25\ {\rm Hz})$ is less than \UpperLimitAlphaZero{} (\UpperLimitAlphaZeroUniform{}) for $\alpha=0$, \UpperLimitAlphaTwoThirds{} (\UpperLimitAlphaTwoThirdsUniform{}) for $\alpha=2/3$, \UpperLimitAlphaThree{} (\UpperLimitAlphaThreeUniform{}) for $\alpha=3$, and \UpperLimitAlphaMarg{} (\UpperLimitAlphaMargUniform{}) when marginalizing over $\alpha$. This represents an improvement by a factor of about \ImprovementAlphaZero{} (\ImprovementAlphaZeroUniform{}) for a flat power law, \ImprovementAlphaTwoThirds{} (\ImprovementAlphaTwoThirdsUniform{}) for a power law of $\alpha=2/3$, and \ImprovementAlphaThree{} (\ImprovementAlphaThreeUniform{}) for a power law of $\alpha=3$.
The improvement for large $\alpha$ is due in part to the improved high-frequency sensitivity of Advanced LIGO in O3; to the addition of the baselines involving Virgo; and to the specific noise realization, in particular the negative point estimate $\alpha=3$ in O3, as seen in Table~\ref{tab:results}.
We find a ${\rm log}_{10}$ Bayes Factor of \SignalToNoiseTensorLike{} when comparing the hypotheses of signal and noise to noise-only when marginalizing over $\alpha$.

\begin{table*}
\renewcommand{\arraystretch}{1.1}
\begin{tabular}{c|c|c|c|c|c|c|c|c}
Power law & $ f_{99\%}^{HL}$ [Hz] & $\hat{C}^{HL}/10^{-9}$  & $ f_{99\%}^{HV}$ [Hz] & $\hat{C}^{HV}/10^{-9}$ & $ f_{99\%}^{LV}$ [Hz] & $\hat{C}^{LV}/10^{-9}$ &$ f^{\rm O1+O2+O3}_{99\%}$ [Hz] &  $\hat{C}^{\rm O1+O2+O3}/10^{-9}$  \\
\hline
\hline
0 & \DeltaFNinetyNineAlphaZeroHL{} & \PtEstErrBarAlphaZeroHL{} & \DeltaFNinetyNineAlphaZeroHV{} & \PtEstErrBarAlphaZeroHV{}  & \DeltaFNinetyNineAlphaZeroLV{} & \PtEstErrBarAlphaZeroLV{}  & \DeltaFNinetyNineAlphaZero{} & \PtEstErrBarAlphaZero{}  \\
2/3 & \DeltaFNinetyNineAlphaTwoThirdsHL{} & \PtEstErrBarAlphaTwoThirdsHL{} & \DeltaFNinetyNineAlphaTwoThirdsHV{} & \PtEstErrBarAlphaTwoThirdsHV{}  & \DeltaFNinetyNineAlphaTwoThirdsLV{} & \PtEstErrBarAlphaTwoThirdsLV{}  & \DeltaFNinetyNineAlphaTwoThirds{} & \PtEstErrBarAlphaTwoThirds \\
3 & \DeltaFNinetyNineAlphaThreeHL{} & \PtEstErrBarAlphaThreeHL{} & \DeltaFNinetyNineAlphaThreeHV{} & \PtEstErrBarAlphaThreeHV{}  & \DeltaFNinetyNineAlphaThreeLV{} & \PtEstErrBarAlphaThreeLV{}  & \DeltaFNinetyNineAlphaThree{} & \PtEstErrBarAlphaThree{}  \\
\end{tabular}
\caption{Search results for an isotropic GWB, using the optimal filter method for power law GWBs with $\alpha=\{0,2/3,3\}$. For each of the three baselines $IJ$, we show the point estimate and 1$\sigma$ uncertainty for the cross-correlation estimate $C_{IJ}$, along with the frequency band from 20 Hz to $f^{IJ}_{99\%}$ containing 99\% of the sensitivity. We see that the HL baseline is the most sensitive, and the HV and LV baselines are more sensitive at higher frequencies, and for larger spectral indices, due to the longer baseline. In the last two columns, we also present the search result combining all three baselines from O3, as well as the O1 and O2 data. As noted in the main text, the point estimates for  the HV and LV are approximately $2\sigma$ away from zero, however this is not consistent with a GWB given the result of the much more sensitive HL baseline.}
\label{tab:results}
\end{table*}

\begin{table*}
\begin{tabular}{l|c c c |c c c}
 &
\multicolumn{3}{c|}{Uniform prior} & 
\multicolumn{3}{c}{Log-uniform prior} \\
\hline
\hline
$\alpha$ &
O3 &
O2 \cite{stoch_O2} &
Improvement  &
O3 &
O2 \cite{stoch_O2} &
Improvement   \\
\hline
0 &
\UpperLimitAlphaZeroUniform & 
\UpperLimitAlphaZeroOTwoUniform & 
\ImprovementAlphaZeroUniform &
\UpperLimitAlphaZero &
\UpperLimitAlphaZeroOTwoLog  &
\ImprovementAlphaZero 
\\
2/3 &
\UpperLimitAlphaTwoThirdsUniform &
\UpperLimitAlphaTwoThirdsOTwoUniform & 
\ImprovementAlphaTwoThirdsUniform &
\UpperLimitAlphaTwoThirds & 
\UpperLimitAlphaTwoThirdsOTwoLog &
\ImprovementAlphaTwoThirds  \\
3 &
\UpperLimitAlphaThreeUniform &
\UpperLimitAlphaThreeOTwoUniform & 
\ImprovementAlphaThreeUniform &
\UpperLimitAlphaThree & 
\UpperLimitAlphaThreeOTwoLog  &
\ImprovementAlphaThree \\
Marg. &
\UpperLimitAlphaMargUniform &
\UpperLimitAlphaMargOTwoUniform & 
\ImprovementAlphaMargUniform & 
\UpperLimitAlphaMarg & 
\UpperLimitAlphaMargOTwoLog &
\ImprovementAlphaMarg \\
\hline
\hline
\end{tabular}
\caption{Upper limits at the 95\% credible level on $\Omega_{\rm ref}$ under the power law model for the GWB.
We show upper limits conditioned on different fixed power law indices $\alpha$, as well as a marginalized limit obtained by integration over $\alpha$, using a Gaussian prior with zero mean and a standard deviation of 3.5.
We show the results using a prior that is uniform in $\Omega_{\rm ref}$, as well as uniform in $\log \Omega_{\rm ref}$. 
As described in the main text, the uniform upper limits are more conservative, while the log uniform priors are more sensitive to weak signals.
We also compare with the upper limits from \cite{stoch_O2}, and give the improvement factor we achieve using O3 data.}
\label{tab:ULs}
\end{table*}

\subsection{Non-GR polarizations}
\label{sec:Non-GR}

We can use our results to constrain modifications to GR by using the SVT-PL model defined in Section~\ref{sec:Methods-PE}. 
This analysis benefits from the inclusion of Virgo data, since adding more detectors to the network can help distinguish between different polarizations, as shown in~\cite{TestingGR_stoch}. 
We note that $\Omega_{\rm GW}$ does not necessarily have the interpretation of an energy density in modified theories of gravity, and it is in general more appropriate to think of these quantities as a measure of the strain power in each polarization~\cite{Isi_Stein_NonGR_Energy}.

We use the log-uniform prior on each strength $\Omega_{\rm ref}^{\rm (p)}$ and the Gaussian prior for each spectral index $\alpha_{\rm p}$, as described in the previous section. 
We show the results in Table~\ref{tab:NonGRULs}.
Marginalizing over the spectral indices for each polarization, we find that the upper limit on a scalar-polarized GWB in this model is $\Omega_{\rm GW}^{\rm (S)}(25\ {\rm Hz})\leq \SLimit{}$, the limit on a vector GWB is $\Omega_{\rm GW}^{\rm (V)}(25\ {\rm Hz})\leq \VLimit{}$, and the limit on a tensor GWB is $\Omega_{\rm GW}^{\rm (T)}(25\ {\rm Hz})\leq \TLimit{}$.
Note that the upper limit on tensor modes in this analysis is slightly different from the upper limit when we consider only GR modes given in the previous section, because of the inclusion of additional parameters.
We compute that the ${\rm log}_{10}$ Bayes factor of the non-GR to GR hypotheses is $\NonGRToGRBF{}$ and the ${\rm log}_{10}$ Bayes factor of the hypothesis that any polarization to be present, to the hypothesis that only noise is present, is $\SignalToNoiseNonGRLike{}$. 
Note that to compute the Bayes factors, we include prior odds between different non-GR hypotheses as described in \cite{TestingGR_stoch}.
This confirms there is no evidence of non-GR polarizations.
The non-detection of scalar and vector polarized GWBs is consistent with predictions of GR. 

\begin{table}
\begin{tabular}{c|c|c|c}
Polarization &  O3  & O2 \cite{stoch_O2} & Improvement \\
\hline
\hline
Tensor & \TLimit & $3.2 \times 10^{-8}$ & \ImprovementTensor \\
Vector & \VLimit & $2.9 \times 10^{-8}$ & \ImprovementVector \\
Scalar & \SLimit & $6.1 \times 10^{-8}$ & \ImprovementScalar  \\
\end{tabular}
\caption{Upper limits at the 95\% credible level on $\Omega_{\rm ref}$ for scalar, vector, and tensor polarizations, along with the improvement of the O3 result over the previous result from O2. We use the log-uniform prior for $\Omega_{\rm ref}$ and a Gaussian prior on the spectral index for each polarization, as described in the main text.}
\label{tab:NonGRULs}
\end{table}

\subsection{Joint fit for GWB and magnetic noise}
\label{sec:JointGWMag}

We extend the standard analysis to do a joint fit allowing for both a GWB with an arbitrary power-law index, as well as an apparent GWB arising from correlated magnetic noise. While we have already seen that correlated magnetic noise is below the O3 sensitivity in Section~\ref{sec:MagneticBudget}, the analysis presented here is complementary because it allows us to simultaneously fit for the presence of both a GWB of astrophysical origin and a correlated magnetic noise component. In future runs, this kind of joint fit will become increasingly important. We use the method described in \cite{Meyers:2020qrb}. 

We evaluate whether correlated magnetic noise is detected by first constructing a likelihood function that includes a model for both the correlated magnetic noise and a power-law GWB, $\Omega_{\rm M}(f|\Theta) = \Omega_{\rm PL}(f|\mathbf{\Theta_{\rm PL}}) + \Omega_{\rm MAG}(f|\mathbf{\Theta}_{\rm MAG})$.
Our model $\Omega_{\rm MAG}(f|\mathbf{\Theta}_{\rm MAG})$  takes the same form as Eq.~\ref{eq:magnetic:model}. However, rather than use the coupling functions measured using magnetic-field injections, we model the coupling functions as power laws, which approximate the frequency dependence of the measurements. The vector $\mathbf{\Theta}_{\rm MAG}$ contains the parameters of the model for the coupling functions $T_{I,J}(f)$, which we take to be a simple power law 
\begin{equation}
|T_I(f)| = \kappa_I \left(\frac{f}{10\ {\rm Hz}}\right)^{-\beta_I}.
\end{equation}
The parameters for the power law GWB are the strength $\Omega_{\rm ref}$ and spectral index $\alpha$. 
We use nested sampling to estimate the model evidences for three separate models: N, MAG, and PL+MAG, using the notation defined in Section~\ref{sec:Methods-PE}.

Our prior distribution for the magnitude $\kappa_I$ is log uniform from $10^{-25}$ to $10^{-22}\ {\rm pT^{-1}}$ for all of the detectors. Our prior on the spectral index $\beta_{I}$ is uniform from $\beta_{I}^{\rm min}$ to $\beta_{I}^{\rm max}$, the minimum and maximum values of the spectral index for the magnetic coupling measured at detector $I$ during the O3 run. For Hanford, Livingston and Virgo, the $\beta$ priors chosen  for the study are (0, 12), (1, 10) and (0, 7), respectively. 
The chosen prior range is large enough to encompass all measured coupling function measurements in O3, including the uncertainties mentioned in Section~\ref{sec:DataQuality}.
We find $\rm{log}_{10} \mathcal{B}_{\textrm{N}}^{\textrm{MAG}} = \SchumannVsNoiseLogBayesFactor$, which indicates that there is no preference for a model with correlated magnetic noise compared to a model with only uncorrelated Gaussian noise. 
We also consider a model with a power-law GWB present, using the log-uniform prior on $\Omega_{\rm ref}$ and Gaussian prior on $\alpha$ as in Section~\ref{sec:UpperLimits}. We find that the Bayes factor between a model with correlated GWB and magnetic noise, to a model with only uncorrelated Gaussian noise, is $\rm{log}_{10} \mathcal{B}_{\textrm{N}}^{\textrm{MAG+PL}} = \SchumannGWVsNoiseLogBayesFactor$, confirming that there is no evidence of a GWB in the data.

\section{Implications for compact binaries}
\label{sec:CBCImplications}

\newcommand{\RateBNSmedian}{320}
\newcommand{\RateBNSupperErr}{490}
\newcommand{\RateBNSlowerErr}{240}
\newcommand{\rateUnits}{\,\mathrm{Gpc}^{-3}\,\mathrm{yr}^{-1}}
\def\MedianRateBBH{{\color{black} \ensuremath{19}}}
\def\UpperErrorRateBBH{{\color{black} \ensuremath{18}}}
\def\LowerErrorRateBBH{{\color{black} \ensuremath{8}}}

\def\MedianOmegaBBH{{\color{black} \ensuremath{5.0}}}
\def\UpperErrorOmegaBBH{{\color{black} \ensuremath{1.7}}}
\def\LowerErrorOmegaBBH{{\color{black} \ensuremath{1.4}}}

\def\MedianOmegaBNS{{\color{black} \ensuremath{2.1}}}
\def\UpperErrorOmegaBNS{{\color{black} \ensuremath{2.9}}}
\def\LowerErrorOmegaBNS{{\color{black} \ensuremath{1.6}}}

\def\UpperLimitOmegaNSBH{{\color{black} \ensuremath{8.4}}}

\def\MedianOmegaTotal{{\color{black} \ensuremath{7.2}}}
\def\UpperErrorOmegaTotal{{\color{black} \ensuremath{3.3}}}
\def\LowerErrorOmegaTotal{{\color{black} \ensuremath{2.3}}}

\def\TotalOmegaUpperLimit{{\color{black} \ensuremath{1.9}}}

With upper limits on the GWB in hand, we now explore the implications of these results for the GWB due to CBCs.
We first compare our upper limits to updated predictions for the energy-density due to CBC sources.
We then combine our limits with the direct detections of CBCs in the local Universe to constrain the merger rate of compact binaries at large redshifts.

\subsection{Fiducial model}
\label{sec:CBCFiducial}

Observations from O3a have significantly increased our knowledge of the compact binary population~\cite{O3aCatalog,O3aRatesAndPop,gw_190412,
gw_190521,gw_190521_implications,gw_190814}.
Here, we update the fiducial model of the GWB due to compact binaries~\cite{gw150914_stoch,stoch_O1,gw170817_stoch,stoch_O2} in accordance with the latest observational and theoretical advances.
The energy-density spectrum due to a particular source class $k$ is
	\begin{equation}
	\label{eq:omg-gw-model}
	\Omega_k(f) = \frac{f}{\rho_c} \int_0^{z_\mathrm{max}} dz \frac{R_k(z) \big\langle (dE_s/df_s)|_{f_s} \big\rangle_k}{(1+z) H(z)},
	\end{equation}
where $R_k(z)$ is the source-frame merger rate per comoving volume of objects of class $k$ and $H(z) = H_0 \sqrt{\Omega_m(1+z)^3 + \Omega_\Lambda}$ is the Hubble parameter, where $\Omega_m$ is the fraction of the critical energy density $\rho_c$ contained in matter and $\Omega_\Lambda$ the fraction contained in the cosmological constant; we take $\Omega_m=1-\Omega_\Lambda=0.3065$~\cite{Planck_2015}.
The quantity $\big\langle \frac{dE_s}{df_s}|_{f_s} \big\rangle_k$ is the source-frame energy radiated by a single source, evaluated at the source frequency $f_s=f(1+z)$ and averaged over the ensemble properties of the given class $k$:
	\begin{equation}
	\left\langle \frac{dE_s}{df_s} \right\rangle_k = \int d\phi \,p_k(\phi) \frac{dE_s}{df_s}(\phi),
	\end{equation}
where $p_k(\phi)$ is the probability distribution of source parameters $\phi$ (e.g. masses, spins, etc.) across class $k$.

We consider here three classes of compact binaries: binary black holes (BBHs), binary neutron stars (BNSs), and neutron-star--black-holes (NSBHs). Except where otherwise stated, we use the same choices for $dE_s/df_s$, $p_k(\phi)$, and $R_k(z)$ as in \cite{stoch_O2}. We note that there are several important astrophysical uncertainties which are not included in our fiducial model, which could potentially have an impact on our predictions. These include the possibility that the initial mass function can lead to a lower number of neutron stars than what we assume \cite{AstroUncertaintyIMF}; indications that the star formation rate may peak at a smaller redshift \cite{AstroUncertaintySFRPeak}; and uncertainty in the metallicity evolution. 

\textit{Binary black holes.}
We assume that BBH formation follows a metallicity-weighted star formation rate (SFR) with a distribution $p(t_d)\sim t_d^{-1}$ of time delays $t_d$ between binary formation and merger, where $50\,\mathrm{Myr}\leq t_d \leq 13.5\,\mathrm{Gyr}$.
We take the SFR from Ref.~\cite{VangioniEA_2015}, and multiply it by the fraction of stellar formation occurring at metallicities $Z<Z_\mathrm{thresh}$.
In Ref.~\cite{stoch_O2}, we adopted $Z_\mathrm{thresh} = 0.5\,Z_\odot$, and applied this threshold only to black holes above $30 M_\odot$.
Here, we adopt a more stringent cutoff $Z_\mathrm{thresh}=0.1\,Z_\odot$~\cite{2019MNRAS.482.5012C,mapelli_properties_2019}.
Moreover, we apply this weighting across the entire mass spectrum, as recent population synthesis studies suggest that the mass spectrum of BH mergers does not evolve appreciably with redshift~\cite{mapelli_properties_2019}.

We additionally update our assumptions regarding the mass and spin distributions of BBHs.
In Ref.~\cite{stoch_O2}, we assumed that BBHs had aligned dimensionless spin magnitudes distributed uniformly between $-1$ to $1$.
It now appears, though, that the BBH population exhibits small effective spins~\cite{Miller:2020zox,O3aRatesAndPop}, and so when computing $\Omega_\mathrm{BBH}(f)$ we now assume that BBHs have negligibly small spin.
We also adopt a close variant of the \textsc{Broken Power Law} model of Ref.~\cite{O3aRatesAndPop} to describe the mass distribution of BBHs (for convenience we assume a sharp low-mass cutoff in the BBH mass spectrum, corresponding to $\delta_m\to 0$ in Eq.~(B6) of \cite{O3aRatesAndPop}).
We do not assume fixed values for the parameters of this model, but include our uncertainty on the BBH mass spectrum as an additional systematic uncertainty in our estimate of $\Omega_\mathrm{BBH}(f)$.
To achieve this, we use GWTC-2~\cite{O3aCatalog} to hierarchically compute a joint posterior on the mass distribution and local merger rate of BBHs, given the assumed redshift distribution described above.
Hierarchical inference is performed following the method discussed in Ref.~\cite{O3aRatesAndPop}.
By evaluating Eq.~\eqref{eq:omg-gw-model} across the resulting ensemble of posterior samples, we subsequently obtain a probability distribution on the energy-density spectrum $\Omega_\mathrm{BBH}(f)$ due to BBH mergers, given our knowledge of the local population.

\begin{figure*}
\includegraphics[width=\textwidth]{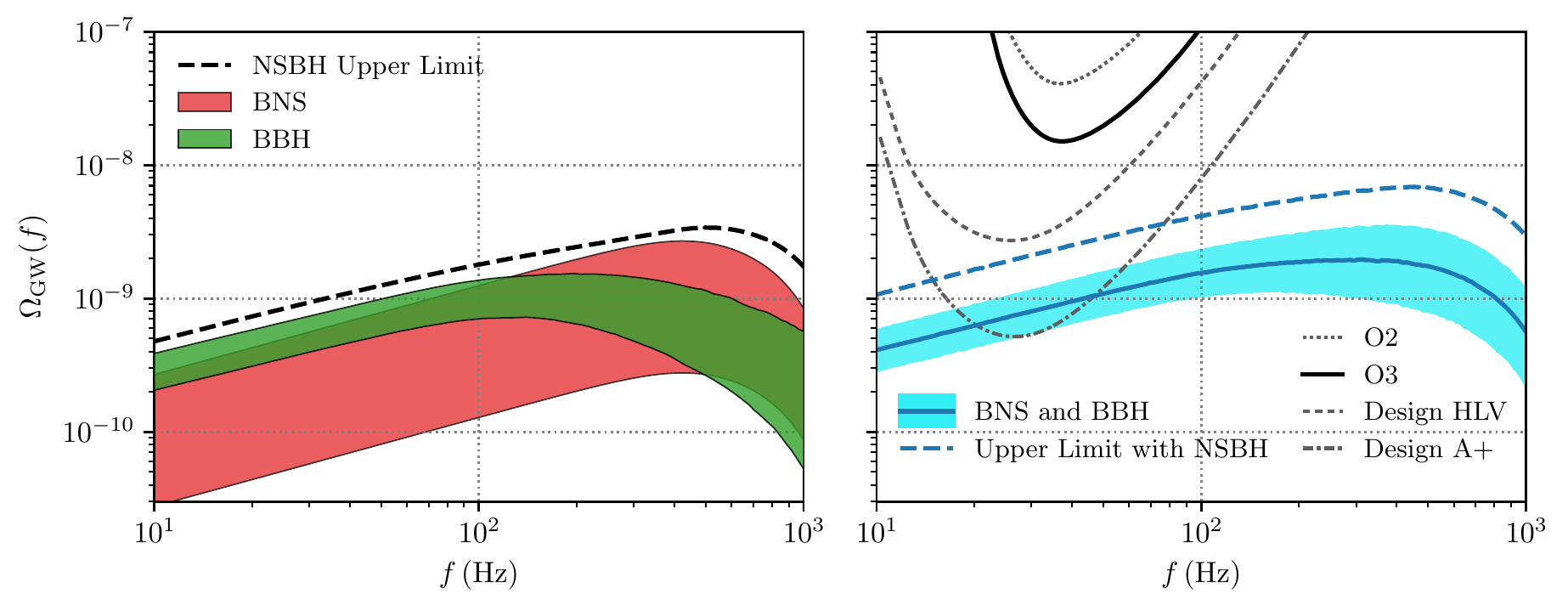}
\caption{Fiducial model predictions for the GWB from BBHs, BNSs, and NSBHs, along with current and projected sensitivity curves. In the left panel we show 90\% credible bands for the GWB contributions from BNS and BBH mergers.
Whereas the BNS uncertainty band illustrates purely the statistical uncertainties in the BNS merger rate, the BBH uncertainty band additionally includes systematic uncertainties in the binary mass distribution, as described in the main text.
As no unambiguous NSBH detections have been made, we only show an upper limit on the possible contribution from such systems.
The right panel compares the combined BBH and BNS energy density spectra, and $2\sigma$ power-law integrated (PI) curves for O2, O3, and projections for the HLV network at design sensitivity, and the A+ detectors. 
The solid blue line shows the median estimate of $\Omega_{\rm BBH+BNS}(f)$ as a function of frequency, while the shaded blue band illustrates 90\% credible uncertainties.
The dashed line, meanwhile, marks our projected upper limit on the total GWB, including our upper limit on the contribution from NSBH mergers. 
}
\label{fig:model-omg}
\end{figure*}

Our updated estimate of $\Omega_\mathrm{BBH}(f)$ is shown in green in Fig.~\ref{fig:model-omg}.
We find $\Omega_{\rm BBH}(25\, {\rm Hz})={\color {black} \MedianOmegaBBH^{+\UpperErrorOmegaBBH}_{-\LowerErrorOmegaBBH} \times 10^{-10}}$.
This uncertainty includes the standard Poisson uncertainty on the local merger rate, which we find to be $R_\mathrm{BBH}(0) = \MedianRateBBH{}^{+\UpperErrorRateBBH{}}_{-\LowerErrorRateBBH{}}\,\mathrm{Gpc}^{-3}\,\mathrm{yr}^{-1}$ (median and symmetric 90\% credible interval) given our fiducial redshift distribution above.
This rate estimate matches that obtained in Ref.~\cite{O3aRatesAndPop} when agnostically allowing the merger rate to evolve with redshift, although in general estimates of $R_\mathrm{BBH}(0)$ may differ under different presumed redshift distributions.
Our estimate of $\Omega_{\rm BBH}(25\, {\rm Hz})$ also reflects, though, the additional \textit{systematic} uncertainty on $\Omega_\mathrm{BBH}(f)$ due to imperfect knowledge of the BBH mass distribution.
This uncertainty on the mass distribution is, for example, responsible for the larger uncertainty in $\Omega_\mathrm{BBH}(f)$ at high frequencies.

\textit{Binary neutron stars.}
As in~\cite{stoch_O2}, we assume that the rate of BNS progenitor formation is proportional to the rate of star formation~\cite{VangioniEA_2015} and that the distribution of time delays $t_d$ between their formation and merger is of the form $p(t_d) \propto t_d^{-1}$ between $20\,\mathrm{Myr}\leq t_d\leq 13.5\,\mathrm{Gyr}$.
The detection of a second binary neutron star merger in O3a, GW190425 \cite{gw_190425}, has decreased uncertainty on the BNS merger rate and demonstrated that at least some neutron star mergers contain significantly heavier masses than expected.
Following~\cite{O3aRatesAndPop}, we assume a uniform distribution of component masses between $1-2.5\,M_\odot$, which yields an estimated present-day merger rate of $R_\mathrm{BNS}(0) = \RateBNSmedian^{+\RateBNSupperErr}_{-\RateBNSlowerErr}\rateUnits$.
When modeling $\Omega_\mathrm{BNS}(f)$, we consider the energy radiated during the inspiral phase only, truncating the BNS energy spectra $\frac{dE_s}{df_s}(\phi)$ at frequencies corresponding to the innermost stable circular orbit.
Our estimate of the BNS GWB is shown in red in Fig.~\ref{fig:model-omg}.
We find $\Omega_{\rm BNS}(25\, {\rm Hz})={\color{black} \MedianOmegaBNS^{+\UpperErrorOmegaBNS}_{-\LowerErrorOmegaBNS} \times 10^{-10}}$.

\textit{Neutron star-black hole binaries.}
To date, Advanced LIGO and Virgo have made no confirmed detections of neutron star-black hole (NSBH) mergers.
Two events, GW190814 and the low-significance candidate GW190426\_152155, have secondary masses constrained below $3\,M_\odot$ with primary masses above $3\,M_\odot$ and so are possibly \textit{consistent} with NSBH systems, but their true physical natures remain unknown~\cite{gw_190814,O3aCatalog}.
In order to forecast the possible contribution of NSBH mergers to the GWB, we therefore use the upper limit on the NSBH merger rate previously adopted in Ref.~\cite{stoch_O2}, again assuming a delta-function mass distribution at $10\,M_\odot + 1.4\,M_\odot$.
We estimate $\Omega_{\rm NSBH}(f)$ using the same redshift distribution as adopted for BBH mergers, and include contributions from the complete inspiral, merger, and ringdown.
This likely results in an \textit{overestimate} of $\Omega_\mathrm{NSBH}(f)$ at high frequencies, since some fraction of NSBH inspirals are expected to end in tidal disruption of the neutron star companion \cite{PhysRevD.83.024005,PhysRevD.82.044049,Kawaguchi:2015bwa}.
The resulting upper limit on $\Omega_\mathrm{NSBH}(f)$ is shown as a dashed black line in Fig.~\ref{fig:model-omg}, with $\Omega_\mathrm{NSBH}(25\,{\rm Hz}) \leq \UpperLimitOmegaNSBH\times10^{-10}$.

\textit{Total CBC GWB.}
In the right-hand side of Fig.~\ref{fig:model-omg} we present an updated estimate of the combined GWB due to BBH and BNS mergers.
Under our model, we predict this combined background to be $\Omega_{\rm BBH+BNS}(25\, {\rm Hz})={\color{black} \MedianOmegaTotal^{+\UpperErrorOmegaTotal}_{-\LowerErrorOmegaTotal} \times 10^{-10}}$.
Combining the upper limit on $\Omega_{\rm NSBH}(f)$ with the upper 95\% credible bound on the contributions from BBH and BNS mergers, we bound the total expected GWB to be $\Omega_{\rm Total}(25\,{\rm Hz}) \leq \TotalOmegaUpperLimit\times 10^{-9}$.
We also show the $2\sigma$ power-law integrated (PI) curves \cite{locus} indicating the integrated sensitivity of the O3 search~\cite{locus}, along with projections for 2 years of the Advanced LIGO-Virgo network at design sensitivity, and the envisioned A+ design sensitivity after 2 years, assuming a 50\% duty cycle.
We use the power spectra available from \cite{Aasi:2013wya,new_design}. 
Previous work has shown that the residual background obtained after subtracting resolvable signals is expected to be within 10\% of the total background for Advanced LIGO and Virgo at design sensitivity, and approximately a factor of 2 smaller for the A+ detectors \cite{Regimbau:2016ike}. 
These curves indicate that by the time the detectors reach the A+ design sensitivity, much of the expected parameter space of the compact binary GWB will be accessible by ground-based detectors.
The continued addition of new instruments to the worldwide detector network, like KAGRA~\cite{kagra_ref} and LIGO-India~\cite{iyer_proposal_2011}, is expected to further improve upon our projected sensitivity.

\subsection{Constraining the BBH merger rate}
\label{sec:CBCMergerRate}

The energy-density spectra in Fig.~\ref{fig:model-omg} show our current best estimates for the GWB under an astrophysically plausible model for the rate density $R_{\rm BBH}(z)$ of BBH mergers of stellar origin.
By combining direct detections of compact binaries with upper limits on the GWB, however, we can alternatively seek to directly measure $R_{\rm BBH}(z)$.
Here, we update constraints on the rate evolution of BBHs from~\cite{Callister:2020arv}, using the latest O3 limits on the GWB and the GWTC-2 ensemble of BBH detections.
We again assume a \textsc{Broken Power Law} form for the mass distribution of BBH mergers, but now adopt a phenomenologically-parametrized form
	\begin{equation}
	R_{\rm BBH}(z) = \mathcal{C}(\lambda_1,\lambda_2,z_{\rm peak}) \frac{R_0 \left(1+z\right)^{\lambda_1}}{1+ \left(\frac{1+z}{1+z_{\rm peak}}\right)^{\lambda_1+\lambda_2}}
	\end{equation}
for their merger rate density.
Under this form, the merger rate evolves as $R_{\rm BBH}(z) \approx \left(1+z\right)^{\lambda_1}$ at $z\lesssim z_{\rm peak}$ and $R_{\rm BBH} \approx \left(1+z\right)^{-\lambda_2}$ at $z\gtrsim z_{\rm peak}$, and at low redshifts $\lambda_1$ can be identified with the parameter $\kappa$ of Ref.~\cite{O3aRatesAndPop}.
The normalization constant $\mathcal{C}(\lambda_1,\lambda_2,z_{\rm peak})$ is defined such that $R_0$ is the local merger rate density of BBHs at $z=0$.

Using the direct BBH detections from GWTC-2 along with the updated GWB search results presented here, we jointly infer the parameters governing both the mass and redshift distributions of BBH mergers.
We adopt the factorized likelihood from Eq.~\ref{eq:factorized-likelihood}, given by the product between the standard GWB likelihood $p(C^{IJ}(f_k)|\mathbf{\Theta}_{\rm BBH})$ under our model for the BBH background, and the likelihood $p(\{d\} | \mathbf{\Theta_{\rm BBH}})$ of having measured data $\{d\}$ associated with the \NumberOfTotalBBHs{} direct BBH detections in GWTC-2 with false alarm rates $<1\,\mathrm{yr}^{-1}$.
This likelihood $p(\{d\} | \mathbf{\Theta_{\rm BBH}})$ for direct detections is evaluated using posterior samples on the parameters of each individual event, as described further in Sect.~4 of~\cite{O3aRatesAndPop}
The direct detection likelihood also corrects for selection biases, such as LIGO and Virgo's higher detection efficiency for higher-mass systems;
we evaluate selection effects using the same injection campaign discussed in~\cite{O3aRatesAndPop}.
Our priors are uniform on $\lambda_1$, $\lambda_2$, and $z_{\rm peak}$, and log-uniform on $R_0$.

\begin{figure}
\includegraphics[width=0.49\textwidth]{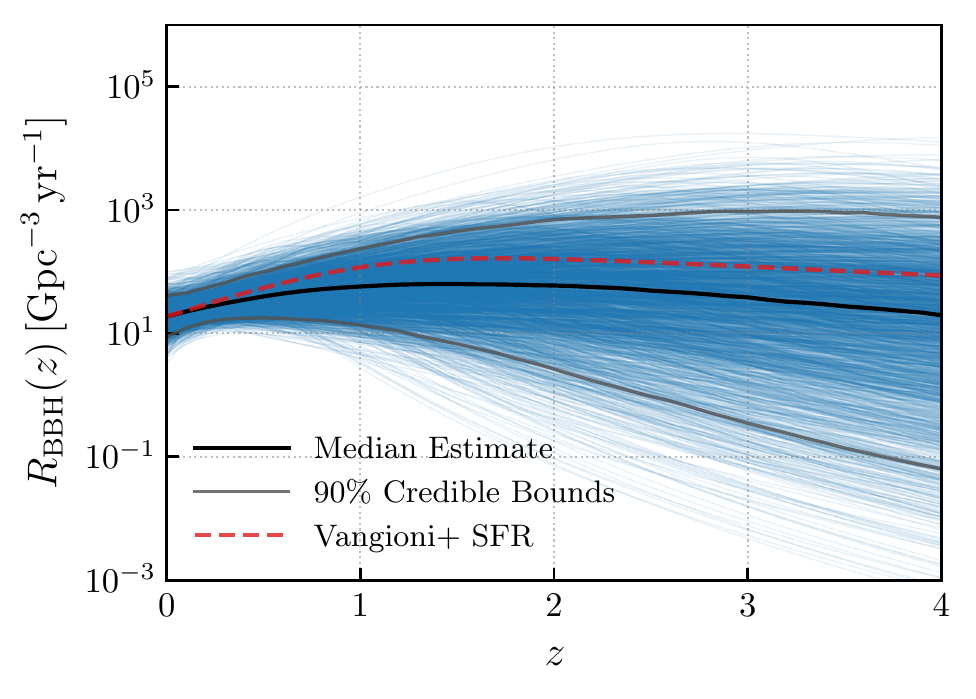}
\caption{
Posterior constraints on the BBH merger rate $R_{\rm BBH}(z)$ as a function of redshift when allowing for a merger rate that peaks and subsequently turns over at high $z$, combining stochastic search results and direct BBH detections.
The black line shows our median estimate of $R_{\rm BBH}(z)$, while solid grey lines denote 90\% credible bounds.
For comparison, the dashed red line is proportional to the rate of cosmic star formation~\cite{VangioniEA_2015}.
At 90\% credibility, the merger rate of BBHs is bounded below $\sim 10^3\,\mathrm{Gpc}^{-3}\,\mathrm{yr}^{-1}$ beyond $z\approx 2$, an order of magnitude improvement relative to O1 and O2~\cite{Callister:2020arv}.
}
\label{fig:r-vs-z}
\end{figure}

The resulting constraints on the BBH merger rate as a function of redshift are shown in Fig.~\ref{fig:r-vs-z}.
Each blue trace represents a single draw from our posterior on the BBH mass distribution and merger rate history.
The black curve marks the median estimated merger rate at a given redshift, while solid grey curves mark our central 90\% credible bound.
From O1 and O2 data, the non-detection of the GWB served to constrain the BBH merger rate to less than $\sim10^4\,\mathrm{Gpc}^{-3}\,\mathrm{yr}^{-1}$ beyond $z\approx 2$ at 90\% credibility~\cite{Callister:2020arv}.
This limit is here improved by a factor of approximately ten.
For reference, the dashed red curve is proportional to the star formation rate model of Ref.~\cite{VangioniEA_2015}.
While the BBH merger rate remains consistent with directly tracing star formation, it likely increases more slowly as a function of redshift, consistent with a non-vanishing time delay distribution between binary formation and merger~\cite{O3aRatesAndPop}.

\begin{figure}
\includegraphics[width=0.49\textwidth]{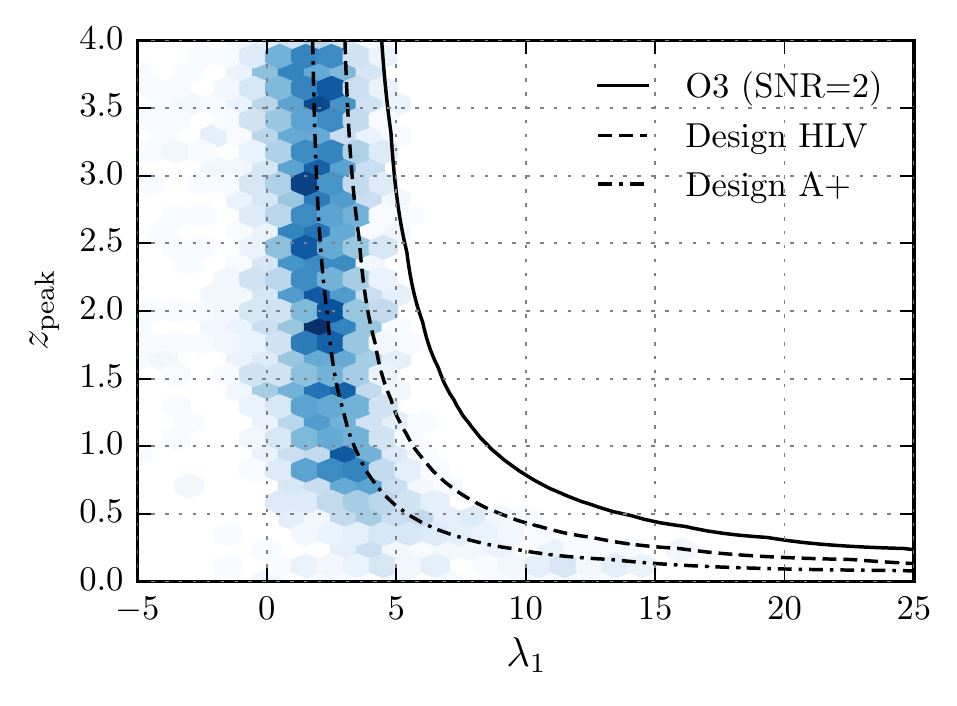}
\caption{2D posterior density for joint CBC-GWB inference on the parameters $\lambda_1$ and $z_{\rm peak}$ describing the BBH redshift distribution, defined in the main text. While the O3 stochastic measurement (solid line) is not competitive with measurements on $z_{peak}$ and $\lambda_1$ obtained from GWTC-2 (shown as a blue posterior density), stochastic background measurements from future observing runs (dashed lines) may be able to put tighter constraints on these parameters in combinations with observations of individual binaries.}
\label{fig:r-posterior}
\end{figure}

While O1/O2 constraints on the behavior of $R_{\rm BBH}(z)$ at redshifts $z\gtrsim 0.5$ were dominated by stochastic search results~\cite{Callister:2020arv}, the results in Fig.~\ref{fig:r-vs-z} from O3 are now due primarily to the direct detections comprising GWTC-2.
The cause for this shift is illustrated in Fig.~\ref{fig:r-posterior}, which shows our joint $\lambda_1-z_{\rm peak}$ posterior (informed by both GWB data and direct BBH detections), marginalized over the remaining parameters governing the BBH mass and redshift distributions.
The solid black contour show the values of $\lambda_1$ and $z_{\rm peak}$ expected to yield a GWB detection with ${\rm SNR} = 2$ in O3; values to the right of this contour can be excluded on the basis of a GWB non-detection.
Direct BBH detections, meanwhile, allow for a measurement of $\lambda_1$, but are not expected to meaningfully constrain $z_{\rm peak}$, which likely lies beyond the horizon of Advanced LIGO and Virgo.
The direct BBH detections in GWTC-1 only allowed for a weak upper limit on $\lambda_1$: $\lambda_1 \leq 13.7$.
The non-detection of the GWB in O2 therefore ruled out a considerable portion of otherwise available parameter space.
Improved measurements due to GWTC-2, though, have revised estimates of $\lambda_1$ downwards to $\lambda_1 = 1.3^{+2.1}_{-2.1}$~\cite{O3aRatesAndPop}, and so present GWB searches cannot further constrain its value.
The results in Fig.~\ref{fig:r-vs-z} are therefore now dominated by direct BBH detections.

With continued data collection, however, the non-detection (or eventual detection) of the GWB may again offer informative constraints on $\lambda_1$ and $z_{\rm peak}$.
As additional direct BBH detections are made, our knowledge of $\lambda_1$ will continue to improve, identifying an increasingly narrow, nearly-vertical contour in the $\lambda_1-z_{\rm peak}$ plane.
Continued time integration in searches for the GWB, meanwhile, will exclude a growing fraction of this plane, ruling out large values of \textit{both} $\lambda_1$ and $z_{\rm peak}$.
In Fig.~\ref{fig:r-posterior}, for example, we show projected exclusion contours corresponding to one year of integration with Advanced LIGO and Virgo, at both their design sensitivity and A+ configurations; both exclusion curves extend into the presently allowed values of $\lambda_1$, where they may again be informative and break the degeneracy between $\lambda_1$ and $z_{\rm peak}$.

\section{Conclusions}
\label{sec:Conclusions}
In this work, we have performed a search for an isotropic GWB using data from Advanced LIGO's and Virgo's first three observing runs. Since we did not find evidence for a background of astrophysical origin, we placed upper limits, improving previous bounds by about a factor of \ImprovementAlphaZero{} for a flat background.

We considered the implications of the results, and by combining the upper limits with measurements from GWTC-2 we have constrained the BBH merger rate as a function of redshift. Our results can be used to constrain additional models such as cosmic strings or phase transitions, using the cross correlation spectra we have made publicly available~\cite{stoch_o3_data_release}. Our results can also be combined with other measurements of the GWB at other frequencies \cite{Lasky:2015lej}.

Moving forward, we expect currently proposed ground-based facilities  such as A+ have the potential to probe a large range of the model space for CBC backgrounds. In order to make full use of the data and confidently claim a detection, it will be important to further develop the methods to handle correlated terrestrial noise sources, such as the magnetic couplings described here.

The authors gratefully acknowledge the support of the United States
National Science Foundation (NSF) for the construction and operation of the
LIGO Laboratory and Advanced LIGO as well as the Science and Technology Facilities Council (STFC) of the
United Kingdom, the Max-Planck-Society (MPS), and the State of
Niedersachsen/Germany for support of the construction of Advanced LIGO 
and construction and operation of the GEO600 detector. 
Additional support for Advanced LIGO was provided by the Australian Research Council.
The authors gratefully acknowledge the Italian Istituto Nazionale di Fisica Nucleare (INFN),  
the French Centre National de la Recherche Scientifique (CNRS) and
the Netherlands Organization for Scientific Research, 
for the construction and operation of the Virgo detector
and the creation and support  of the EGO consortium. 
The authors also gratefully acknowledge research support from these agencies as well as by 
the Council of Scientific and Industrial Research of India, 
the Department of Science and Technology, India,
the Science \& Engineering Research Board (SERB), India,
the Ministry of Human Resource Development, India,
the Spanish Agencia Estatal de Investigaci\'on,
the Vicepresid\`encia i Conselleria d'Innovaci\'o, Recerca i Turisme and the Conselleria d'Educaci\'o i Universitat del Govern de les Illes Balears,
the Conselleria d'Innovaci\'o, Universitats, Ci\`encia i Societat Digital de la Generalitat Valenciana and
the CERCA Programme Generalitat de Catalunya, Spain,
the National Science Centre of Poland and the Foundation for Polish Science (FNP),
the Swiss National Science Foundation (SNSF),
the Russian Foundation for Basic Research, 
the Russian Science Foundation,
the European Commission,
the European Regional Development Funds (ERDF),
the Royal Society, 
the Scottish Funding Council, 
the Scottish Universities Physics Alliance, 
the Hungarian Scientific Research Fund (OTKA),
the French Lyon Institute of Origins (LIO),
the Belgian Fonds de la Recherche Scientifique (FRS-FNRS), 
Actions de Recherche Concertées (ARC) and
Fonds Wetenschappelijk Onderzoek – Vlaanderen (FWO), Belgium,
the Paris \^{I}le-de-France Region, 
the National Research, Development and Innovation Office Hungary (NKFIH), 
the National Research Foundation of Korea,
the Natural Science and Engineering Research Council Canada,
Canadian Foundation for Innovation (CFI),
the Brazilian Ministry of Science, Technology, and Innovations,
the International Center for Theoretical Physics South American Institute for Fundamental Research (ICTP-SAIFR), 
the Research Grants Council of Hong Kong,
the National Natural Science Foundation of China (NSFC),
the Leverhulme Trust, 
the Research Corporation, 
the Ministry of Science and Technology (MOST), Taiwan,
the United States Department of Energy,
and
the Kavli Foundation.
The authors gratefully acknowledge the support of the NSF, STFC, INFN and CNRS for provision of computational resources.

This work was supported by MEXT, JSPS Leading-edge Research Infrastructure Program, JSPS Grant-in-Aid for Specially Promoted Research 26000005, JSPS Grant-in-Aid for Scientific Research on Innovative Areas 2905: JP17H06358, JP17H06361 and JP17H06364, JSPS Core-to-Core Program A. Advanced Research Networks, JSPS Grant-in-Aid for Scientific Research (S) 17H06133, the joint research program of the Institute for Cosmic Ray Research, University of Tokyo, National Research Foundation (NRF) and Computing Infrastructure Project of KISTI-GSDC in Korea, Academia Sinica (AS), AS Grid Center (ASGC) and the Ministry of Science and Technology (MoST) in Taiwan under grants including AS-CDA-105-M06, Advanced Technology Center (ATC) of NAOJ, and Mechanical Engineering Center of KEK. 

 All plots have been
  prepared using Matplotlib \citep{2007CSE.....9...90H}.

{\it We would like to thank all of the essential workers who put their health at risk during the COVID-19 pandemic, without whom we would not have been able to complete this work.}

This document has been assigned the number LIGO-DCC-P2000314.

\bibliographystyle{apsrev4-1}
\bibliography{refs}

\begin{thebibliography}{122}%
\makeatletter
\providecommand \@ifxundefined [1]{%
 \@ifx{#1\undefined}
}%
\providecommand \@ifnum [1]{%
 \ifnum #1\expandafter \@firstoftwo
 \else \expandafter \@secondoftwo
 \fi
}%
\providecommand \@ifx [1]{%
 \ifx #1\expandafter \@firstoftwo
 \else \expandafter \@secondoftwo
 \fi
}%
\providecommand \natexlab [1]{#1}%
\providecommand \enquote  [1]{``#1''}%
\providecommand \bibnamefont  [1]{#1}%
\providecommand \bibfnamefont [1]{#1}%
\providecommand \citenamefont [1]{#1}%
\providecommand \href@noop [0]{\@secondoftwo}%
\providecommand \href [0]{\begingroup \@sanitize@url \@href}%
\providecommand \@href[1]{\@@startlink{#1}\@@href}%
\providecommand \@@href[1]{\endgroup#1\@@endlink}%
\providecommand \@sanitize@url [0]{\catcode `\\12\catcode `\$12\catcode
  `\&12\catcode `\#12\catcode `\^12\catcode `\_12\catcode `\%12\relax}%
\providecommand \@@startlink[1]{}%
\providecommand \@@endlink[0]{}%
\providecommand \url  [0]{\begingroup\@sanitize@url \@url }%
\providecommand \@url [1]{\endgroup\@href {#1}{\urlprefix }}%
\providecommand \urlprefix  [0]{URL }%
\providecommand \Eprint [0]{\href }%
\providecommand \doibase [0]{http://dx.doi.org/}%
\providecommand \selectlanguage [0]{\@gobble}%
\providecommand \bibinfo  [0]{\@secondoftwo}%
\providecommand \bibfield  [0]{\@secondoftwo}%
\providecommand \translation [1]{[#1]}%
\providecommand \BibitemOpen [0]{}%
\providecommand \bibitemStop [0]{}%
\providecommand \bibitemNoStop [0]{.\EOS\space}%
\providecommand \EOS [0]{\spacefactor3000\relax}%
\providecommand \BibitemShut  [1]{\csname bibitem#1\endcsname}%
\let\auto@bib@innerbib\@empty
\bibitem [{\citenamefont {Cornish}\ and\ \citenamefont
  {Romano}(2015)}]{Cornish:2015pda}%
  \BibitemOpen
  \bibfield  {author} {\bibinfo {author} {\bibfnamefont {N.~J.}\ \bibnamefont
  {Cornish}}\ and\ \bibinfo {author} {\bibfnamefont {J.~D.}\ \bibnamefont
  {Romano}},\ }\href {\doibase 10.1103/PhysRevD.92.042001} {\bibfield
  {journal} {\bibinfo  {journal} {Phys. Rev. D}\ }\textbf {\bibinfo {volume}
  {92}},\ \bibinfo {pages} {042001} (\bibinfo {year} {2015})},\ \Eprint
  {http://arxiv.org/abs/1505.08084} {arXiv:1505.08084 [gr-qc]} \BibitemShut
  {NoStop}%
\bibitem [{\citenamefont {Rosado}(2011)}]{PhysRevD.84.084004}%
  \BibitemOpen
  \bibfield  {author} {\bibinfo {author} {\bibfnamefont {P.~A.}\ \bibnamefont
  {Rosado}},\ }\href {\doibase 10.1103/PhysRevD.84.084004} {\bibfield
  {journal} {\bibinfo  {journal} {Phys. Rev. D}\ }\textbf {\bibinfo {volume}
  {84}},\ \bibinfo {pages} {084004} (\bibinfo {year} {2011})}\BibitemShut
  {NoStop}%
\bibitem [{\citenamefont {{Zhu}}\ \emph
  {et~al.}(2011{\natexlab{a}})\citenamefont {{Zhu}}, \citenamefont {{Howell}},
  \citenamefont {{Regimbau}}, \citenamefont {{Blair}},\ and\ \citenamefont
  {{Zhu}}}]{2011ApJ...739...86Z}%
  \BibitemOpen
  \bibfield  {author} {\bibinfo {author} {\bibfnamefont {X.-J.}\ \bibnamefont
  {{Zhu}}}, \bibinfo {author} {\bibfnamefont {E.}~\bibnamefont {{Howell}}},
  \bibinfo {author} {\bibfnamefont {T.}~\bibnamefont {{Regimbau}}}, \bibinfo
  {author} {\bibfnamefont {D.}~\bibnamefont {{Blair}}}, \ and\ \bibinfo
  {author} {\bibfnamefont {Z.-H.}\ \bibnamefont {{Zhu}}},\ }\href {\doibase
  10.1088/0004-637X/739/2/86} {\bibfield  {journal} {\bibinfo  {journal}
  {Astrophys. J.}\ }\textbf {\bibinfo {volume} {739}},\ \bibinfo {eid} {86}
  (\bibinfo {year} {2011}{\natexlab{a}})}\BibitemShut {NoStop}%
\bibitem [{\citenamefont {Marassi}\ \emph {et~al.}(2011)\citenamefont
  {Marassi}, \citenamefont {Schneider}, \citenamefont {Corvino}, \citenamefont
  {Ferrari},\ and\ \citenamefont {Zwart}}]{PhysRevD.84.124037}%
  \BibitemOpen
  \bibfield  {author} {\bibinfo {author} {\bibfnamefont {S.}~\bibnamefont
  {Marassi}}, \bibinfo {author} {\bibfnamefont {R.}~\bibnamefont {Schneider}},
  \bibinfo {author} {\bibfnamefont {G.}~\bibnamefont {Corvino}}, \bibinfo
  {author} {\bibfnamefont {V.}~\bibnamefont {Ferrari}}, \ and\ \bibinfo
  {author} {\bibfnamefont {S.~P.}\ \bibnamefont {Zwart}},\ }\href {\doibase
  10.1103/PhysRevD.84.124037} {\bibfield  {journal} {\bibinfo  {journal} {Phys.
  Rev. D}\ }\textbf {\bibinfo {volume} {84}},\ \bibinfo {pages} {124037}
  (\bibinfo {year} {2011})}\BibitemShut {NoStop}%
\bibitem [{\citenamefont {Wu}\ \emph {et~al.}(2012)\citenamefont {Wu},
  \citenamefont {Mandic},\ and\ \citenamefont {Regimbau}}]{PhysRevD.85.104024}%
  \BibitemOpen
  \bibfield  {author} {\bibinfo {author} {\bibfnamefont {C.}~\bibnamefont
  {Wu}}, \bibinfo {author} {\bibfnamefont {V.}~\bibnamefont {Mandic}}, \ and\
  \bibinfo {author} {\bibfnamefont {T.}~\bibnamefont {Regimbau}},\ }\href
  {\doibase 10.1103/PhysRevD.85.104024} {\bibfield  {journal} {\bibinfo
  {journal} {Phys. Rev. D}\ }\textbf {\bibinfo {volume} {85}},\ \bibinfo
  {pages} {104024} (\bibinfo {year} {2012})}\BibitemShut {NoStop}%
\bibitem [{\citenamefont {{Zhu}}\ \emph {et~al.}(2013)\citenamefont {{Zhu}},
  \citenamefont {{Howell}}, \citenamefont {{Blair}},\ and\ \citenamefont
  {{Zhu}}}]{2013MNRAS.431..882Z}%
  \BibitemOpen
  \bibfield  {author} {\bibinfo {author} {\bibfnamefont {X.-J.}\ \bibnamefont
  {{Zhu}}}, \bibinfo {author} {\bibfnamefont {E.~J.}\ \bibnamefont {{Howell}}},
  \bibinfo {author} {\bibfnamefont {D.~G.}\ \bibnamefont {{Blair}}}, \ and\
  \bibinfo {author} {\bibfnamefont {Z.-H.}\ \bibnamefont {{Zhu}}},\ }\href
  {\doibase 10.1093/mnras/stt207} {\bibfield  {journal} {\bibinfo  {journal}
  {Mon. Not. R. Ast. Soc.}\ }\textbf {\bibinfo {volume} {431}},\ \bibinfo
  {pages} {882} (\bibinfo {year} {2013})}\BibitemShut {NoStop}%
\bibitem [{\citenamefont {Buonanno}\ \emph {et~al.}(2005)\citenamefont
  {Buonanno}, \citenamefont {Sigl}, \citenamefont {Raffelt}, \citenamefont
  {Janka},\ and\ \citenamefont {Muller}}]{Buonanno:2004tp}%
  \BibitemOpen
  \bibfield  {author} {\bibinfo {author} {\bibfnamefont {A.}~\bibnamefont
  {Buonanno}}, \bibinfo {author} {\bibfnamefont {G.}~\bibnamefont {Sigl}},
  \bibinfo {author} {\bibfnamefont {G.~G.}\ \bibnamefont {Raffelt}}, \bibinfo
  {author} {\bibfnamefont {H.-T.}\ \bibnamefont {Janka}}, \ and\ \bibinfo
  {author} {\bibfnamefont {E.}~\bibnamefont {Muller}},\ }\href {\doibase
  10.1103/PhysRevD.72.084001} {\bibfield  {journal} {\bibinfo  {journal} {Phys.
  Rev.}\ }\textbf {\bibinfo {volume} {D72}},\ \bibinfo {pages} {084001}
  (\bibinfo {year} {2005})},\ \Eprint {http://arxiv.org/abs/astro-ph/0412277}
  {arXiv:astro-ph/0412277 [astro-ph]} \BibitemShut {NoStop}%
\bibitem [{\citenamefont {{Howell}}\ \emph {et~al.}(2004)\citenamefont
  {{Howell}}, \citenamefont {{Coward}}, \citenamefont {{Burman}}, \citenamefont
  {{Blair}},\ and\ \citenamefont {{Gilmore}}}]{2004MNRAS.351.1237H}%
  \BibitemOpen
  \bibfield  {author} {\bibinfo {author} {\bibfnamefont {E.}~\bibnamefont
  {{Howell}}}, \bibinfo {author} {\bibfnamefont {D.}~\bibnamefont {{Coward}}},
  \bibinfo {author} {\bibfnamefont {R.}~\bibnamefont {{Burman}}}, \bibinfo
  {author} {\bibfnamefont {D.}~\bibnamefont {{Blair}}}, \ and\ \bibinfo
  {author} {\bibfnamefont {J.}~\bibnamefont {{Gilmore}}},\ }\href {\doibase
  10.1111/j.1365-2966.2004.07863.x} {\bibfield  {journal} {\bibinfo  {journal}
  {\mnras}\ }\textbf {\bibinfo {volume} {351}},\ \bibinfo {pages} {1237}
  (\bibinfo {year} {2004})}\BibitemShut {NoStop}%
\bibitem [{\citenamefont {Sandick}\ \emph
  {et~al.}(2006{\natexlab{a}})\citenamefont {Sandick}, \citenamefont {Olive},
  \citenamefont {Daigne},\ and\ \citenamefont {Vangioni}}]{Sandick:2006sm}%
  \BibitemOpen
  \bibfield  {author} {\bibinfo {author} {\bibfnamefont {P.}~\bibnamefont
  {Sandick}}, \bibinfo {author} {\bibfnamefont {K.~A.}\ \bibnamefont {Olive}},
  \bibinfo {author} {\bibfnamefont {F.}~\bibnamefont {Daigne}}, \ and\ \bibinfo
  {author} {\bibfnamefont {E.}~\bibnamefont {Vangioni}},\ }\href {\doibase
  10.1103/PhysRevD.73.104024} {\bibfield  {journal} {\bibinfo  {journal} {Phys.
  Rev.}\ }\textbf {\bibinfo {volume} {D73}},\ \bibinfo {pages} {104024}
  (\bibinfo {year} {2006}{\natexlab{a}})},\ \Eprint
  {http://arxiv.org/abs/astro-ph/0603544} {arXiv:astro-ph/0603544 [astro-ph]}
  \BibitemShut {NoStop}%
\bibitem [{\citenamefont {{Marassi}}\ \emph {et~al.}(2009)\citenamefont
  {{Marassi}}, \citenamefont {{Schneider}},\ and\ \citenamefont
  {{Ferrari}}}]{2009MNRAS.398..293M}%
  \BibitemOpen
  \bibfield  {author} {\bibinfo {author} {\bibfnamefont {S.}~\bibnamefont
  {{Marassi}}}, \bibinfo {author} {\bibfnamefont {R.}~\bibnamefont
  {{Schneider}}}, \ and\ \bibinfo {author} {\bibfnamefont {V.}~\bibnamefont
  {{Ferrari}}},\ }\href {\doibase 10.1111/j.1365-2966.2009.15120.x} {\bibfield
  {journal} {\bibinfo  {journal} {Mon. Not. R. Ast. Soc.}\ }\textbf {\bibinfo
  {volume} {398}},\ \bibinfo {pages} {293} (\bibinfo {year}
  {2009})}\BibitemShut {NoStop}%
\bibitem [{\citenamefont {{Zhu}}\ \emph {et~al.}(2010)\citenamefont {{Zhu}},
  \citenamefont {{Howell}},\ and\ \citenamefont
  {{Blair}}}]{2010MNRAS.409L.132Z}%
  \BibitemOpen
  \bibfield  {author} {\bibinfo {author} {\bibfnamefont {X.-J.}\ \bibnamefont
  {{Zhu}}}, \bibinfo {author} {\bibfnamefont {E.}~\bibnamefont {{Howell}}}, \
  and\ \bibinfo {author} {\bibfnamefont {D.}~\bibnamefont {{Blair}}},\ }\href
  {\doibase 10.1111/j.1745-3933.2010.00965.x} {\bibfield  {journal} {\bibinfo
  {journal} {Mon. Not. R. Ast. Soc.}\ }\textbf {\bibinfo {volume} {409}},\
  \bibinfo {pages} {L132} (\bibinfo {year} {2010})}\BibitemShut {NoStop}%
\bibitem [{\citenamefont {Ferrari}\ \emph {et~al.}(1999)\citenamefont
  {Ferrari}, \citenamefont {Matarrese},\ and\ \citenamefont
  {Schneider}}]{Ferrari:1998jf}%
  \BibitemOpen
  \bibfield  {author} {\bibinfo {author} {\bibfnamefont {V.}~\bibnamefont
  {Ferrari}}, \bibinfo {author} {\bibfnamefont {S.}~\bibnamefont {Matarrese}},
  \ and\ \bibinfo {author} {\bibfnamefont {R.}~\bibnamefont {Schneider}},\
  }\href {\doibase 10.1046/j.1365-8711.1999.02207.x} {\bibfield  {journal}
  {\bibinfo  {journal} {Mon. Not. Roy. Astron. Soc.}\ }\textbf {\bibinfo
  {volume} {303}},\ \bibinfo {pages} {258} (\bibinfo {year} {1999})},\ \Eprint
  {http://arxiv.org/abs/astro-ph/9806357} {arXiv:astro-ph/9806357 [astro-ph]}
  \BibitemShut {NoStop}%
\bibitem [{\citenamefont {Regimbau}\ and\ \citenamefont
  {de~Freitas~Pacheco}(2001)}]{Regimbau:2001kx}%
  \BibitemOpen
  \bibfield  {author} {\bibinfo {author} {\bibfnamefont {T.}~\bibnamefont
  {Regimbau}}\ and\ \bibinfo {author} {\bibfnamefont {J.~A.}\ \bibnamefont
  {de~Freitas~Pacheco}},\ }\href {\doibase 10.1051/0004-6361:20011005}
  {\bibfield  {journal} {\bibinfo  {journal} {Astron. Astrophys.}\ }\textbf
  {\bibinfo {volume} {376}},\ \bibinfo {pages} {381} (\bibinfo {year}
  {2001})},\ \Eprint {http://arxiv.org/abs/astro-ph/0105260}
  {arXiv:astro-ph/0105260 [astro-ph]} \BibitemShut {NoStop}%
\bibitem [{\citenamefont {{Howell}}\ \emph {et~al.}(2011)\citenamefont
  {{Howell}}, \citenamefont {{Regimbau}}, \citenamefont {{Corsi}},
  \citenamefont {{Coward}},\ and\ \citenamefont
  {{Burman}}}]{2011MNRAS.410.2123H}%
  \BibitemOpen
  \bibfield  {author} {\bibinfo {author} {\bibfnamefont {E.}~\bibnamefont
  {{Howell}}}, \bibinfo {author} {\bibfnamefont {T.}~\bibnamefont
  {{Regimbau}}}, \bibinfo {author} {\bibfnamefont {A.}~\bibnamefont {{Corsi}}},
  \bibinfo {author} {\bibfnamefont {D.}~\bibnamefont {{Coward}}}, \ and\
  \bibinfo {author} {\bibfnamefont {R.}~\bibnamefont {{Burman}}},\ }\href
  {\doibase 10.1111/j.1365-2966.2010.17585.x} {\bibfield  {journal} {\bibinfo
  {journal} {\mnras}\ }\textbf {\bibinfo {volume} {410}},\ \bibinfo {pages}
  {2123} (\bibinfo {year} {2011})},\ \Eprint {http://arxiv.org/abs/1008.3941}
  {arXiv:1008.3941 [astro-ph.HE]} \BibitemShut {NoStop}%
\bibitem [{\citenamefont {{Zhu}}\ \emph
  {et~al.}(2011{\natexlab{b}})\citenamefont {{Zhu}}, \citenamefont {{Fan}},\
  and\ \citenamefont {{Zhu}}}]{2011ApJ...729...59Z}%
  \BibitemOpen
  \bibfield  {author} {\bibinfo {author} {\bibfnamefont {X.-J.}\ \bibnamefont
  {{Zhu}}}, \bibinfo {author} {\bibfnamefont {X.-L.}\ \bibnamefont {{Fan}}}, \
  and\ \bibinfo {author} {\bibfnamefont {Z.-H.}\ \bibnamefont {{Zhu}}},\ }\href
  {\doibase 10.1088/0004-637X/729/1/59} {\bibfield  {journal} {\bibinfo
  {journal} {\apj}\ }\textbf {\bibinfo {volume} {729}},\ \bibinfo {eid} {59}
  (\bibinfo {year} {2011}{\natexlab{b}})}\BibitemShut {NoStop}%
\bibitem [{\citenamefont {{Marassi}}\ \emph {et~al.}(2011)\citenamefont
  {{Marassi}}, \citenamefont {{Ciolfi}}, \citenamefont {{Schneider}},
  \citenamefont {{Stella}},\ and\ \citenamefont
  {{Ferrari}}}]{2011MNRAS.411.2549M}%
  \BibitemOpen
  \bibfield  {author} {\bibinfo {author} {\bibfnamefont {S.}~\bibnamefont
  {{Marassi}}}, \bibinfo {author} {\bibfnamefont {R.}~\bibnamefont {{Ciolfi}}},
  \bibinfo {author} {\bibfnamefont {R.}~\bibnamefont {{Schneider}}}, \bibinfo
  {author} {\bibfnamefont {L.}~\bibnamefont {{Stella}}}, \ and\ \bibinfo
  {author} {\bibfnamefont {V.}~\bibnamefont {{Ferrari}}},\ }\href {\doibase
  10.1111/j.1365-2966.2010.17861.x} {\bibfield  {journal} {\bibinfo  {journal}
  {\mnras}\ }\textbf {\bibinfo {volume} {411}},\ \bibinfo {pages} {2549}
  (\bibinfo {year} {2011})},\ \Eprint {http://arxiv.org/abs/1009.1240}
  {arXiv:1009.1240} \BibitemShut {NoStop}%
\bibitem [{\citenamefont {Rosado}(2012)}]{Rosado:2012bk}%
  \BibitemOpen
  \bibfield  {author} {\bibinfo {author} {\bibfnamefont {P.~A.}\ \bibnamefont
  {Rosado}},\ }\href {\doibase 10.1103/PhysRevD.86.104007} {\bibfield
  {journal} {\bibinfo  {journal} {Phys. Rev.}\ }\textbf {\bibinfo {volume}
  {D86}},\ \bibinfo {pages} {104007} (\bibinfo {year} {2012})},\ \Eprint
  {http://arxiv.org/abs/1206.1330} {arXiv:1206.1330 [gr-qc]} \BibitemShut
  {NoStop}%
\bibitem [{\citenamefont {{Wu}}\ \emph {et~al.}(2013)\citenamefont {{Wu}},
  \citenamefont {{Mandic}},\ and\ \citenamefont {{Regimbau}}}]{WuEA_2013}%
  \BibitemOpen
  \bibfield  {author} {\bibinfo {author} {\bibfnamefont {C.-J.}\ \bibnamefont
  {{Wu}}}, \bibinfo {author} {\bibfnamefont {V.}~\bibnamefont {{Mandic}}}, \
  and\ \bibinfo {author} {\bibfnamefont {T.}~\bibnamefont {{Regimbau}}},\
  }\href {\doibase 10.1103/PhysRevD.87.042002} {\bibfield  {journal} {\bibinfo
  {journal} {\prd}\ }\textbf {\bibinfo {volume} {87}},\ \bibinfo {eid} {042002}
  (\bibinfo {year} {2013})}\BibitemShut {NoStop}%
\bibitem [{\citenamefont {Lasky}\ \emph {et~al.}(2013)\citenamefont {Lasky},
  \citenamefont {Bennett},\ and\ \citenamefont {Melatos}}]{PhysRevD.87.063004}%
  \BibitemOpen
  \bibfield  {author} {\bibinfo {author} {\bibfnamefont {P.~D.}\ \bibnamefont
  {Lasky}}, \bibinfo {author} {\bibfnamefont {M.~F.}\ \bibnamefont {Bennett}},
  \ and\ \bibinfo {author} {\bibfnamefont {A.}~\bibnamefont {Melatos}},\ }\href
  {\doibase 10.1103/PhysRevD.87.063004} {\bibfield  {journal} {\bibinfo
  {journal} {Phys. Rev. D}\ }\textbf {\bibinfo {volume} {87}},\ \bibinfo
  {pages} {063004} (\bibinfo {year} {2013})}\BibitemShut {NoStop}%
\bibitem [{\citenamefont {{Crocker}}\ \emph {et~al.}(2015)\citenamefont
  {{Crocker}}, \citenamefont {{Mandic}}, \citenamefont {{Regimbau}},
  \citenamefont {{Belczynski}}, \citenamefont {{Gladysz}}, \citenamefont
  {{Olive}}, \citenamefont {{Prestegard}},\ and\ \citenamefont
  {{Vangioni}}}]{2015PhRvD..92f3005C}%
  \BibitemOpen
  \bibfield  {author} {\bibinfo {author} {\bibfnamefont {K.}~\bibnamefont
  {{Crocker}}}, \bibinfo {author} {\bibfnamefont {V.}~\bibnamefont {{Mandic}}},
  \bibinfo {author} {\bibfnamefont {T.}~\bibnamefont {{Regimbau}}}, \bibinfo
  {author} {\bibfnamefont {K.}~\bibnamefont {{Belczynski}}}, \bibinfo {author}
  {\bibfnamefont {W.}~\bibnamefont {{Gladysz}}}, \bibinfo {author}
  {\bibfnamefont {K.}~\bibnamefont {{Olive}}}, \bibinfo {author} {\bibfnamefont
  {T.}~\bibnamefont {{Prestegard}}}, \ and\ \bibinfo {author} {\bibfnamefont
  {E.}~\bibnamefont {{Vangioni}}},\ }\href {\doibase
  10.1103/PhysRevD.92.063005} {\bibfield  {journal} {\bibinfo  {journal}
  {\prd}\ }\textbf {\bibinfo {volume} {92}},\ \bibinfo {eid} {063005} (\bibinfo
  {year} {2015})},\ \Eprint {http://arxiv.org/abs/1506.02631} {arXiv:1506.02631
  [gr-qc]} \BibitemShut {NoStop}%
\bibitem [{\citenamefont {{Crocker}}\ \emph {et~al.}(2017)\citenamefont
  {{Crocker}}, \citenamefont {{Prestegard}}, \citenamefont {{Mandic}},
  \citenamefont {{Regimbau}}, \citenamefont {{Olive}},\ and\ \citenamefont
  {{Vangioni}}}]{2017PhRvD..95f3015C}%
  \BibitemOpen
  \bibfield  {author} {\bibinfo {author} {\bibfnamefont {K.}~\bibnamefont
  {{Crocker}}}, \bibinfo {author} {\bibfnamefont {T.}~\bibnamefont
  {{Prestegard}}}, \bibinfo {author} {\bibfnamefont {V.}~\bibnamefont
  {{Mandic}}}, \bibinfo {author} {\bibfnamefont {T.}~\bibnamefont
  {{Regimbau}}}, \bibinfo {author} {\bibfnamefont {K.}~\bibnamefont {{Olive}}},
  \ and\ \bibinfo {author} {\bibfnamefont {E.}~\bibnamefont {{Vangioni}}},\
  }\href {\doibase 10.1103/PhysRevD.95.063015} {\bibfield  {journal} {\bibinfo
  {journal} {\prd}\ }\textbf {\bibinfo {volume} {95}},\ \bibinfo {eid} {063015}
  (\bibinfo {year} {2017})},\ \Eprint {http://arxiv.org/abs/1701.02638}
  {arXiv:1701.02638} \BibitemShut {NoStop}%
\bibitem [{\citenamefont {{Kibble}}(1976)}]{1976JPhA....9.1387K}%
  \BibitemOpen
  \bibfield  {author} {\bibinfo {author} {\bibfnamefont {T.~W.~B.}\
  \bibnamefont {{Kibble}}},\ }\href {\doibase 10.1088/0305-4470/9/8/029}
  {\bibfield  {journal} {\bibinfo  {journal} {Journal of Physics A Mathematical
  General}\ }\textbf {\bibinfo {volume} {9}},\ \bibinfo {pages} {1387}
  (\bibinfo {year} {1976})}\BibitemShut {NoStop}%
\bibitem [{\citenamefont {{Sarangi}}\ and\ \citenamefont
  {{Tye}}(2002)}]{2002PhLB..536..185S}%
  \BibitemOpen
  \bibfield  {author} {\bibinfo {author} {\bibfnamefont {S.}~\bibnamefont
  {{Sarangi}}}\ and\ \bibinfo {author} {\bibfnamefont {S.-H.~H.}\ \bibnamefont
  {{Tye}}},\ }\href {\doibase 10.1016/S0370-2693(02)01824-5} {\bibfield
  {journal} {\bibinfo  {journal} {Physics Letters B}\ }\textbf {\bibinfo
  {volume} {536}},\ \bibinfo {pages} {185} (\bibinfo {year}
  {2002})}\BibitemShut {NoStop}%
\bibitem [{\citenamefont {{Damour}}\ and\ \citenamefont
  {{Vilenkin}}(2005)}]{2005PhRvD..71f3510D}%
  \BibitemOpen
  \bibfield  {author} {\bibinfo {author} {\bibfnamefont {T.}~\bibnamefont
  {{Damour}}}\ and\ \bibinfo {author} {\bibfnamefont {A.}~\bibnamefont
  {{Vilenkin}}},\ }\href {\doibase 10.1103/PhysRevD.71.063510} {\bibfield
  {journal} {\bibinfo  {journal} {Phys. Rev. D}\ }\textbf {\bibinfo {volume}
  {71}},\ \bibinfo {eid} {063510} (\bibinfo {year} {2005})}\BibitemShut
  {NoStop}%
\bibitem [{\citenamefont {{Siemens}}\ \emph {et~al.}(2007)\citenamefont
  {{Siemens}}, \citenamefont {{Mandic}},\ and\ \citenamefont
  {{Creighton}}}]{2007PhRvL..98k1101S}%
  \BibitemOpen
  \bibfield  {author} {\bibinfo {author} {\bibfnamefont {X.}~\bibnamefont
  {{Siemens}}}, \bibinfo {author} {\bibfnamefont {V.}~\bibnamefont {{Mandic}}},
  \ and\ \bibinfo {author} {\bibfnamefont {J.}~\bibnamefont {{Creighton}}},\
  }\href {\doibase 10.1103/PhysRevLett.98.111101} {\bibfield  {journal}
  {\bibinfo  {journal} {Physical Review Letters}\ }\textbf {\bibinfo {volume}
  {98}},\ \bibinfo {eid} {111101} (\bibinfo {year} {2007})}\BibitemShut
  {NoStop}%
\bibitem [{\citenamefont {Abbott}\ \emph
  {et~al.}(2018{\natexlab{a}})\citenamefont {Abbott} \emph
  {et~al.}}]{O1_cosmic_strings}%
  \BibitemOpen
  \bibfield  {author} {\bibinfo {author} {\bibfnamefont {B.}~\bibnamefont
  {Abbott}} \emph {et~al.} (\bibinfo {collaboration} {LIGO Scientific
  Collaboration and Virgo Collaboration}),\ }\href {\doibase
  10.1103/PhysRevD.97.102002} {\bibfield  {journal} {\bibinfo  {journal} {Phys.
  Rev.}\ }\textbf {\bibinfo {volume} {D97}},\ \bibinfo {pages} {102002}
  (\bibinfo {year} {2018}{\natexlab{a}})},\ \Eprint
  {http://arxiv.org/abs/1712.01168} {arXiv:1712.01168 [gr-qc]} \BibitemShut
  {NoStop}%
\bibitem [{\citenamefont {{Sasaki}}\ \emph {et~al.}(2016)\citenamefont
  {{Sasaki}}, \citenamefont {{Suyama}}, \citenamefont {{Tanaka}},\ and\
  \citenamefont {{Yokoyama}}}]{SasakiEA_2016}%
  \BibitemOpen
  \bibfield  {author} {\bibinfo {author} {\bibfnamefont {M.}~\bibnamefont
  {{Sasaki}}}, \bibinfo {author} {\bibfnamefont {T.}~\bibnamefont {{Suyama}}},
  \bibinfo {author} {\bibfnamefont {T.}~\bibnamefont {{Tanaka}}}, \ and\
  \bibinfo {author} {\bibfnamefont {S.}~\bibnamefont {{Yokoyama}}},\ }\href
  {\doibase 10.1103/PhysRevLett.117.061101} {\bibfield  {journal} {\bibinfo
  {journal} {Phys.~Rev.~Lett.}\ }\textbf {\bibinfo {volume} {117}},\ \bibinfo
  {eid} {061101} (\bibinfo {year} {2016})}\BibitemShut {NoStop}%
\bibitem [{\citenamefont {{Mandic}}\ \emph {et~al.}(2016)\citenamefont
  {{Mandic}}, \citenamefont {{Bird}},\ and\ \citenamefont
  {{Cholis}}}]{MandicEA_2016}%
  \BibitemOpen
  \bibfield  {author} {\bibinfo {author} {\bibfnamefont {V.}~\bibnamefont
  {{Mandic}}}, \bibinfo {author} {\bibfnamefont {S.}~\bibnamefont {{Bird}}}, \
  and\ \bibinfo {author} {\bibfnamefont {I.}~\bibnamefont {{Cholis}}},\ }\href
  {\doibase 10.1103/PhysRevLett.117.201102} {\bibfield  {journal} {\bibinfo
  {journal} {Phys.~Rev.~Lett.}\ }\textbf {\bibinfo {volume} {117}},\ \bibinfo
  {eid} {201102} (\bibinfo {year} {2016})}\BibitemShut {NoStop}%
\bibitem [{\citenamefont {Wang}\ \emph {et~al.}(2018)\citenamefont {Wang},
  \citenamefont {Wang}, \citenamefont {Huang},\ and\ \citenamefont
  {Li}}]{Wang:2016ana}%
  \BibitemOpen
  \bibfield  {author} {\bibinfo {author} {\bibfnamefont {S.}~\bibnamefont
  {Wang}}, \bibinfo {author} {\bibfnamefont {Y.-F.}\ \bibnamefont {Wang}},
  \bibinfo {author} {\bibfnamefont {Q.-G.}\ \bibnamefont {Huang}}, \ and\
  \bibinfo {author} {\bibfnamefont {T.~G.~F.}\ \bibnamefont {Li}},\ }\href
  {\doibase 10.1103/PhysRevLett.120.191102} {\bibfield  {journal} {\bibinfo
  {journal} {Phys. Rev. Lett.}\ }\textbf {\bibinfo {volume} {120}},\ \bibinfo
  {pages} {191102} (\bibinfo {year} {2018})},\ \Eprint
  {http://arxiv.org/abs/1610.08725} {arXiv:1610.08725 [astro-ph.CO]}
  \BibitemShut {NoStop}%
\bibitem [{\citenamefont {Brito}\ \emph
  {et~al.}(2017{\natexlab{a}})\citenamefont {Brito}, \citenamefont {Ghosh},
  \citenamefont {Barausse}, \citenamefont {Berti}, \citenamefont {Cardoso},
  \citenamefont {Dvorkin}, \citenamefont {Klein},\ and\ \citenamefont
  {Pani}}]{Brito:2017wnc}%
  \BibitemOpen
  \bibfield  {author} {\bibinfo {author} {\bibfnamefont {R.}~\bibnamefont
  {Brito}}, \bibinfo {author} {\bibfnamefont {S.}~\bibnamefont {Ghosh}},
  \bibinfo {author} {\bibfnamefont {E.}~\bibnamefont {Barausse}}, \bibinfo
  {author} {\bibfnamefont {E.}~\bibnamefont {Berti}}, \bibinfo {author}
  {\bibfnamefont {V.}~\bibnamefont {Cardoso}}, \bibinfo {author} {\bibfnamefont
  {I.}~\bibnamefont {Dvorkin}}, \bibinfo {author} {\bibfnamefont
  {A.}~\bibnamefont {Klein}}, \ and\ \bibinfo {author} {\bibfnamefont
  {P.}~\bibnamefont {Pani}},\ }\href {\doibase 10.1103/PhysRevLett.119.131101}
  {\bibfield  {journal} {\bibinfo  {journal} {Phys. Rev. Lett.}\ }\textbf
  {\bibinfo {volume} {119}},\ \bibinfo {pages} {131101} (\bibinfo {year}
  {2017}{\natexlab{a}})},\ \Eprint {http://arxiv.org/abs/1706.05097}
  {arXiv:1706.05097 [gr-qc]} \BibitemShut {NoStop}%
\bibitem [{\citenamefont {Brito}\ \emph
  {et~al.}(2017{\natexlab{b}})\citenamefont {Brito}, \citenamefont {Ghosh},
  \citenamefont {Barausse}, \citenamefont {Berti}, \citenamefont {Cardoso},
  \citenamefont {Dvorkin}, \citenamefont {Klein},\ and\ \citenamefont
  {Pani}}]{Brito:2017zvb}%
  \BibitemOpen
  \bibfield  {author} {\bibinfo {author} {\bibfnamefont {R.}~\bibnamefont
  {Brito}}, \bibinfo {author} {\bibfnamefont {S.}~\bibnamefont {Ghosh}},
  \bibinfo {author} {\bibfnamefont {E.}~\bibnamefont {Barausse}}, \bibinfo
  {author} {\bibfnamefont {E.}~\bibnamefont {Berti}}, \bibinfo {author}
  {\bibfnamefont {V.}~\bibnamefont {Cardoso}}, \bibinfo {author} {\bibfnamefont
  {I.}~\bibnamefont {Dvorkin}}, \bibinfo {author} {\bibfnamefont
  {A.}~\bibnamefont {Klein}}, \ and\ \bibinfo {author} {\bibfnamefont
  {P.}~\bibnamefont {Pani}},\ }\href {\doibase 10.1103/PhysRevD.96.064050}
  {\bibfield  {journal} {\bibinfo  {journal} {Phys. Rev.}\ }\textbf {\bibinfo
  {volume} {D96}},\ \bibinfo {pages} {064050} (\bibinfo {year}
  {2017}{\natexlab{b}})},\ \Eprint {http://arxiv.org/abs/1706.06311}
  {arXiv:1706.06311 [gr-qc]} \BibitemShut {NoStop}%
\bibitem [{\citenamefont {Fan}\ and\ \citenamefont {Chen}(2018)}]{Fan:2017cfw}%
  \BibitemOpen
  \bibfield  {author} {\bibinfo {author} {\bibfnamefont {X.-L.}\ \bibnamefont
  {Fan}}\ and\ \bibinfo {author} {\bibfnamefont {Y.-B.}\ \bibnamefont {Chen}},\
  }\href {\doibase 10.1103/PhysRevD.98.044020} {\bibfield  {journal} {\bibinfo
  {journal} {Phys. Rev.}\ }\textbf {\bibinfo {volume} {D98}},\ \bibinfo {pages}
  {044020} (\bibinfo {year} {2018})},\ \Eprint
  {http://arxiv.org/abs/1712.00784} {arXiv:1712.00784 [gr-qc]} \BibitemShut
  {NoStop}%
\bibitem [{\citenamefont {Tsukada}\ \emph {et~al.}(2019)\citenamefont
  {Tsukada}, \citenamefont {Callister}, \citenamefont {Matas},\ and\
  \citenamefont {Meyers}}]{Tsukada:2018mbp}%
  \BibitemOpen
  \bibfield  {author} {\bibinfo {author} {\bibfnamefont {L.}~\bibnamefont
  {Tsukada}}, \bibinfo {author} {\bibfnamefont {T.}~\bibnamefont {Callister}},
  \bibinfo {author} {\bibfnamefont {A.}~\bibnamefont {Matas}}, \ and\ \bibinfo
  {author} {\bibfnamefont {P.}~\bibnamefont {Meyers}},\ }\href {\doibase
  10.1103/PhysRevD.99.103015} {\bibfield  {journal} {\bibinfo  {journal} {Phys.
  Rev.}\ }\textbf {\bibinfo {volume} {D99}},\ \bibinfo {pages} {103015}
  (\bibinfo {year} {2019})},\ \Eprint {http://arxiv.org/abs/1812.09622}
  {arXiv:1812.09622 [astro-ph.HE]} \BibitemShut {NoStop}%
\bibitem [{\citenamefont {Lopez}\ and\ \citenamefont
  {Freese}(2015)}]{Lopez:2013mqa}%
  \BibitemOpen
  \bibfield  {author} {\bibinfo {author} {\bibfnamefont {A.}~\bibnamefont
  {Lopez}}\ and\ \bibinfo {author} {\bibfnamefont {K.}~\bibnamefont {Freese}},\
  }\href {\doibase 10.1088/1475-7516/2015/01/037} {\bibfield  {journal}
  {\bibinfo  {journal} {JCAP}\ }\textbf {\bibinfo {volume} {1501}},\ \bibinfo
  {pages} {037} (\bibinfo {year} {2015})},\ \Eprint
  {http://arxiv.org/abs/1305.5855} {arXiv:1305.5855 [astro-ph.HE]} \BibitemShut
  {NoStop}%
\bibitem [{\citenamefont {Dev}\ and\ \citenamefont
  {Mazumdar}(2016)}]{Dev:2016feu}%
  \BibitemOpen
  \bibfield  {author} {\bibinfo {author} {\bibfnamefont {P.~S.~B.}\
  \bibnamefont {Dev}}\ and\ \bibinfo {author} {\bibfnamefont {A.}~\bibnamefont
  {Mazumdar}},\ }\href {\doibase 10.1103/PhysRevD.93.104001} {\bibfield
  {journal} {\bibinfo  {journal} {Phys. Rev. D}\ }\textbf {\bibinfo {volume}
  {93}},\ \bibinfo {pages} {104001} (\bibinfo {year} {2016})},\ \Eprint
  {http://arxiv.org/abs/1602.04203} {arXiv:1602.04203 [hep-ph]} \BibitemShut
  {NoStop}%
\bibitem [{\citenamefont {Marzola}\ \emph {et~al.}(2017)\citenamefont
  {Marzola}, \citenamefont {Racioppi},\ and\ \citenamefont
  {Vaskonen}}]{Marzola:2017jzl}%
  \BibitemOpen
  \bibfield  {author} {\bibinfo {author} {\bibfnamefont {L.}~\bibnamefont
  {Marzola}}, \bibinfo {author} {\bibfnamefont {A.}~\bibnamefont {Racioppi}}, \
  and\ \bibinfo {author} {\bibfnamefont {V.}~\bibnamefont {Vaskonen}},\ }\href
  {\doibase 10.1140/epjc/s10052-017-4996-1} {\bibfield  {journal} {\bibinfo
  {journal} {Eur. Phys. J. C}\ }\textbf {\bibinfo {volume} {77}},\ \bibinfo
  {pages} {484} (\bibinfo {year} {2017})},\ \Eprint
  {http://arxiv.org/abs/1704.01034} {arXiv:1704.01034 [hep-ph]} \BibitemShut
  {NoStop}%
\bibitem [{\citenamefont {Von~Harling}\ \emph {et~al.}(2020)\citenamefont
  {Von~Harling}, \citenamefont {Pomarol}, \citenamefont {Pujolàs},\ and\
  \citenamefont {Rompineve}}]{vonHarling:2019gme}%
  \BibitemOpen
  \bibfield  {author} {\bibinfo {author} {\bibfnamefont {B.}~\bibnamefont
  {Von~Harling}}, \bibinfo {author} {\bibfnamefont {A.}~\bibnamefont
  {Pomarol}}, \bibinfo {author} {\bibfnamefont {O.}~\bibnamefont {Pujolàs}}, \
  and\ \bibinfo {author} {\bibfnamefont {F.}~\bibnamefont {Rompineve}},\ }\href
  {\doibase 10.1007/JHEP04(2020)195} {\bibfield  {journal} {\bibinfo  {journal}
  {JHEP}\ }\textbf {\bibinfo {volume} {04}},\ \bibinfo {pages} {195} (\bibinfo
  {year} {2020})},\ \Eprint {http://arxiv.org/abs/1912.07587} {arXiv:1912.07587
  [hep-ph]} \BibitemShut {NoStop}%
\bibitem [{\citenamefont {{Starobinski{\v \i}}}(1979)}]{1979JETPL..30..682S}%
  \BibitemOpen
  \bibfield  {author} {\bibinfo {author} {\bibfnamefont {A.~A.}\ \bibnamefont
  {{Starobinski{\v \i}}}},\ }\href@noop {} {\bibfield  {journal} {\bibinfo
  {journal} {Soviet Journal of Experimental and Theoretical Physics Letters}\
  }\textbf {\bibinfo {volume} {30}},\ \bibinfo {pages} {682} (\bibinfo {year}
  {1979})}\BibitemShut {NoStop}%
\bibitem [{\citenamefont {{Turner}}(1997)}]{1997PhRvD..55..435T}%
  \BibitemOpen
  \bibfield  {author} {\bibinfo {author} {\bibfnamefont {M.~S.}\ \bibnamefont
  {{Turner}}},\ }\href {\doibase 10.1103/PhysRevD.55.R435} {\bibfield
  {journal} {\bibinfo  {journal} {\prd}\ }\textbf {\bibinfo {volume} {55}},\
  \bibinfo {pages} {435} (\bibinfo {year} {1997})}\BibitemShut {NoStop}%
\bibitem [{\citenamefont {{Bar-Kana}}(1994)}]{1994PhRvD..50.1157B}%
  \BibitemOpen
  \bibfield  {author} {\bibinfo {author} {\bibfnamefont {R.}~\bibnamefont
  {{Bar-Kana}}},\ }\href {\doibase 10.1103/PhysRevD.50.1157} {\bibfield
  {journal} {\bibinfo  {journal} {Phys. Rev. D}\ }\textbf {\bibinfo {volume}
  {50}},\ \bibinfo {pages} {1157} (\bibinfo {year} {1994})}\BibitemShut
  {NoStop}%
\bibitem [{\citenamefont {{Easther}}\ and\ \citenamefont
  {{Lim}}(2006)}]{2006JCAP...04..010E}%
  \BibitemOpen
  \bibfield  {author} {\bibinfo {author} {\bibfnamefont {R.}~\bibnamefont
  {{Easther}}}\ and\ \bibinfo {author} {\bibfnamefont {E.~A.}\ \bibnamefont
  {{Lim}}},\ }\href {\doibase 10.1088/1475-7516/2006/04/010} {\bibfield
  {journal} {\bibinfo  {journal} {JCAP}\ }\textbf {\bibinfo {volume} {4}},\
  \bibinfo {eid} {010} (\bibinfo {year} {2006})}\BibitemShut {NoStop}%
\bibitem [{\citenamefont {{Easther}}\ \emph {et~al.}(2007)\citenamefont
  {{Easther}}, \citenamefont {{Giblin}},\ and\ \citenamefont
  {{Lim}}}]{2007PhRvL..99v1301E}%
  \BibitemOpen
  \bibfield  {author} {\bibinfo {author} {\bibfnamefont {R.}~\bibnamefont
  {{Easther}}}, \bibinfo {author} {\bibfnamefont {J.~T.}\ \bibnamefont
  {{Giblin}}, \bibfnamefont {Jr.}}, \ and\ \bibinfo {author} {\bibfnamefont
  {E.~A.}\ \bibnamefont {{Lim}}},\ }\href {\doibase
  10.1103/PhysRevLett.99.221301} {\bibfield  {journal} {\bibinfo  {journal}
  {Physical Review Letters}\ }\textbf {\bibinfo {volume} {99}},\ \bibinfo {eid}
  {221301} (\bibinfo {year} {2007})}\BibitemShut {NoStop}%
\bibitem [{\citenamefont {Abbott}\ \emph
  {et~al.}(2019{\natexlab{a}})\citenamefont {Abbott} \emph
  {et~al.}}]{stoch_O2}%
  \BibitemOpen
  \bibfield  {author} {\bibinfo {author} {\bibfnamefont {B.}~\bibnamefont
  {Abbott}} \emph {et~al.} (\bibinfo {collaboration} {LIGO Scientific
  Collaboration, Virgo Collaboration}),\ }\href {\doibase
  10.1103/PhysRevD.100.061101} {\bibfield  {journal} {\bibinfo  {journal}
  {Phys. Rev. D}\ }\textbf {\bibinfo {volume} {100}},\ \bibinfo {pages}
  {061101} (\bibinfo {year} {2019}{\natexlab{a}})},\ \Eprint
  {http://arxiv.org/abs/1903.02886} {arXiv:1903.02886 [gr-qc]} \BibitemShut
  {NoStop}%
\bibitem [{\citenamefont {Abbott}\ \emph
  {et~al.}(2019{\natexlab{b}})\citenamefont {Abbott} \emph
  {et~al.}}]{stoch_dir_O2}%
  \BibitemOpen
  \bibfield  {author} {\bibinfo {author} {\bibfnamefont {B.}~\bibnamefont
  {Abbott}} \emph {et~al.} (\bibinfo {collaboration} {LIGO Scientific
  Collaboration, Virgo Collaboration}),\ }\href {\doibase
  10.1103/PhysRevD.100.062001} {\bibfield  {journal} {\bibinfo  {journal}
  {Phys. Rev. D}\ }\textbf {\bibinfo {volume} {100}},\ \bibinfo {pages}
  {062001} (\bibinfo {year} {2019}{\natexlab{b}})},\ \Eprint
  {http://arxiv.org/abs/1903.08844} {arXiv:1903.08844 [gr-qc]} \BibitemShut
  {NoStop}%
\bibitem [{\citenamefont {Callister}\ \emph {et~al.}(2017)\citenamefont
  {Callister}, \citenamefont {Biscoveanu}, \citenamefont {Christensen},
  \citenamefont {Isi}, \citenamefont {Matas}, \citenamefont {Minazzoli},
  \citenamefont {Regimbau}, \citenamefont {Sakellariadou}, \citenamefont
  {Tasson},\ and\ \citenamefont {Thrane}}]{TestingGR_stoch}%
  \BibitemOpen
  \bibfield  {author} {\bibinfo {author} {\bibfnamefont {T.}~\bibnamefont
  {Callister}}, \bibinfo {author} {\bibfnamefont {A.~S.}\ \bibnamefont
  {Biscoveanu}}, \bibinfo {author} {\bibfnamefont {N.}~\bibnamefont
  {Christensen}}, \bibinfo {author} {\bibfnamefont {M.}~\bibnamefont {Isi}},
  \bibinfo {author} {\bibfnamefont {A.}~\bibnamefont {Matas}}, \bibinfo
  {author} {\bibfnamefont {O.}~\bibnamefont {Minazzoli}}, \bibinfo {author}
  {\bibfnamefont {T.}~\bibnamefont {Regimbau}}, \bibinfo {author}
  {\bibfnamefont {M.}~\bibnamefont {Sakellariadou}}, \bibinfo {author}
  {\bibfnamefont {J.}~\bibnamefont {Tasson}}, \ and\ \bibinfo {author}
  {\bibfnamefont {E.}~\bibnamefont {Thrane}},\ }\href {\doibase
  10.1103/PhysRevX.7.041058} {\bibfield  {journal} {\bibinfo  {journal} {Phys.
  Rev. X}\ }\textbf {\bibinfo {volume} {7}},\ \bibinfo {pages} {041058}
  (\bibinfo {year} {2017})},\ \Eprint {http://arxiv.org/abs/1704.08373}
  {arXiv:1704.08373} \BibitemShut {NoStop}%
\bibitem [{\citenamefont {Abbott}\ \emph
  {et~al.}(2018{\natexlab{b}})\citenamefont {Abbott} \emph
  {et~al.}}]{stoch_nongr_O1}%
  \BibitemOpen
  \bibfield  {author} {\bibinfo {author} {\bibfnamefont {B.~P.}\ \bibnamefont
  {Abbott}} \emph {et~al.} (\bibinfo {collaboration} {LIGO Scientific
  Collaboration and Virgo Collaboration}),\ }\href {\doibase
  10.1103/PhysRevLett.120.201102} {\bibfield  {journal} {\bibinfo  {journal}
  {Phys. Rev. Lett.}\ }\textbf {\bibinfo {volume} {120}},\ \bibinfo {pages}
  {201102} (\bibinfo {year} {2018}{\natexlab{b}})},\ \Eprint
  {http://arxiv.org/abs/1802.10194} {arXiv:1802.10194 [gr-qc]} \BibitemShut
  {NoStop}%
\bibitem [{\citenamefont {Romano}\ and\ \citenamefont
  {Cornish}(2017)}]{RomanoCornish}%
  \BibitemOpen
  \bibfield  {author} {\bibinfo {author} {\bibfnamefont {J.~D.}\ \bibnamefont
  {Romano}}\ and\ \bibinfo {author} {\bibfnamefont {N.~J.}\ \bibnamefont
  {Cornish}},\ }\href {\doibase 10.1007/s41114-017-0004-1} {\bibfield
  {journal} {\bibinfo  {journal} {Living Rev. Rel.}\ }\textbf {\bibinfo
  {volume} {20}},\ \bibinfo {pages} {2} (\bibinfo {year} {2017})},\ \Eprint
  {http://arxiv.org/abs/1608.06889} {arXiv:1608.06889 [gr-qc]} \BibitemShut
  {NoStop}%
\bibitem [{\citenamefont {Christensen}(2018)}]{Christensen_2018}%
  \BibitemOpen
  \bibfield  {author} {\bibinfo {author} {\bibfnamefont {N.}~\bibnamefont
  {Christensen}},\ }\href {\doibase 10.1088/1361-6633/aae6b5} {\bibfield
  {journal} {\bibinfo  {journal} {Reports on Progress in Physics}\ }\textbf
  {\bibinfo {volume} {82}},\ \bibinfo {pages} {016903} (\bibinfo {year}
  {2018})}\BibitemShut {NoStop}%
\bibitem [{\citenamefont {Abbott}\ \emph
  {et~al.}(2019{\natexlab{c}})\citenamefont {Abbott} \emph
  {et~al.}}]{Abbott:2019ebz}%
  \BibitemOpen
  \bibfield  {author} {\bibinfo {author} {\bibfnamefont {R.}~\bibnamefont
  {Abbott}} \emph {et~al.} (\bibinfo {collaboration} {LIGO Scientific
  Collaboration, Virgo Collaboration}),\ }\href@noop {} {\  (\bibinfo {year}
  {2019}{\natexlab{c}})},\ \Eprint {http://arxiv.org/abs/1912.11716}
  {arXiv:1912.11716 [gr-qc]} \BibitemShut {NoStop}%
\bibitem [{\citenamefont {Renzini}\ and\ \citenamefont
  {Contaldi}(2018)}]{Renzini:2018vkx}%
  \BibitemOpen
  \bibfield  {author} {\bibinfo {author} {\bibfnamefont {A.}~\bibnamefont
  {Renzini}}\ and\ \bibinfo {author} {\bibfnamefont {C.}~\bibnamefont
  {Contaldi}},\ }\href {\doibase 10.1093/mnras/sty2546} {\bibfield  {journal}
  {\bibinfo  {journal} {Mon. Not. Roy. Astron. Soc.}\ }\textbf {\bibinfo
  {volume} {481}},\ \bibinfo {pages} {4650} (\bibinfo {year} {2018})},\ \Eprint
  {http://arxiv.org/abs/1806.11360} {arXiv:1806.11360 [astro-ph.IM]}
  \BibitemShut {NoStop}%
\bibitem [{\citenamefont {Renzini}\ and\ \citenamefont
  {Contaldi}(2019{\natexlab{a}})}]{Renzini:2018nee}%
  \BibitemOpen
  \bibfield  {author} {\bibinfo {author} {\bibfnamefont {A.~I.}\ \bibnamefont
  {Renzini}}\ and\ \bibinfo {author} {\bibfnamefont {C.~R.}\ \bibnamefont
  {Contaldi}},\ }\href {\doibase 10.1103/PhysRevLett.122.081102} {\bibfield
  {journal} {\bibinfo  {journal} {Phys. Rev. Lett.}\ }\textbf {\bibinfo
  {volume} {122}},\ \bibinfo {pages} {081102} (\bibinfo {year}
  {2019}{\natexlab{a}})},\ \Eprint {http://arxiv.org/abs/1811.12922}
  {arXiv:1811.12922 [astro-ph.CO]} \BibitemShut {NoStop}%
\bibitem [{\citenamefont {Renzini}\ and\ \citenamefont
  {Contaldi}(2019{\natexlab{b}})}]{Renzini:2019vmt}%
  \BibitemOpen
  \bibfield  {author} {\bibinfo {author} {\bibfnamefont {A.}~\bibnamefont
  {Renzini}}\ and\ \bibinfo {author} {\bibfnamefont {C.}~\bibnamefont
  {Contaldi}},\ }\href {\doibase 10.1103/PhysRevD.100.063527} {\bibfield
  {journal} {\bibinfo  {journal} {Phys. Rev. D}\ }\textbf {\bibinfo {volume}
  {100}},\ \bibinfo {pages} {063527} (\bibinfo {year} {2019}{\natexlab{b}})},\
  \Eprint {http://arxiv.org/abs/1907.10329} {arXiv:1907.10329 [gr-qc]}
  \BibitemShut {NoStop}%
\bibitem [{\citenamefont {Smith}\ and\ \citenamefont
  {Thrane}(2018)}]{tbs_methods}%
  \BibitemOpen
  \bibfield  {author} {\bibinfo {author} {\bibfnamefont {R.}~\bibnamefont
  {Smith}}\ and\ \bibinfo {author} {\bibfnamefont {E.}~\bibnamefont {Thrane}},\
  }\href {\doibase 10.1103/PhysRevX.8.021019} {\bibfield  {journal} {\bibinfo
  {journal} {Phys. Rev.}\ }\textbf {\bibinfo {volume} {X8}},\ \bibinfo {pages}
  {021019} (\bibinfo {year} {2018})},\ \Eprint
  {http://arxiv.org/abs/1712.00688} {arXiv:1712.00688 [gr-qc]} \BibitemShut
  {NoStop}%
\bibitem [{\citenamefont {Aasi}\ \emph {et~al.}(2015)\citenamefont {Aasi} \emph
  {et~al.}}]{aLIGO_2015}%
  \BibitemOpen
  \bibfield  {author} {\bibinfo {author} {\bibfnamefont {J.}~\bibnamefont
  {Aasi}} \emph {et~al.} (\bibinfo {collaboration} {LIGO Scientific
  Collaboration}),\ }\href {\doibase 10.1088/0264-9381/32/7/074001} {\bibfield
  {journal} {\bibinfo  {journal} {Classical and Quantum Gravity}\ }\textbf
  {\bibinfo {volume} {32}},\ \bibinfo {pages} {074001} (\bibinfo {year}
  {2015})}\BibitemShut {NoStop}%
\bibitem [{\citenamefont {{Acernese}}\ \emph {et~al.}(2015)\citenamefont
  {{Acernese}} \emph {et~al.}}]{aVirgo_2015}%
  \BibitemOpen
  \bibfield  {author} {\bibinfo {author} {\bibfnamefont {F.}~\bibnamefont
  {{Acernese}}} \emph {et~al.},\ }\href {\doibase
  10.1088/0264-9381/32/2/024001} {\bibfield  {journal} {\bibinfo  {journal}
  {Classical and Quantum Gravity}\ }\textbf {\bibinfo {volume} {32}},\ \bibinfo
  {eid} {024001} (\bibinfo {year} {2015})}\BibitemShut {NoStop}%
\bibitem [{\citenamefont {Jeffreys}(1946)}]{JeffreysPrior}%
  \BibitemOpen
  \bibfield  {author} {\bibinfo {author} {\bibfnamefont {H.}~\bibnamefont
  {Jeffreys}},\ }\href {\doibase 10.1098/rspa.1946.0056} {\bibfield  {journal}
  {\bibinfo  {journal} {Proc. R. Soc. Lond. A}\ }\textbf {\bibinfo {volume}
  {186}},\ \bibinfo {pages} {453} (\bibinfo {year} {1946})}\BibitemShut
  {NoStop}%
\bibitem [{\citenamefont {Abbott}\ \emph {et~al.}()\citenamefont {Abbott} \emph
  {et~al.}}]{stoch_o3_data_release}%
  \BibitemOpen
  \bibfield  {author} {\bibinfo {author} {\bibfnamefont {R.}~\bibnamefont
  {Abbott}} \emph {et~al.} (\bibinfo {collaboration} {LIGO Scientific
  Collaboration, Virgo Collaboration}),\ }\href@noop {} {}\bibinfo
  {howpublished} {{https://dcc.ligo.org/G2001287/public}}\BibitemShut {NoStop}%
\bibitem [{\citenamefont {{Allen}}\ and\ \citenamefont
  {{Romano}}(1999)}]{Allen_Romano_1999}%
  \BibitemOpen
  \bibfield  {author} {\bibinfo {author} {\bibfnamefont {B.}~\bibnamefont
  {{Allen}}}\ and\ \bibinfo {author} {\bibfnamefont {J.~D.}\ \bibnamefont
  {{Romano}}},\ }\href {\doibase 10.1103/PhysRevD.59.102001} {\bibfield
  {journal} {\bibinfo  {journal} {\prd}\ }\textbf {\bibinfo {volume} {59}},\
  \bibinfo {eid} {102001} (\bibinfo {year} {1999})}\BibitemShut {NoStop}%
\bibitem [{\citenamefont {{K. Riles and J. Zweizig}}(2021)}]{gatingDocument}%
  \BibitemOpen
  \bibfield  {author} {\bibinfo {author} {\bibnamefont {{K. Riles and J.
  Zweizig}}},\ }\href@noop {} {}\bibinfo {howpublished}
  {\url{https://dcc.ligo.org/T2000384/public}} (\bibinfo {year}
  {2021})\BibitemShut {NoStop}%
\bibitem [{\citenamefont {{A. Matas, I. Dvorkin, T. Regimbau, and A.
  Romero}}(2021)}]{y_stochasticGatingDocument}%
  \BibitemOpen
  \bibfield  {author} {\bibinfo {author} {\bibnamefont {{A. Matas, I. Dvorkin,
  T. Regimbau, and A. Romero}}},\ }\href@noop {} {}\bibinfo {howpublished}
  {\url{https://dcc.ligo.org/P2000546/public}} (\bibinfo {year}
  {2021})\BibitemShut {NoStop}%
\bibitem [{\citenamefont {Meyers}\ \emph {et~al.}(2020)\citenamefont {Meyers},
  \citenamefont {Martinovic}, \citenamefont {Christensen},\ and\ \citenamefont
  {Sakellariadou}}]{Meyers:2020qrb}%
  \BibitemOpen
  \bibfield  {author} {\bibinfo {author} {\bibfnamefont {P.~M.}\ \bibnamefont
  {Meyers}}, \bibinfo {author} {\bibfnamefont {K.}~\bibnamefont {Martinovic}},
  \bibinfo {author} {\bibfnamefont {N.}~\bibnamefont {Christensen}}, \ and\
  \bibinfo {author} {\bibfnamefont {M.}~\bibnamefont {Sakellariadou}},\ }\href
  {\doibase 10.1103/PhysRevD.102.102005} {\bibfield  {journal} {\bibinfo
  {journal} {Phys. Rev. D}\ }\textbf {\bibinfo {volume} {102}},\ \bibinfo
  {pages} {102005} (\bibinfo {year} {2020})},\ \Eprint
  {http://arxiv.org/abs/2008.00789} {arXiv:2008.00789 [gr-qc]} \BibitemShut
  {NoStop}%
\bibitem [{\citenamefont {{Abbott}}\ \emph {et~al.}(2016)\citenamefont
  {{Abbott}} \emph {et~al.}}]{gw150914_stoch}%
  \BibitemOpen
  \bibfield  {author} {\bibinfo {author} {\bibfnamefont {B.~P.}\ \bibnamefont
  {{Abbott}}} \emph {et~al.} (\bibinfo {collaboration} {LIGO Scientific
  Collaboration and Virgo Collaboration}),\ }\href {\doibase
  10.1103/PhysRevLett.116.131102} {\bibfield  {journal} {\bibinfo  {journal}
  {\prl}\ }\textbf {\bibinfo {volume} {116}},\ \bibinfo {eid} {131102}
  (\bibinfo {year} {2016})}\BibitemShut {NoStop}%
\bibitem [{\citenamefont {Abbott}\ \emph
  {et~al.}(2018{\natexlab{c}})\citenamefont {Abbott} \emph
  {et~al.}}]{gw170817_stoch}%
  \BibitemOpen
  \bibfield  {author} {\bibinfo {author} {\bibfnamefont {B.~P.}\ \bibnamefont
  {Abbott}} \emph {et~al.} (\bibinfo {collaboration} {LIGO Scientific
  Collaboration and Virgo Collaboration}),\ }\href {\doibase
  10.1103/PhysRevLett.120.091101} {\bibfield  {journal} {\bibinfo  {journal}
  {Phys. Rev. Lett.}\ }\textbf {\bibinfo {volume} {120}},\ \bibinfo {pages}
  {091101} (\bibinfo {year} {2018}{\natexlab{c}})},\ \Eprint
  {http://arxiv.org/abs/1710.05837} {arXiv:1710.05837 [gr-qc]} \BibitemShut
  {NoStop}%
\bibitem [{\citenamefont {Regimbau}\ \emph {et~al.}(2012)\citenamefont
  {Regimbau} \emph {et~al.}}]{Regimbau:2012ir}%
  \BibitemOpen
  \bibfield  {author} {\bibinfo {author} {\bibfnamefont {T.}~\bibnamefont
  {Regimbau}} \emph {et~al.},\ }\href {\doibase 10.1103/PhysRevD.86.122001}
  {\bibfield  {journal} {\bibinfo  {journal} {Phys. Rev. D}\ }\textbf {\bibinfo
  {volume} {86}},\ \bibinfo {pages} {122001} (\bibinfo {year} {2012})},\
  \Eprint {http://arxiv.org/abs/1201.3563} {arXiv:1201.3563 [gr-qc]}
  \BibitemShut {NoStop}%
\bibitem [{\citenamefont {Regimbau}\ \emph {et~al.}(2014)\citenamefont
  {Regimbau}, \citenamefont {Meacher},\ and\ \citenamefont
  {Coughlin}}]{Regimbau:2014uia}%
  \BibitemOpen
  \bibfield  {author} {\bibinfo {author} {\bibfnamefont {T.}~\bibnamefont
  {Regimbau}}, \bibinfo {author} {\bibfnamefont {D.}~\bibnamefont {Meacher}}, \
  and\ \bibinfo {author} {\bibfnamefont {M.}~\bibnamefont {Coughlin}},\ }\href
  {\doibase 10.1103/PhysRevD.89.084046} {\bibfield  {journal} {\bibinfo
  {journal} {Phys. Rev. D}\ }\textbf {\bibinfo {volume} {89}},\ \bibinfo
  {pages} {084046} (\bibinfo {year} {2014})},\ \Eprint
  {http://arxiv.org/abs/1404.1134} {arXiv:1404.1134 [astro-ph.CO]} \BibitemShut
  {NoStop}%
\bibitem [{\citenamefont {Meacher}\ \emph {et~al.}(2015)\citenamefont
  {Meacher}, \citenamefont {Coughlin}, \citenamefont {Morris}, \citenamefont
  {Regimbau}, \citenamefont {Christensen}, \citenamefont {Kandhasamy},
  \citenamefont {Mandic}, \citenamefont {Romano},\ and\ \citenamefont
  {Thrane}}]{Meacher:2015iua}%
  \BibitemOpen
  \bibfield  {author} {\bibinfo {author} {\bibfnamefont {D.}~\bibnamefont
  {Meacher}}, \bibinfo {author} {\bibfnamefont {M.}~\bibnamefont {Coughlin}},
  \bibinfo {author} {\bibfnamefont {S.}~\bibnamefont {Morris}}, \bibinfo
  {author} {\bibfnamefont {T.}~\bibnamefont {Regimbau}}, \bibinfo {author}
  {\bibfnamefont {N.}~\bibnamefont {Christensen}}, \bibinfo {author}
  {\bibfnamefont {S.}~\bibnamefont {Kandhasamy}}, \bibinfo {author}
  {\bibfnamefont {V.}~\bibnamefont {Mandic}}, \bibinfo {author} {\bibfnamefont
  {J.~D.}\ \bibnamefont {Romano}}, \ and\ \bibinfo {author} {\bibfnamefont
  {E.}~\bibnamefont {Thrane}},\ }\href {\doibase 10.1103/PhysRevD.92.063002}
  {\bibfield  {journal} {\bibinfo  {journal} {Phys. Rev. D}\ }\textbf {\bibinfo
  {volume} {92}},\ \bibinfo {pages} {063002} (\bibinfo {year} {2015})},\
  \Eprint {http://arxiv.org/abs/1506.06744} {arXiv:1506.06744 [astro-ph.HE]}
  \BibitemShut {NoStop}%
\bibitem [{\citenamefont {Abbott}\ \emph
  {et~al.}(2020{\natexlab{a}})\citenamefont {Abbott} \emph
  {et~al.}}]{O3aCatalog}%
  \BibitemOpen
  \bibfield  {author} {\bibinfo {author} {\bibfnamefont {R.}~\bibnamefont
  {Abbott}} \emph {et~al.} (\bibinfo {collaboration} {LIGO Scientific,
  Virgo}),\ }\href@noop {} {\  (\bibinfo {year} {2020}{\natexlab{a}})},\
  \Eprint {http://arxiv.org/abs/2010.14527} {arXiv:2010.14527 [gr-qc]}
  \BibitemShut {NoStop}%
\bibitem [{\citenamefont {Abbott}\ \emph
  {et~al.}(2020{\natexlab{b}})\citenamefont {Abbott} \emph
  {et~al.}}]{gw_190412}%
  \BibitemOpen
  \bibfield  {author} {\bibinfo {author} {\bibfnamefont {R.}~\bibnamefont
  {Abbott}} \emph {et~al.} (\bibinfo {collaboration} {LIGO Scientific
  Collaboration, Virgo Collaboration}),\ }\href {\doibase
  10.1103/PhysRevD.102.043015} {\bibfield  {journal} {\bibinfo  {journal}
  {Phys. Rev. D}\ }\textbf {\bibinfo {volume} {102}},\ \bibinfo {pages}
  {043015} (\bibinfo {year} {2020}{\natexlab{b}})},\ \Eprint
  {http://arxiv.org/abs/2004.08342} {arXiv:2004.08342 [astro-ph.HE]}
  \BibitemShut {NoStop}%
\bibitem [{\citenamefont {Abbott}\ \emph
  {et~al.}(2020{\natexlab{c}})\citenamefont {Abbott} \emph
  {et~al.}}]{gw_190425}%
  \BibitemOpen
  \bibfield  {author} {\bibinfo {author} {\bibfnamefont {B.}~\bibnamefont
  {Abbott}} \emph {et~al.} (\bibinfo {collaboration} {LIGO Scientific,
  Virgo}),\ }\href {\doibase 10.3847/2041-8213/ab75f5} {\bibfield  {journal}
  {\bibinfo  {journal} {Astrophys. J. Lett.}\ }\textbf {\bibinfo {volume}
  {892}},\ \bibinfo {pages} {L3} (\bibinfo {year} {2020}{\natexlab{c}})},\
  \Eprint {http://arxiv.org/abs/2001.01761} {arXiv:2001.01761 [astro-ph.HE]}
  \BibitemShut {NoStop}%
\bibitem [{\citenamefont {Abbott}\ \emph
  {et~al.}(2020{\natexlab{d}})\citenamefont {Abbott} \emph
  {et~al.}}]{gw_190521}%
  \BibitemOpen
  \bibfield  {author} {\bibinfo {author} {\bibfnamefont {R.}~\bibnamefont
  {Abbott}} \emph {et~al.} (\bibinfo {collaboration} {LIGO Scientific,
  Virgo}),\ }\href {\doibase 10.1103/PhysRevLett.125.101102} {\bibfield
  {journal} {\bibinfo  {journal} {Phys. Rev. Lett.}\ }\textbf {\bibinfo
  {volume} {125}},\ \bibinfo {pages} {101102} (\bibinfo {year}
  {2020}{\natexlab{d}})},\ \Eprint {http://arxiv.org/abs/2009.01075}
  {arXiv:2009.01075 [gr-qc]} \BibitemShut {NoStop}%
\bibitem [{\citenamefont {Abbott}\ \emph
  {et~al.}(2020{\natexlab{e}})\citenamefont {Abbott} \emph
  {et~al.}}]{gw_190521_implications}%
  \BibitemOpen
  \bibfield  {author} {\bibinfo {author} {\bibfnamefont {R.}~\bibnamefont
  {Abbott}} \emph {et~al.} (\bibinfo {collaboration} {LIGO Scientific,
  Virgo}),\ }\href {\doibase 10.3847/2041-8213/aba493} {\bibfield  {journal}
  {\bibinfo  {journal} {Astrophys. J. Lett.}\ }\textbf {\bibinfo {volume}
  {900}},\ \bibinfo {pages} {L13} (\bibinfo {year} {2020}{\natexlab{e}})},\
  \Eprint {http://arxiv.org/abs/2009.01190} {arXiv:2009.01190 [astro-ph.HE]}
  \BibitemShut {NoStop}%
\bibitem [{\citenamefont {Abbott}\ \emph
  {et~al.}(2020{\natexlab{f}})\citenamefont {Abbott} \emph
  {et~al.}}]{gw_190814}%
  \BibitemOpen
  \bibfield  {author} {\bibinfo {author} {\bibfnamefont {R.}~\bibnamefont
  {Abbott}} \emph {et~al.} (\bibinfo {collaboration} {LIGO Scientific
  Collaboration, Virgo Collaboration}),\ }\href {\doibase
  10.3847/2041-8213/ab960f} {\bibfield  {journal} {\bibinfo  {journal}
  {Astrophys. J.}\ }\textbf {\bibinfo {volume} {896}},\ \bibinfo {pages} {L44}
  (\bibinfo {year} {2020}{\natexlab{f}})},\ \Eprint
  {http://arxiv.org/abs/2006.12611} {arXiv:2006.12611 [astro-ph.HE]}
  \BibitemShut {NoStop}%
\bibitem [{\citenamefont {Barsotti}\ \emph
  {et~al.}({\natexlab{a}})\citenamefont {Barsotti}, \citenamefont {McCuller},
  \citenamefont {Evans},\ and\ \citenamefont {Fritschel}}]{aplus_design}%
  \BibitemOpen
  \bibfield  {author} {\bibinfo {author} {\bibfnamefont {L.}~\bibnamefont
  {Barsotti}}, \bibinfo {author} {\bibfnamefont {L.}~\bibnamefont {McCuller}},
  \bibinfo {author} {\bibfnamefont {M.}~\bibnamefont {Evans}}, \ and\ \bibinfo
  {author} {\bibfnamefont {P.}~\bibnamefont {Fritschel}},\ }\href@noop {}
  {}\bibinfo {howpublished} {\url{https://dcc.ligo.org/LIGO-T1800042/public}}
  ({\natexlab{a}})\BibitemShut {NoStop}%
\bibitem [{\citenamefont {Callister}\ \emph {et~al.}(2020)\citenamefont
  {Callister}, \citenamefont {Fishbach}, \citenamefont {Holz},\ and\
  \citenamefont {Farr}}]{Callister:2020arv}%
  \BibitemOpen
  \bibfield  {author} {\bibinfo {author} {\bibfnamefont {T.}~\bibnamefont
  {Callister}}, \bibinfo {author} {\bibfnamefont {M.}~\bibnamefont {Fishbach}},
  \bibinfo {author} {\bibfnamefont {D.}~\bibnamefont {Holz}}, \ and\ \bibinfo
  {author} {\bibfnamefont {W.}~\bibnamefont {Farr}},\ }\href {\doibase
  10.3847/2041-8213/ab9743} {\bibfield  {journal} {\bibinfo  {journal}
  {Astrophys. J.}\ }\textbf {\bibinfo {volume} {896}},\ \bibinfo {pages} {L32}
  (\bibinfo {year} {2020})},\ \Eprint {http://arxiv.org/abs/2003.12152}
  {arXiv:2003.12152 [astro-ph.HE]} \BibitemShut {NoStop}%
\bibitem [{\citenamefont {Abbott}\ \emph
  {et~al.}(2020{\natexlab{g}})\citenamefont {Abbott} \emph
  {et~al.}}]{O3aRatesAndPop}%
  \BibitemOpen
  \bibfield  {author} {\bibinfo {author} {\bibfnamefont {R.}~\bibnamefont
  {Abbott}} \emph {et~al.} (\bibinfo {collaboration} {LIGO Scientific,
  Virgo}),\ }\href@noop {} {\  (\bibinfo {year} {2020}{\natexlab{g}})},\
  \Eprint {http://arxiv.org/abs/2010.14533} {arXiv:2010.14533 [astro-ph.HE]}
  \BibitemShut {NoStop}%
\bibitem [{\citenamefont {{Ade}}\ \emph {et~al.}(2016)\citenamefont {{Ade}}
  \emph {et~al.}}]{Planck_2015}%
  \BibitemOpen
  \bibfield  {author} {\bibinfo {author} {\bibfnamefont {P.~A.~R.}\
  \bibnamefont {{Ade}}} \emph {et~al.},\ }\href {\doibase
  10.1051/0004-6361/201525830} {\bibfield  {journal} {\bibinfo  {journal}
  {\aap}\ }\textbf {\bibinfo {volume} {594}},\ \bibinfo {eid} {A13} (\bibinfo
  {year} {2016})}\BibitemShut {NoStop}%
\bibitem [{\citenamefont {Christensen}(1992)}]{christensen92}%
  \BibitemOpen
  \bibfield  {author} {\bibinfo {author} {\bibfnamefont {N.}~\bibnamefont
  {Christensen}},\ }\href {\doibase 10.1103/PhysRevD.46.5250} {\bibfield
  {journal} {\bibinfo  {journal} {Phys. Rev. D}\ }\textbf {\bibinfo {volume}
  {46}},\ \bibinfo {pages} {5250} (\bibinfo {year} {1992})}\BibitemShut
  {NoStop}%
\bibitem [{\citenamefont {{Mingarelli}}\ \emph {et~al.}(2019)\citenamefont
  {{Mingarelli}}, \citenamefont {{Taylor}}, \citenamefont {{Sathyaprakash}},\
  and\ \citenamefont {{Farr}}}]{Mingarelli2019}%
  \BibitemOpen
  \bibfield  {author} {\bibinfo {author} {\bibfnamefont {C.~M.~F.}\
  \bibnamefont {{Mingarelli}}}, \bibinfo {author} {\bibfnamefont {S.~R.}\
  \bibnamefont {{Taylor}}}, \bibinfo {author} {\bibfnamefont {B.~S.}\
  \bibnamefont {{Sathyaprakash}}}, \ and\ \bibinfo {author} {\bibfnamefont
  {W.~M.}\ \bibnamefont {{Farr}}},\ }\href@noop {} {\bibfield  {journal}
  {\bibinfo  {journal} {arXiv e-prints}\ } (\bibinfo {year} {2019})},\ \Eprint
  {http://arxiv.org/abs/1911.09745} {arXiv:1911.09745 [gr-qc]} \BibitemShut
  {NoStop}%
\bibitem [{\citenamefont {Mandic}\ \emph {et~al.}(2012)\citenamefont {Mandic},
  \citenamefont {Thrane}, \citenamefont {Giampanis},\ and\ \citenamefont
  {Regimbau}}]{StochPE}%
  \BibitemOpen
  \bibfield  {author} {\bibinfo {author} {\bibfnamefont {V.}~\bibnamefont
  {Mandic}}, \bibinfo {author} {\bibfnamefont {E.}~\bibnamefont {Thrane}},
  \bibinfo {author} {\bibfnamefont {S.}~\bibnamefont {Giampanis}}, \ and\
  \bibinfo {author} {\bibfnamefont {T.}~\bibnamefont {Regimbau}},\ }\href
  {\doibase 10.1103/PhysRevLett.109.171102} {\bibfield  {journal} {\bibinfo
  {journal} {Phys. Rev. Lett.}\ }\textbf {\bibinfo {volume} {109}},\ \bibinfo
  {pages} {171102} (\bibinfo {year} {2012})}\BibitemShut {NoStop}%
\bibitem [{\citenamefont {Matas}\ and\ \citenamefont
  {Romano}(2020)}]{Matas:2020roi}%
  \BibitemOpen
  \bibfield  {author} {\bibinfo {author} {\bibfnamefont {A.}~\bibnamefont
  {Matas}}\ and\ \bibinfo {author} {\bibfnamefont {J.~D.}\ \bibnamefont
  {Romano}},\ }\href@noop {} {\  (\bibinfo {year} {2020})},\ \Eprint
  {http://arxiv.org/abs/2012.00907} {arXiv:2012.00907 [gr-qc]} \BibitemShut
  {NoStop}%
\bibitem [{\citenamefont {Lasky}\ \emph {et~al.}(2016)\citenamefont {Lasky}
  \emph {et~al.}}]{Lasky:2015lej}%
  \BibitemOpen
  \bibfield  {author} {\bibinfo {author} {\bibfnamefont {P.~D.}\ \bibnamefont
  {Lasky}} \emph {et~al.},\ }\href {\doibase 10.1103/PhysRevX.6.011035}
  {\bibfield  {journal} {\bibinfo  {journal} {Phys. Rev.}\ }\textbf {\bibinfo
  {volume} {X6}},\ \bibinfo {pages} {011035} (\bibinfo {year} {2016})},\
  \Eprint {http://arxiv.org/abs/1511.05994} {arXiv:1511.05994 [astro-ph.CO]}
  \BibitemShut {NoStop}%
\bibitem [{lem()}]{lemi}%
  \BibitemOpen
  \href@noop {} {}\bibinfo {howpublished}
  {\url{http://www.lemisensors.com}}\BibitemShut {NoStop}%
\bibitem [{met()}]{metronix}%
  \BibitemOpen
  \href@noop {} {}\bibinfo {howpublished}
  {\url{https://www.geo-metronix.de/mtxgeo/index.php/mfs-06e-overview}}\BibitemShut
  {NoStop}%
\bibitem [{x_s()}]{x_stochastic_public}%
  \BibitemOpen
  \href@noop {} {}\bibinfo {howpublished}
  {\url{https://git.ligo.org/stochastic-public/stochastic/}}\BibitemShut
  {NoStop}%
\bibitem [{\citenamefont {MATLAB}(2020)}]{MATLAB:2020}%
  \BibitemOpen
  \bibfield  {author} {\bibinfo {author} {\bibnamefont {MATLAB}},\ }\href@noop
  {} {\emph {\bibinfo {title} {9.8.0.1323502 (R2020a)}}}\ (\bibinfo
  {publisher} {The MathWorks Inc.},\ \bibinfo {address} {Natick,
  Massachusetts},\ \bibinfo {year} {2020})\BibitemShut {NoStop}%
\bibitem [{\citenamefont {Biwer}\ \emph {et~al.}(2017)\citenamefont {Biwer}
  \emph {et~al.}}]{Biwer:2016oyg}%
  \BibitemOpen
  \bibfield  {author} {\bibinfo {author} {\bibfnamefont {C.}~\bibnamefont
  {Biwer}} \emph {et~al.},\ }\href {\doibase 10.1103/PhysRevD.95.062002}
  {\bibfield  {journal} {\bibinfo  {journal} {Phys. Rev. D}\ }\textbf {\bibinfo
  {volume} {95}},\ \bibinfo {pages} {062002} (\bibinfo {year} {2017})},\
  \Eprint {http://arxiv.org/abs/1612.07864} {arXiv:1612.07864 [astro-ph.IM]}
  \BibitemShut {NoStop}%
\bibitem [{\citenamefont {Abbott}\ \emph
  {et~al.}(2018{\natexlab{d}})\citenamefont {Abbott} \emph
  {et~al.}}]{O1_CBC_DQ}%
  \BibitemOpen
  \bibfield  {author} {\bibinfo {author} {\bibfnamefont {B.~P.}\ \bibnamefont
  {Abbott}} \emph {et~al.} (\bibinfo {collaboration} {LIGO Scientific
  Collaboration and Virgo Collaboration}),\ }\href {\doibase
  10.1088/1361-6382/aaaafa} {\bibfield  {journal} {\bibinfo  {journal} {Class.
  Quant. Grav.}\ }\textbf {\bibinfo {volume} {35}},\ \bibinfo {pages} {065010}
  (\bibinfo {year} {2018}{\natexlab{d}})},\ \Eprint
  {http://arxiv.org/abs/1710.02185} {arXiv:1710.02185 [gr-qc]} \BibitemShut
  {NoStop}%
\bibitem [{\citenamefont {Covas}\ \emph {et~al.}(2018)\citenamefont {Covas}
  \emph {et~al.}}]{CWStochDetchar}%
  \BibitemOpen
  \bibfield  {author} {\bibinfo {author} {\bibfnamefont {P.}~\bibnamefont
  {Covas}} \emph {et~al.} (\bibinfo {collaboration} {LSC Instrument Authors}),\
  }\href {\doibase 10.1103/PhysRevD.97.082002} {\bibfield  {journal} {\bibinfo
  {journal} {Phys. Rev.}\ }\textbf {\bibinfo {volume} {D97}},\ \bibinfo {pages}
  {082002} (\bibinfo {year} {2018})},\ \Eprint
  {http://arxiv.org/abs/1801.07204} {arXiv:1801.07204 [astro-ph.IM]}
  \BibitemShut {NoStop}%
\bibitem [{\citenamefont {Schumann.}(1952)}]{Schumann_theory}%
  \BibitemOpen
  \bibfield  {author} {\bibinfo {author} {\bibfnamefont {W.}~\bibnamefont
  {Schumann.}},\ }\href@noop {} {\bibfield  {journal} {\bibinfo  {journal}
  {Zeitschrift f{\"u}r Naturforschung A}\ }\textbf {\bibinfo {volume} {7}},\
  \bibinfo {pages} {250} (\bibinfo {year} {1952})}\BibitemShut {NoStop}%
\bibitem [{\citenamefont {Coughlin}\ \emph {et~al.}(2018)\citenamefont
  {Coughlin} \emph {et~al.}}]{Schumann_4}%
  \BibitemOpen
  \bibfield  {author} {\bibinfo {author} {\bibfnamefont {M.~W.}\ \bibnamefont
  {Coughlin}} \emph {et~al.},\ }\href {\doibase 10.1103/PhysRevD.97.102007}
  {\bibfield  {journal} {\bibinfo  {journal} {Phys. Rev.}\ }\textbf {\bibinfo
  {volume} {D97}},\ \bibinfo {pages} {102007} (\bibinfo {year} {2018})},\
  \Eprint {http://arxiv.org/abs/1802.00885} {arXiv:1802.00885 [gr-qc]}
  \BibitemShut {NoStop}%
\bibitem [{\citenamefont {Thrane}\ \emph {et~al.}(2013)\citenamefont {Thrane},
  \citenamefont {Christensen},\ and\ \citenamefont {Schofield}}]{Schumann_1}%
  \BibitemOpen
  \bibfield  {author} {\bibinfo {author} {\bibfnamefont {E.}~\bibnamefont
  {Thrane}}, \bibinfo {author} {\bibfnamefont {N.}~\bibnamefont {Christensen}},
  \ and\ \bibinfo {author} {\bibfnamefont {R.}~\bibnamefont {Schofield}},\
  }\href {\doibase 10.1103/PhysRevD.87.123009} {\bibfield  {journal} {\bibinfo
  {journal} {Phys. Rev.}\ }\textbf {\bibinfo {volume} {D87}},\ \bibinfo {pages}
  {123009} (\bibinfo {year} {2013})},\ \Eprint {http://arxiv.org/abs/1303.2613}
  {arXiv:1303.2613 [astro-ph.IM]} \BibitemShut {NoStop}%
\bibitem [{\citenamefont {Thrane}\ \emph {et~al.}(2014)\citenamefont {Thrane},
  \citenamefont {Christensen}, \citenamefont {Schofield},\ and\ \citenamefont
  {Effler}}]{Schumann_2}%
  \BibitemOpen
  \bibfield  {author} {\bibinfo {author} {\bibfnamefont {E.}~\bibnamefont
  {Thrane}}, \bibinfo {author} {\bibfnamefont {N.}~\bibnamefont {Christensen}},
  \bibinfo {author} {\bibfnamefont {R.~M.~S.}\ \bibnamefont {Schofield}}, \
  and\ \bibinfo {author} {\bibfnamefont {A.}~\bibnamefont {Effler}},\ }\href
  {\doibase 10.1103/PhysRevD.90.023013} {\bibfield  {journal} {\bibinfo
  {journal} {Phys. Rev.}\ }\textbf {\bibinfo {volume} {D90}},\ \bibinfo {pages}
  {023013} (\bibinfo {year} {2014})},\ \Eprint {http://arxiv.org/abs/1406.2367}
  {arXiv:1406.2367 [astro-ph.IM]} \BibitemShut {NoStop}%
\bibitem [{\citenamefont {{Abbott}}\ \emph {et~al.}(2017)\citenamefont
  {{Abbott}} \emph {et~al.}}]{stoch_O1}%
  \BibitemOpen
  \bibfield  {author} {\bibinfo {author} {\bibfnamefont {B.~P.}\ \bibnamefont
  {{Abbott}}} \emph {et~al.} (\bibinfo {collaboration} {LIGO Scientific
  Collaboration and Virgo Collaboration}),\ }\href {\doibase
  10.1103/PhysRevLett.118.121101} {\bibfield  {journal} {\bibinfo  {journal}
  {Phys.~Rev.~Lett.}\ }\textbf {\bibinfo {volume} {118}},\ \bibinfo {eid}
  {121101} (\bibinfo {year} {2017})}\BibitemShut {NoStop}%
\bibitem [{\citenamefont {Merfeld}\ \emph {et~al.}()\citenamefont {Merfeld}
  \emph {et~al.}}]{alog:WeeklyMagneticInjections}%
  \BibitemOpen
  \bibfield  {author} {\bibinfo {author} {\bibfnamefont {K.}~\bibnamefont
  {Merfeld}} \emph {et~al.},\ }\href@noop {} {\enquote {\bibinfo {title}
  {{aLIGO LHO Logbook}},}\ }\bibinfo {howpublished}
  {\url{https://alog.ligo-wa.caltech.edu/aLOG/index.php?callRep=48212}}\BibitemShut
  {NoStop}%
\bibitem [{\citenamefont {Nguyen}\ \emph {et~al.}({\natexlab{a}})\citenamefont
  {Nguyen} \emph {et~al.}}]{z_PEMCouplingUncertainty}%
  \BibitemOpen
  \bibfield  {author} {\bibinfo {author} {\bibfnamefont {P.}~\bibnamefont
  {Nguyen}} \emph {et~al.},\ }\href@noop {} {\enquote {\bibinfo {title} {{aLIGO
  LHO Logbook}},}\ }\bibinfo {howpublished}
  {\url{https://alog.ligo-wa.caltech.edu/aLOG/index.php?callRep=57672}}
  ({\natexlab{a}})\BibitemShut {NoStop}%
\bibitem [{\citenamefont {{Thrane}}\ and\ \citenamefont
  {{Romano}}(2013)}]{locus}%
  \BibitemOpen
  \bibfield  {author} {\bibinfo {author} {\bibfnamefont {E.}~\bibnamefont
  {{Thrane}}}\ and\ \bibinfo {author} {\bibfnamefont {J.~D.}\ \bibnamefont
  {{Romano}}},\ }\href {\doibase 10.1103/PhysRevD.88.124032} {\bibfield
  {journal} {\bibinfo  {journal} {\prd}\ }\textbf {\bibinfo {volume} {88}},\
  \bibinfo {eid} {124032} (\bibinfo {year} {2013})}\BibitemShut {NoStop}%
\bibitem [{\citenamefont {Nguyen}\ \emph {et~al.}({\natexlab{b}})\citenamefont
  {Nguyen} \emph {et~al.}}]{q_alog:LargeCoil}%
  \BibitemOpen
  \bibfield  {author} {\bibinfo {author} {\bibfnamefont {P.}~\bibnamefont
  {Nguyen}} \emph {et~al.},\ }\href@noop {} {\enquote {\bibinfo {title} {{aLIGO
  LHO Logbook}},}\ }\bibinfo {howpublished}
  {\url{https://alog.ligo-wa.caltech.edu/aLOG/index.php?callRep=43406}}
  ({\natexlab{b}})\BibitemShut {NoStop}%
\bibitem [{\citenamefont {Michaloliakos}\ \emph {et~al.}()\citenamefont
  {Michaloliakos} \emph {et~al.}}]{alog:MagneticCouplingResonances}%
  \BibitemOpen
  \bibfield  {author} {\bibinfo {author} {\bibfnamefont {I.}~\bibnamefont
  {Michaloliakos}} \emph {et~al.},\ }\href@noop {} {\enquote {\bibinfo {title}
  {{aLIGO LHO Logbook}},}\ }\bibinfo {howpublished}
  {\url{https://alog.ligo-wa.caltech.edu/aLOG/index.php?callRep=56295}}\BibitemShut
  {NoStop}%
\bibitem [{\citenamefont {{Regimbau}}(2011)}]{Regimbau_2011}%
  \BibitemOpen
  \bibfield  {author} {\bibinfo {author} {\bibfnamefont {T.}~\bibnamefont
  {{Regimbau}}},\ }\href {\doibase 10.1088/1674-4527/11/4/001} {\bibfield
  {journal} {\bibinfo  {journal} {Res.~Astron.~Astrophys.}\ }\textbf {\bibinfo
  {volume} {11}},\ \bibinfo {pages} {369} (\bibinfo {year} {2011})}\BibitemShut
  {NoStop}%
\bibitem [{\citenamefont {P\'erigois}\ \emph {et~al.}(2020)\citenamefont
  {P\'erigois}, \citenamefont {Belczynski}, \citenamefont {Bulik},\ and\
  \citenamefont {Regimbau}}]{Perigois:2020ymr}%
  \BibitemOpen
  \bibfield  {author} {\bibinfo {author} {\bibfnamefont {C.}~\bibnamefont
  {P\'erigois}}, \bibinfo {author} {\bibfnamefont {C.}~\bibnamefont
  {Belczynski}}, \bibinfo {author} {\bibfnamefont {T.}~\bibnamefont {Bulik}}, \
  and\ \bibinfo {author} {\bibfnamefont {T.}~\bibnamefont {Regimbau}},\
  }\href@noop {} {\  (\bibinfo {year} {2020})},\ \Eprint
  {http://arxiv.org/abs/2008.04890} {arXiv:2008.04890 [astro-ph.CO]}
  \BibitemShut {NoStop}%
\bibitem [{\citenamefont {Ezquiaga}\ and\ \citenamefont
  {Holz}(2020)}]{Ezquiaga:2020tns}%
  \BibitemOpen
  \bibfield  {author} {\bibinfo {author} {\bibfnamefont {J.~M.}\ \bibnamefont
  {Ezquiaga}}\ and\ \bibinfo {author} {\bibfnamefont {D.~E.}\ \bibnamefont
  {Holz}},\ }\href@noop {} {\  (\bibinfo {year} {2020})},\ \Eprint
  {http://arxiv.org/abs/2006.02211} {arXiv:2006.02211 [astro-ph.HE]}
  \BibitemShut {NoStop}%
\bibitem [{\citenamefont {Sandick}\ \emph
  {et~al.}(2006{\natexlab{b}})\citenamefont {Sandick}, \citenamefont {Olive},
  \citenamefont {Daigne},\ and\ \citenamefont {Vangioni}}]{PhysRevD.73.104024}%
  \BibitemOpen
  \bibfield  {author} {\bibinfo {author} {\bibfnamefont {P.}~\bibnamefont
  {Sandick}}, \bibinfo {author} {\bibfnamefont {K.~A.}\ \bibnamefont {Olive}},
  \bibinfo {author} {\bibfnamefont {F.}~\bibnamefont {Daigne}}, \ and\ \bibinfo
  {author} {\bibfnamefont {E.}~\bibnamefont {Vangioni}},\ }\href {\doibase
  10.1103/PhysRevD.73.104024} {\bibfield  {journal} {\bibinfo  {journal} {Phys.
  Rev. D}\ }\textbf {\bibinfo {volume} {73}},\ \bibinfo {pages} {104024}
  (\bibinfo {year} {2006}{\natexlab{b}})}\BibitemShut {NoStop}%
\bibitem [{\citenamefont {Regimbau}\ \emph {et~al.}(2017)\citenamefont
  {Regimbau}, \citenamefont {Evans}, \citenamefont {Christensen}, \citenamefont
  {Katsavounidis}, \citenamefont {Sathyaprakash},\ and\ \citenamefont
  {Vitale}}]{Regimbau:2016ike}%
  \BibitemOpen
  \bibfield  {author} {\bibinfo {author} {\bibfnamefont {T.}~\bibnamefont
  {Regimbau}}, \bibinfo {author} {\bibfnamefont {M.}~\bibnamefont {Evans}},
  \bibinfo {author} {\bibfnamefont {N.}~\bibnamefont {Christensen}}, \bibinfo
  {author} {\bibfnamefont {E.}~\bibnamefont {Katsavounidis}}, \bibinfo {author}
  {\bibfnamefont {B.}~\bibnamefont {Sathyaprakash}}, \ and\ \bibinfo {author}
  {\bibfnamefont {S.}~\bibnamefont {Vitale}},\ }\href {\doibase
  10.1103/PhysRevLett.118.151105} {\bibfield  {journal} {\bibinfo  {journal}
  {Phys. Rev. Lett.}\ }\textbf {\bibinfo {volume} {118}},\ \bibinfo {pages}
  {151105} (\bibinfo {year} {2017})},\ \Eprint
  {http://arxiv.org/abs/1611.08943} {arXiv:1611.08943 [astro-ph.CO]}
  \BibitemShut {NoStop}%
\bibitem [{\citenamefont {Sachdev}\ \emph {et~al.}(2020)\citenamefont
  {Sachdev}, \citenamefont {Regimbau},\ and\ \citenamefont
  {Sathyaprakash}}]{Sachdev:2020bkk}%
  \BibitemOpen
  \bibfield  {author} {\bibinfo {author} {\bibfnamefont {S.}~\bibnamefont
  {Sachdev}}, \bibinfo {author} {\bibfnamefont {T.}~\bibnamefont {Regimbau}}, \
  and\ \bibinfo {author} {\bibfnamefont {B.}~\bibnamefont {Sathyaprakash}},\
  }\href {\doibase 10.1103/PhysRevD.102.024051} {\bibfield  {journal} {\bibinfo
   {journal} {Phys. Rev. D}\ }\textbf {\bibinfo {volume} {102}},\ \bibinfo
  {pages} {024051} (\bibinfo {year} {2020})},\ \Eprint
  {http://arxiv.org/abs/2002.05365} {arXiv:2002.05365 [gr-qc]} \BibitemShut
  {NoStop}%
\bibitem [{\citenamefont {Martinovic}\ \emph {et~al.}(2020)\citenamefont
  {Martinovic}, \citenamefont {Meyers}, \citenamefont {Sakellariadou},\ and\
  \citenamefont {Christensen}}]{Martinovic:2020hru}%
  \BibitemOpen
  \bibfield  {author} {\bibinfo {author} {\bibfnamefont {K.}~\bibnamefont
  {Martinovic}}, \bibinfo {author} {\bibfnamefont {P.~M.}\ \bibnamefont
  {Meyers}}, \bibinfo {author} {\bibfnamefont {M.}~\bibnamefont
  {Sakellariadou}}, \ and\ \bibinfo {author} {\bibfnamefont {N.}~\bibnamefont
  {Christensen}},\ }\href@noop {} {\  (\bibinfo {year} {2020})},\ \Eprint
  {http://arxiv.org/abs/2011.05697} {arXiv:2011.05697 [gr-qc]} \BibitemShut
  {NoStop}%
\bibitem [{\citenamefont {{Whelan}}\ \emph {et~al.}(2014)\citenamefont
  {{Whelan}}, \citenamefont {{Robinson}}, \citenamefont {{Romano}},\ and\
  \citenamefont {{Thrane}}}]{StochCalUncertainty}%
  \BibitemOpen
  \bibfield  {author} {\bibinfo {author} {\bibfnamefont {J.~T.}\ \bibnamefont
  {{Whelan}}}, \bibinfo {author} {\bibfnamefont {E.~L.}\ \bibnamefont
  {{Robinson}}}, \bibinfo {author} {\bibfnamefont {J.~D.}\ \bibnamefont
  {{Romano}}}, \ and\ \bibinfo {author} {\bibfnamefont {E.~H.}\ \bibnamefont
  {{Thrane}}},\ }\href {\doibase 10.1088/1742-6596/484/1/012027} {\bibfield
  {journal} {\bibinfo  {journal} {Journal of Physics Conference Series}\
  }\textbf {\bibinfo {volume} {484}},\ \bibinfo {eid} {012027} (\bibinfo {year}
  {2014})}\BibitemShut {NoStop}%
\bibitem [{\citenamefont {Sun}\ \emph {et~al.}(2020)\citenamefont {Sun} \emph
  {et~al.}}]{Sun:2020wke}%
  \BibitemOpen
  \bibfield  {author} {\bibinfo {author} {\bibfnamefont {L.}~\bibnamefont
  {Sun}} \emph {et~al.},\ }\href@noop {} {\  (\bibinfo {year} {2020})},\
  \Eprint {http://arxiv.org/abs/2005.02531} {arXiv:2005.02531 [astro-ph.IM]}
  \BibitemShut {NoStop}%
\bibitem [{\citenamefont {Isi}\ and\ \citenamefont
  {Stein}(2018)}]{Isi_Stein_NonGR_Energy}%
  \BibitemOpen
  \bibfield  {author} {\bibinfo {author} {\bibfnamefont {M.}~\bibnamefont
  {Isi}}\ and\ \bibinfo {author} {\bibfnamefont {L.~C.}\ \bibnamefont
  {Stein}},\ }\href {\doibase 10.1103/PhysRevD.98.104025} {\bibfield  {journal}
  {\bibinfo  {journal} {Phys. Rev.}\ }\textbf {\bibinfo {volume} {D98}},\
  \bibinfo {pages} {104025} (\bibinfo {year} {2018})},\ \Eprint
  {http://arxiv.org/abs/1807.02123} {arXiv:1807.02123 [gr-qc]} \BibitemShut
  {NoStop}%
\bibitem [{\citenamefont {{Chru\'{}sli\'{}nska, M.}}\ \emph
  {et~al.}(2020)\citenamefont {{Chru\'{}sli\'{}nska, M.}}, \citenamefont
  {{Jer\'abkov\'a, T.}}, \citenamefont {{Nelemans, G.}},\ and\ \citenamefont
  {{Yan, Z.}}}]{AstroUncertaintyIMF}%
  \BibitemOpen
  \bibfield  {author} {\bibinfo {author} {\bibnamefont {{Chru\'{}sli\'{}nska,
  M.}}}, \bibinfo {author} {\bibnamefont {{Jer\'abkov\'a, T.}}}, \bibinfo
  {author} {\bibnamefont {{Nelemans, G.}}}, \ and\ \bibinfo {author}
  {\bibnamefont {{Yan, Z.}}},\ }\href {\doibase 10.1051/0004-6361/202037688}
  {\bibfield  {journal} {\bibinfo  {journal} {A\&A}\ }\textbf {\bibinfo
  {volume} {636}},\ \bibinfo {pages} {A10} (\bibinfo {year}
  {2020})}\BibitemShut {NoStop}%
\bibitem [{\citenamefont {{L\'opez Fern\'andez, R.}}\ \emph
  {et~al.}(2018)\citenamefont {{L\'opez Fern\'andez, R.}}, \citenamefont
  {{Gonz\'alez Delgado, R. M.}}, \citenamefont {{P\'erez, E.}}, \citenamefont
  {{Garc\'{\i}a-Benito, R.}}, \citenamefont {{Cid Fernandes, R.}},
  \citenamefont {{Schoenell, W.}}, \citenamefont {{S\'anchez, S. F.}},
  \citenamefont {{Gallazzi, A.}}, \citenamefont {{S\'anchez-Bl\'azquez, P.}},
  \citenamefont {{Vale Asari, N.}},\ and\ \citenamefont {{Walcher, C.
  J.}}}]{AstroUncertaintySFRPeak}%
  \BibitemOpen
  \bibfield  {author} {\bibinfo {author} {\bibnamefont {{L\'opez Fern\'andez,
  R.}}}, \bibinfo {author} {\bibnamefont {{Gonz\'alez Delgado, R. M.}}},
  \bibinfo {author} {\bibnamefont {{P\'erez, E.}}}, \bibinfo {author}
  {\bibnamefont {{Garc\'{\i}a-Benito, R.}}}, \bibinfo {author} {\bibnamefont
  {{Cid Fernandes, R.}}}, \bibinfo {author} {\bibnamefont {{Schoenell, W.}}},
  \bibinfo {author} {\bibnamefont {{S\'anchez, S. F.}}}, \bibinfo {author}
  {\bibnamefont {{Gallazzi, A.}}}, \bibinfo {author} {\bibnamefont
  {{S\'anchez-Bl\'azquez, P.}}}, \bibinfo {author} {\bibnamefont {{Vale Asari,
  N.}}}, \ and\ \bibinfo {author} {\bibnamefont {{Walcher, C. J.}}},\ }\href
  {\doibase 10.1051/0004-6361/201732358} {\bibfield  {journal} {\bibinfo
  {journal} {A\&A}\ }\textbf {\bibinfo {volume} {615}},\ \bibinfo {pages} {A27}
  (\bibinfo {year} {2018})}\BibitemShut {NoStop}%
\bibitem [{\citenamefont {{Vangioni}}\ \emph {et~al.}(2015)\citenamefont
  {{Vangioni}} \emph {et~al.}}]{VangioniEA_2015}%
  \BibitemOpen
  \bibfield  {author} {\bibinfo {author} {\bibfnamefont {E.}~\bibnamefont
  {{Vangioni}}} \emph {et~al.},\ }\href {\doibase 10.1093/mnras/stu2600}
  {\bibfield  {journal} {\bibinfo  {journal} {\mnras}\ }\textbf {\bibinfo
  {volume} {447}},\ \bibinfo {pages} {2575} (\bibinfo {year}
  {2015})}\BibitemShut {NoStop}%
\bibitem [{\citenamefont {{Chruslinska}}\ \emph {et~al.}(2019)\citenamefont
  {{Chruslinska}}, \citenamefont {{Nelemans}},\ and\ \citenamefont
  {{Belczynski}}}]{2019MNRAS.482.5012C}%
  \BibitemOpen
  \bibfield  {author} {\bibinfo {author} {\bibfnamefont {M.}~\bibnamefont
  {{Chruslinska}}}, \bibinfo {author} {\bibfnamefont {G.}~\bibnamefont
  {{Nelemans}}}, \ and\ \bibinfo {author} {\bibfnamefont {K.}~\bibnamefont
  {{Belczynski}}},\ }\href {\doibase 10.1093/mnras/sty3087} {\bibfield
  {journal} {\bibinfo  {journal} {\mnras}\ }\textbf {\bibinfo {volume} {482}},\
  \bibinfo {pages} {5012} (\bibinfo {year} {2019})},\ \Eprint
  {http://arxiv.org/abs/1811.03565} {arXiv:1811.03565 [astro-ph.HE]}
  \BibitemShut {NoStop}%
\bibitem [{\citenamefont {Mapelli}\ \emph {et~al.}(2019)\citenamefont
  {Mapelli}, \citenamefont {Giacobbo}, \citenamefont {Santoliquido},\ and\
  \citenamefont {Artale}}]{mapelli_properties_2019}%
  \BibitemOpen
  \bibfield  {author} {\bibinfo {author} {\bibfnamefont {M.}~\bibnamefont
  {Mapelli}}, \bibinfo {author} {\bibfnamefont {N.}~\bibnamefont {Giacobbo}},
  \bibinfo {author} {\bibfnamefont {F.}~\bibnamefont {Santoliquido}}, \ and\
  \bibinfo {author} {\bibfnamefont {M.~C.}\ \bibnamefont {Artale}},\ }\href
  {\doibase 10.1093/mnras/stz1150} {\bibfield  {journal} {\bibinfo  {journal}
  {Monthly Notices of the Royal Astronomical Society}\ } (\bibinfo {year}
  {2019}),\ 10.1093/mnras/stz1150},\ \bibinfo {note} {arXiv:
  1902.01419}\BibitemShut {NoStop}%
\bibitem [{\citenamefont {Miller}\ \emph {et~al.}(2020)\citenamefont {Miller},
  \citenamefont {Callister},\ and\ \citenamefont {Farr}}]{Miller:2020zox}%
  \BibitemOpen
  \bibfield  {author} {\bibinfo {author} {\bibfnamefont {S.}~\bibnamefont
  {Miller}}, \bibinfo {author} {\bibfnamefont {T.~A.}\ \bibnamefont
  {Callister}}, \ and\ \bibinfo {author} {\bibfnamefont {W.}~\bibnamefont
  {Farr}},\ }\href {\doibase 10.3847/1538-4357/ab80c0} {\bibfield  {journal}
  {\bibinfo  {journal} {Astrophys. J.}\ }\textbf {\bibinfo {volume} {895}},\
  \bibinfo {pages} {128} (\bibinfo {year} {2020})},\ \Eprint
  {http://arxiv.org/abs/2001.06051} {arXiv:2001.06051 [astro-ph.HE]}
  \BibitemShut {NoStop}%
\bibitem [{\citenamefont {Foucart}\ \emph {et~al.}(2011)\citenamefont
  {Foucart}, \citenamefont {Duez}, \citenamefont {Kidder},\ and\ \citenamefont
  {Teukolsky}}]{PhysRevD.83.024005}%
  \BibitemOpen
  \bibfield  {author} {\bibinfo {author} {\bibfnamefont {F.}~\bibnamefont
  {Foucart}}, \bibinfo {author} {\bibfnamefont {M.~D.}\ \bibnamefont {Duez}},
  \bibinfo {author} {\bibfnamefont {L.~E.}\ \bibnamefont {Kidder}}, \ and\
  \bibinfo {author} {\bibfnamefont {S.~A.}\ \bibnamefont {Teukolsky}},\ }\href
  {\doibase 10.1103/PhysRevD.83.024005} {\bibfield  {journal} {\bibinfo
  {journal} {Phys. Rev. D}\ }\textbf {\bibinfo {volume} {83}},\ \bibinfo
  {pages} {024005} (\bibinfo {year} {2011})}\BibitemShut {NoStop}%
\bibitem [{\citenamefont {Kyutoku}\ \emph {et~al.}(2010)\citenamefont
  {Kyutoku}, \citenamefont {Shibata},\ and\ \citenamefont
  {Taniguchi}}]{PhysRevD.82.044049}%
  \BibitemOpen
  \bibfield  {author} {\bibinfo {author} {\bibfnamefont {K.}~\bibnamefont
  {Kyutoku}}, \bibinfo {author} {\bibfnamefont {M.}~\bibnamefont {Shibata}}, \
  and\ \bibinfo {author} {\bibfnamefont {K.}~\bibnamefont {Taniguchi}},\ }\href
  {\doibase 10.1103/PhysRevD.82.044049} {\bibfield  {journal} {\bibinfo
  {journal} {Phys. Rev. D}\ }\textbf {\bibinfo {volume} {82}},\ \bibinfo
  {pages} {044049} (\bibinfo {year} {2010})}\BibitemShut {NoStop}%
\bibitem [{\citenamefont {Kawaguchi}\ \emph {et~al.}(2015)\citenamefont
  {Kawaguchi}, \citenamefont {Kyutoku}, \citenamefont {Nakano}, \citenamefont
  {Okawa}, \citenamefont {Shibata},\ and\ \citenamefont
  {Taniguchi}}]{Kawaguchi:2015bwa}%
  \BibitemOpen
  \bibfield  {author} {\bibinfo {author} {\bibfnamefont {K.}~\bibnamefont
  {Kawaguchi}}, \bibinfo {author} {\bibfnamefont {K.}~\bibnamefont {Kyutoku}},
  \bibinfo {author} {\bibfnamefont {H.}~\bibnamefont {Nakano}}, \bibinfo
  {author} {\bibfnamefont {H.}~\bibnamefont {Okawa}}, \bibinfo {author}
  {\bibfnamefont {M.}~\bibnamefont {Shibata}}, \ and\ \bibinfo {author}
  {\bibfnamefont {K.}~\bibnamefont {Taniguchi}},\ }\href {\doibase
  10.1103/PhysRevD.92.024014} {\bibfield  {journal} {\bibinfo  {journal} {Phys.
  Rev. D}\ }\textbf {\bibinfo {volume} {92}},\ \bibinfo {pages} {024014}
  (\bibinfo {year} {2015})},\ \Eprint {http://arxiv.org/abs/1506.05473}
  {arXiv:1506.05473 [astro-ph.HE]} \BibitemShut {NoStop}%
\bibitem [{\citenamefont {Abbott}\ \emph
  {et~al.}(2018{\natexlab{e}})\citenamefont {Abbott} \emph
  {et~al.}}]{Aasi:2013wya}%
  \BibitemOpen
  \bibfield  {author} {\bibinfo {author} {\bibfnamefont {B.}~\bibnamefont
  {Abbott}} \emph {et~al.} (\bibinfo {collaboration} {KAGRA Collabration, LIGO
  Scientific Collaboration, and VIRGO Collaboration}),\ }\href {\doibase
  10.1007/s41114-018-0012-9} {\bibfield  {journal} {\bibinfo  {journal} {Living
  Rev. Rel.}\ }\textbf {\bibinfo {volume} {21}},\ \bibinfo {pages} {3}
  (\bibinfo {year} {2018}{\natexlab{e}})},\ \Eprint
  {http://arxiv.org/abs/1304.0670} {arXiv:1304.0670 [gr-qc]} \BibitemShut
  {NoStop}%
\bibitem [{\citenamefont {Barsotti}\ \emph
  {et~al.}({\natexlab{b}})\citenamefont {Barsotti}, \citenamefont {Fritschel},
  \citenamefont {Evans},\ and\ \citenamefont {Gras}}]{new_design}%
  \BibitemOpen
  \bibfield  {author} {\bibinfo {author} {\bibfnamefont {L.}~\bibnamefont
  {Barsotti}}, \bibinfo {author} {\bibfnamefont {P.}~\bibnamefont {Fritschel}},
  \bibinfo {author} {\bibfnamefont {M.}~\bibnamefont {Evans}}, \ and\ \bibinfo
  {author} {\bibfnamefont {S.}~\bibnamefont {Gras}},\ }\href@noop {} {}\bibinfo
  {howpublished} {\url{https://dcc.ligo.org/T1800044-v5/public}}
  ({\natexlab{b}})\BibitemShut {NoStop}%
\bibitem [{\citenamefont {{Akutsu}}\ \emph {et~al.}(2020)\citenamefont
  {{Akutsu}}, \citenamefont {{Ando}}, \citenamefont {{Arai}}, \citenamefont
  {{Arai}}, \citenamefont {{Araki}} \emph {et~al.}}]{kagra_ref}%
  \BibitemOpen
  \bibfield  {author} {\bibinfo {author} {\bibfnamefont {T.}~\bibnamefont
  {{Akutsu}}}, \bibinfo {author} {\bibfnamefont {M.}~\bibnamefont {{Ando}}},
  \bibinfo {author} {\bibfnamefont {K.}~\bibnamefont {{Arai}}}, \bibinfo
  {author} {\bibfnamefont {Y.}~\bibnamefont {{Arai}}}, \bibinfo {author}
  {\bibfnamefont {S.}~\bibnamefont {{Araki}}},  \emph {et~al.},\ }\href@noop {}
  {\bibfield  {journal} {\bibinfo  {journal} {arXiv e-prints}\ ,\ \bibinfo
  {eid} {arXiv:2005.05574}} (\bibinfo {year} {2020})},\ \Eprint
  {http://arxiv.org/abs/2005.05574} {arXiv:2005.05574 [physics.ins-det]}
  \BibitemShut {NoStop}%
\bibitem [{\citenamefont {Iyer}\ \emph {et~al.}(2011)\citenamefont {Iyer},
  \citenamefont {Souradeep}, \citenamefont {Unnikrishnan}, \citenamefont
  {Dhurandhar}, \citenamefont {Raja}, \citenamefont {Kumar},\ and\
  \citenamefont {Sengupta}}]{iyer_proposal_2011}%
  \BibitemOpen
  \bibfield  {author} {\bibinfo {author} {\bibfnamefont {B.}~\bibnamefont
  {Iyer}}, \bibinfo {author} {\bibfnamefont {T.}~\bibnamefont {Souradeep}},
  \bibinfo {author} {\bibfnamefont {C.~S.}\ \bibnamefont {Unnikrishnan}},
  \bibinfo {author} {\bibfnamefont {S.}~\bibnamefont {Dhurandhar}}, \bibinfo
  {author} {\bibfnamefont {S.}~\bibnamefont {Raja}}, \bibinfo {author}
  {\bibfnamefont {A.}~\bibnamefont {Kumar}}, \ and\ \bibinfo {author}
  {\bibfnamefont {A.}~\bibnamefont {Sengupta}},\ }\href
  {https://dcc.ligo.org/M1100296/} {\bibfield  {journal} {\bibinfo  {journal}
  {LIGO-India Technical Report No. LIGO-M1100296}\ } (\bibinfo {year}
  {2011})}\BibitemShut {NoStop}%
\bibitem [{\citenamefont {{Hunter}}(2007)}]{2007CSE.....9...90H}%
  \BibitemOpen
  \bibfield  {author} {\bibinfo {author} {\bibfnamefont {J.~D.}\ \bibnamefont
  {{Hunter}}},\ }\href {\doibase 10.1109/MCSE.2007.55} {\bibfield  {journal}
  {\bibinfo  {journal} {CSE}\ }\textbf {\bibinfo {volume} {9}},\ \bibinfo
  {pages} {90} (\bibinfo {year} {2007})}\BibitemShut {NoStop}%
\end{thebibliography}%

\iftoggle{endauthorlist}{
  \let\author\myauthor
  \let\affiliation\myaffiliation
  \let\maketitle\mymaketitle
  \title{The LIGO Scientific Collaboration, Virgo Collaboration, and KAGRA Collaboration  }
  \pacs{}


\author{R.~Abbott}
\affiliation{LIGO Laboratory, California Institute of Technology, Pasadena, CA 91125, USA}
\author{T.~D.~Abbott}
\affiliation{Louisiana State University, Baton Rouge, LA 70803, USA}
\author{S.~Abraham}
\affiliation{Inter-University Centre for Astronomy and Astrophysics, Pune 411007, India}
\author{F.~Acernese}
\affiliation{Dipartimento di Farmacia, Universit\`a di Salerno, I-84084 Fisciano, Salerno, Italy  }
\affiliation{INFN, Sezione di Napoli, Complesso Universitario di Monte S.Angelo, I-80126 Napoli, Italy  }
\author{K.~Ackley}
\affiliation{OzGrav, School of Physics \& Astronomy, Monash University, Clayton 3800, Victoria, Australia}
\author{A.~Adams}
\affiliation{Christopher Newport University, Newport News, VA 23606, USA}
\author{C.~Adams}
\affiliation{LIGO Livingston Observatory, Livingston, LA 70754, USA}
\author{R.~X.~Adhikari}
\affiliation{LIGO Laboratory, California Institute of Technology, Pasadena, CA 91125, USA}
\author{V.~B.~Adya}
\affiliation{OzGrav, Australian National University, Canberra, Australian Capital Territory 0200, Australia}
\author{C.~Affeldt}
\affiliation{Max Planck Institute for Gravitational Physics (Albert Einstein Institute), D-30167 Hannover, Germany}
\affiliation{Leibniz Universit\"at Hannover, D-30167 Hannover, Germany}
% ...

\collaboration{The LIGO Scientific Collaboration, the Virgo Collaboration, and the KAGRA Collaboration}



  \newpage
  \maketitle
}

\end{document}